\documentclass[amsmath,amssymb,11pt]{article}
\usepackage{jheppub2}  
\pdfoutput=1

\usepackage{graphicx}     
\usepackage{multirow}
\usepackage{booktabs}
\usepackage{float}        
\usepackage{subcaption}   
\usepackage{caption}      
\usepackage[utf8]{inputenc} 
\usepackage{hyperref}     
\usepackage{amsmath, amssymb, amsthm} 

\usepackage{xcolor}
\definecolor{cherryblossompink}{rgb}{1.0, 0.72, 0.77}
\definecolor{lightblue}{rgb}{0.68, 0.85, 0.9}

\usepackage{esint}
\usepackage{mathrsfs}
\usepackage{mathtools}
\usepackage{amsfonts}
\usepackage{lmodern}
\usepackage{cancel}
\usepackage{accents}
\usepackage{soul}
\usepackage[titletoc,title]{appendix}
\setstcolor{red}

\usepackage{tikz}
\usetikzlibrary{positioning}
\usetikzlibrary{intersections}
\usetikzlibrary{fadings} 
\usetikzlibrary{arrows.meta} 
\usetikzlibrary{arrows}
\usetikzlibrary{decorations.pathmorphing}
\usetikzlibrary{decorations.pathreplacing,decorations.markings}
\usetikzlibrary{backgrounds,automata}

\newcommand{\doubletilde}[1]{\accentset{\approx}{#1}}

\newcommand{\beq}{\begin{equation}}
\newcommand{\eeq}{\end{equation}}
\newcommand{\be}{\begin{equation}}
\newcommand{\ee}{\end{equation}}
\newcommand{\bea}{\begin{eqnarray}}
\newcommand{\eea}{\end{eqnarray}}
\newcommand*{\affmark}[1][*]{\textsuperscript{#1}}

\title{Stochastic dark matter: Covariant Brownian motion from Planckian discreteness}
\author{Emma Albertini,\affmark[1]}
\emailAdd{emma.albertini17@imperial.ac.uk}
\author{Arad Nasiri,\affmark[1]}
\emailAdd{a.nasiri21@imperial.ac.uk}

\author{Emanuele Panella\affmark[2]}
\emailAdd{emanuele.panella.21@ucl.ac.uk}

\affiliation{\affmark[1]
Blackett Laboratory, Imperial College London, SW7 2AZ, UK\\
 \affmark[2] Department of Physics and Astronomy, University College London,\\
Gower Street, London, WC1E 6BT, UK\\}

\abstract{Quantum gravity has long remained elusive from an observational standpoint. Developing effective cosmological models motivated by the fundamental aspects of quantum gravity is crucial for bridging theory with observations. One key aspect is the granularity of spacetime, which suggests that free particles would deviate from classical geodesics by following a covariant Brownian motion. This notion is further supported by swerves models in causal set theory, a discrete approach to quantum gravity. At an effective level, such deviations are described by a stochastic correction to the geodesic equation. We show that the form of this correction is strictly restricted by covariance and the mass-shell condition. Under minimal coupling to curvature, the resulting covariant Brownian motion is unique. The process is equivalently described by a covariant diffusion equation for the distribution of massive particles in their relativistic phase space. When applied to dark matter particles, covariant Brownian motion results in spontaneous warming at late times, suppressing the matter power spectrum at small scales in a time-dependent manner. Using bounds on the diffusion rate from CMB and growth history measurements of $f\sigma_8$, we show that the model offers a resolution to the $S_8$ tension. Future studies on the model's behavior at non-linear cosmological scales will provide further constraints and, therefore, critical tests for the viability of stochastic dark matter.}

\begin{document}

\maketitle

\section{Introduction}

The search for a quantum theory of gravity hinges on a deeper understanding of the nature of spacetime on the one hand and uncovering its quantum dynamics on the other. Regarding kinematics, a key concept in several quantum gravity proposals is the idea of spacetime discreteness~\citep{williams2006discrete}. In these theories, the smooth spacetime of general relativity is viewed as an effective, coarse-grained approximation of a fundamental discrete structure. Several theoretically compelling reasons support this view. Apart from the immediate reason of the mere existence of Planck length as a distinguished physical scale, discreteness provides a natural explanation for the entropy of black holes based on either counting the microscopic discrete states \citep{krasnov1997counting,machet2021horizon} or computing the entanglement entropy \citep{sorkin2018entanglement}. Moreover, in sum-over-histories quantization, a countable summation over kinematically discrete degrees of freedom is well-defined mathematically and more tractable than continuum path integrals \citep{loll2019quantum,loomis2017suppression}. A discrete spacetime also naturally imposes a UV cutoff in field theories, turning perturbation series into finite sums \citep{dable2020algebraic,albertini2024correlators}.

However, a major challenge in the search for quantum gravity in general and discrete spacetime in particular is the lack of direct empirical evidence. The Planck energy scale is by many orders of magnitude beyond the reach of current collider experiments. Yet, small quantum gravity effects could accumulate and amplify over cosmic distances or across the age of the universe leaving observable traces. In the era of precision cosmology, this opens a unique window for quantum gravity phenomenology \citep{addazi2022quantum}. In addition to open theoretical problems like the nature of dark matter and dark energy, current cosmology faces tensions between different measurements of $H_0$ and $S_8$ within the standard $\Lambda$CDM model \citep{riess2022comprehensive,di2021realm,khalife2024review,amon2022dark,secco2022dark,li2023kids,anchordoqui2021dissecting,vagnozzi2023seven,abdalla2022cosmology}.  If these discrepancies are signs of new physics, they provide clear observable targets for well-motivated phenomenological models arising from quantum gravity to provide alternative explanations. Here, our primary focus will be on the $S_8$ tension—a discrepancy that arises when comparing the variance of matter overdensities at the scale of $8h^{-1} \text{Mpc}$ as measured from CMB \citep{aghanim2020planck} with measurements at lower redshifts from probes such as weak lensing \citep{amon2022dark,secco2022dark,li2023kids}, galaxy power spectrum \citep{ivanov2020cosmological,ivanov2021cosmological,philcox2022boss}, galaxy cluster counts \citep{pratt2019galaxy}, or redshift-space distortion \citep{benisty2021quantifying,nunes2021arbitrating}.\\

How could the discreteness of spacetime have any observable signature? Consider the analogy with the atomic theory of matter. Brownian motion, linked to kinetic theory and the diffusion equation by Einstein in 1905, was key evidence for the atomic nature of matter. The stochastic motion of test particles, whether understood through the behavior of a single one or the diffusion of many, can be effectively described by just two free parameters: the diffusion coefficient and the friction coefficient. No detailed knowledge of the microscopic dynamics of the background fluid is needed. The inherent stochasticity of the Brownian motion reflects our ignorance of the detailed microscopic configuration of the fluid's atoms or molecules. 

Similarly, one can propose the Brownian motion of massive particles in the vacuum of general relativity as a manifestation of underlying spacetime discreteness. In a smooth spacetime manifold, free massive particles follow timelike geodesics. However, if this manifold is merely the continuum limit of a fundamentally discrete spacetime, the geodesic equation would be modified, causing the particles to ‘swerve’. Pictorially, as the particles jump from one spacetime atom to another, they will almost surely not find a spacetime atom exactly along the continuum timelike geodesic and will therefore end up on the next closest point. This leads to diffusion around the continuum-limit timelike geodesic. In 2003, Dowker et al. proposed a concrete model in the framework of causal set theory, a Lorentz-preserving discrete approach to quantum gravity, to explore this idea \cite{dowker2004quantum}. Later in 2008, Dowker et al. added two intrinsic models on a causal set that encompassed the idea \citep{philpott2009energy}. These models are collectively called the \textit{swerves} models. They all feature a non-local scale, called the forgetting time~$\tau_f$. This is the amount of proper time that a massive particle “remembers" from its past motion, so as to approximately maintain its motion on a geodesic. It is distinct from other typical non-local scales in the sense that the geodesic motion is recovered in the $\tau_f \to \infty$ limit. These models, however, should be taken with a grain of salt, as the authors themselves noted that a point particle is not an adequate approximation when quantum gravity effects are significant. Rather, they serve as a first approximation to the more complicated problem of the stochastic dynamics of a quantum field due to spacetime discreteness. Regardless of the exact fundamental discrete degrees of freedom or their interaction with quantum matter, the effective description of a single particle's motion or the distribution of many particles in the continuum approximation should depend only on a few parameters.\\

One of the earliest works on relativistic Brownian motion was by Dudley in 1965~\citep{dudley1966lorentz}, where it was shown that no Lorentz-invariant stochastic process exists in Minkowski spacetime. One way to understand this is to note that the metric $\eta_{\mu\nu}$ that appears in the flat Laplacian is not positive semi-definite, meaning that it cannot be used as the diffusion tensor for a stochastic process. The resolution is to include the momentum variables in the state of the particle and define the diffusion process on the mass shell in momentum space instead. This allowed the proposal of a diffusion equation on the mass shell valid for a spatially uniform distribution of particles. In \citep{dowker2004quantum}, Dowker et al. used Sorkin's 1985 formalism of stochastic evolution on the manifold of states \cite{sorkin1986stochastic} to derive a Lorentz invariant diffusion equation for relativistic particles on the mass shell, with only one free parameter: the diffusion constant~$\kappa$. This generalized Dudley's equation to spatially inhomogeneous distributions. They highlighted the role of proper time elapsed along the worldline of the particle as the evolution parameter for the distribution function. They also placed an upper bound on the diffusion parameter from cold Hydrogen gas in laboratory settings. In \citep{kaloper2006low}, Kaloper and Mattingly found that the observations of cold molecular clouds in the Milky Way, Bose-Einstein condensate experiments with 87-Rubidium, and stability of old heavy nuclei put similar bounds on the diffusion parameter. Later in \citep{philpott2009energy}, Dowker et al. elaborated on why global Lorentz symmetry is so restrictive, essentially making the Lorentz invariant diffusion equation unique. The authors also considered massless particles, where key differences arise compared to massive particles when Lorentz invariance is enforced. For massless particles, the diffusion does not happen on the whole light cone. Instead, the distribution broadens and drifts in one dimension along the original momentum, without changing direction. The two effects are controlled by two independent free parameters. To test this, they applied the massless diffusion equation to CMB photons, adding a Hubble friction term by hand to extend it to FRW spacetime. Using the blackbody spectrum of CMB photons, they placed upper bounds on the two free parameters.
Later in \cite{Contaldi_2010}, Contaldi et al. used the CMB polarization power spectrum to set upper bounds on a Lorentz invariant diffusion model of photon polarization. \\

In this paper, we delve into both foundational and cosmological aspects of quantum-gravity-motivated stochastic motion. In the first part, we generalize previous work to curved spacetime from both the distribution and single-particle perspectives. To this end, we formulate the notion of general covariance and admissible coordinate transformations on the phase space of massive particles in curved spacetime. The Fokker-Planck equation, in its general form,  governs the deterministic evolution of the distribution of particles undergoing stochastic motion. Demanding the Fokker-Planck equation to be covariant under admissible coordinate transformations, we derive a \textit{covariant diffusion equation} in phase space. This equation extends the Lorentz invariant diffusion equation of \citep{dowker2004quantum} to arbitrary curved spacetimes and can also be seen as the generalization of the Boltzmann equation to incorporate diffusion on the mass shell. This derivation is detailed in section \ref{SorkinStochastic}.

An alternative perspective on diffusion processes is through Brownian motion of individual particles where the equations of motion are no longer deterministic but are governed by Langevin or stochastic differential equations (SDE). In section \ref{SDE_section}, we begin with a stochastic correction to the geodesic equation. Guided by first principles including general covariance, mass conservation, and the physical requirement that particles do not travel back in time, we derive a \textit{covariant Brownian motion}. In Appendix \ref{app:unravelling}, we show that these two equations—the covariant diffusion equation and the covariant Brownian motion—are equivalent and describe the same underlying stochastic process. The SDE formulation has the virtue of revealing the extent to which this process is unique. While Lorentz invariance ensures uniqueness in flat spacetime, curvature could couple to the Brownian motion. Nevertheless, under minimal coupling, the equations retain their uniqueness, governed solely by the diffusion parameter $\kappa$.

We highlight that although our motivation stems from considerations in causal set theory, the resulting covariant diffusion equation is independent of causal sets. Similar to the Brownian motion in fluids, the covariant Brownian motion is an effective theory that is agnostic to the UV-complete theory of quantum gravity. Due to its uniqueness, the minimally coupled covariant diffusion can describe any Lorentz-preserving stochastic correction to the motion of point particles, regardless of the underlying microscopic description—be it a covariant effect from loop quantum gravity, quantum foam, spacetime uncertainty, or spontaneous collapse models. This versatility makes it a powerful phenomenological model, serving as the effective description of a range of possible quantum gravity effects. 
\\

In the second part of the paper, we examine the cosmological implications. For the covariant Brownian motion to have an observable effect, the particles must experience the longest possible proper time. Therefore, we aim to apply the covariant diffusion equation to massive particles in cosmology. While in principle, all massive particles are subject to the covariant Brownian motion, each with its own diffusion constant, our focus is on dark matter particles, the most dominant matter species. This leads to a model of stochastic dark matter. Since the covariant Brownian motion spontaneously heats up the particles, we anticipate stochastic dark matter to have implications for the $S_8$ tension.

The covariant diffusion equation generalizes the Boltzmann equation, modifying both the background and perturbation equations for the stochastic dark matter. In section \ref{GBH_section}, we outline the use of the generalized Boltzmann hierarchy for numerically solving these equations. Intuitively, we expect the diffusion of dark matter particles to counteract clustering and reduce late-time structure growth at small scales. Using the semi-analytic solutions to the distribution function in section \ref{section_semi_analytic}, we confirm this intuition. We show that for certain values of the diffusion parameter, the effect remains negligible in the background, yet small-scale perturbations can be significantly damped. Consequently, the matter power spectrum is suppressed at small scales in a redshift-dependent manner, with stronger suppression at lower redshifts. 

A question left unresolved in previous works is the source of energy-momentum non-conservation in swerves models and their effective Lorentz invariant diffusion equation. To address this, we take a pragmatic approach by introducing an imperfect dark energy fluid with heat flux to maintain consistency within Einstein equations, as detailed in section \ref{DEmaintext}.  

Finally, in section \ref{mcmc_section}, we show that the specificity of the diffusion to small scales and low redshifts enables stochastic dark matter to satisfy CMB constraints as well as constraints on the expansion and growth histories from BAO, Ly-$\alpha$, and uncalibrated SN Ia while reducing and broadening the $S_8$ posterior from CMB to make it consistent with weak lensing measurements. Interestingly, using an $S_8$ prior from the Dark Energy Survey (DES) introduces a $1.7\sigma$ preference for a non-zero diffusion rate. We end with a discussion on possible implications and future directions.

\section{Foundation}

\subsection{From Fokker-Planck to Covariant Diffusion Equation}
\label{SorkinStochastic}

As mentioned in the introduction, our motivation is the stochastic motion of point particles due to spacetime discreteness. However, all our treatment is an effective description in the continuum, independent of the microscopic details of any discrete theory of quantum gravity. Our key physical principle is spacetime coordinate independence, and local Lorentz invariance on the tangent space at each spacetime point. This is unlike many Lorentz-invariance-violating models in quantum gravity phenomenology \citep{addazi2022quantum}. In addition, as in swerves models \citep{philpott2009energy}, the quantum gravity description of the stochastic motion most probably involves a non-locality scale $\tau_f$. In our effective description, however, we assume sufficient coarse-graining, so that we are working at scales where the dynamics is Markovian or local. Our discussion parallels the arguments in \citep{philpott2009energy} but is generalized to arbitrary curved spacetimes.

Stochastic processes can be described in terms of the evolution law for their distribution in phase space. Let the coordinates on the phase space $\mathcal{S}$ be $Z^A$ and $\rho(Z,\tau)$ be the distribution density of particles in this phase space, where $\tau$ is the time evolution parameter of $\rho$, which in this work we take to be the proper time of the particles. The most general evolution of the density $\rho$, assuming continuity in phase space and Markovianity, is governed by the Fokker-Planck equation
\begin{equation}
\label{eq:Fokker-Planck}
     \partial_\tau \rho=-\partial_A \Big(v^A\rho\Big)+\partial_A\partial_B\bigg(K^{AB}\rho\bigg),
\end{equation}
where the first term on the right accounts for the transport of the particles according to some least-action principle, and the second term accounts for diffusion around their least-action trajectories. $v^A(Z)$ and $K^{AB}(Z)$ 
are one-index and two-index objects living in the phase space that have to be determined for any particular dynamics. It is a standard result of stochastic processes that $v^A(Z)$ and $K^{AB}(Z)$ are related to the mean and covariance of the motion of individual particles \citep{garcia2007introduction}:
\begin{align}
\label{jump_moment_eqs}
   v^A &= \lim_{\delta \tau \to 0}\frac{1}{\delta\tau}\Big\langle \delta Z^A\Big\rangle \notag \\
   K^{AB} &=\lim_{\delta \tau \to 0} \frac{1}{2\delta\tau}\Big\langle \delta Z^A \delta Z^B\Big\rangle.
\end{align}
See an alternative derivation based on \citep{sorkin1986stochastic} in Appendix \ref{app:equiv_itoFP}.

The total particle-number conservation is implicitly assumed in \eqref{eq:Fokker-Planck} and follows from the conservation of probability for the forward Kolmogorov equation since, when normalized by the total number of particles, $\rho$ equivalently acts as a probability distribution on phase space. Taking $v^A=0$ and the diffusion tensor to be the identity matrix, \eqref{eq:Fokker-Planck} is just the familiar heat equation. A derivation of the above Fokker-Planck equation based on first physical principles is outlined in Appendix \ref{appendix_stochasticev}.

Looking at the equation \eqref{eq:Fokker-Planck}, a critical question that comes into mind is what are the transformation rules for $v^A$ and $K^{AB}$ under coordinate transformations in phase space $Z^A\rightarrow Z'^M(Z)$. This can be answered by noting that $\rho$ is a scalar density, as discussed in Appendix \ref{appendix_stochasticev}. As noted by Graham \citep{graham1977path} and Sorkin~\citep{sorkin1986stochastic} this implies that $v^A$ is a pseudo-vector. When transformed, it mixes with $K^{AB}$:
\begin{equation}
\label{v_trans_rule}
    v'^M=\frac{\partial Z'^M}{\partial Z^A}v^A+\frac{\partial^2 Z'^M}{\partial Z^A\partial Z^B}K^{AB}.
\end{equation}
On the other hand, surprisingly, $K^{AB}$ turns out to be a tensor on the phase space:
\begin{equation}
\label{K_trans_rule}
K'^{MN}=\frac{\partial Z'^M}{\partial Z^A}\frac{\partial Z'^N}{\partial Z^B}K^{AB}.
\end{equation}
The transformation properties of the drift and diffusion tensor provide a key guiding principle when searching for a covariant description of diffusion in spacetime. Indeed, the derivation reduces to identifying the phase space of massive particles on a curved spacetime and then finding the most general pseudo-vector $v^{A}$ and tensor $K^{AB}$ that live in this phase space.

Before tackling this problem, it is helpful to replace $v^A$ with an actual vector on the phase space. The extra term in~\eqref{v_trans_rule} has similarities with the transformation rule of the partial derivatives of a metric tensor. Indeed, it turns out that if we assume $\hat{g}_{AB}$ to be a given metric on the phase space with determinant $\hat{g}$, then 
\begin{equation}
\label{u_vector}
    u^A=v^A-\frac{1}{\sqrt{|\hat{g}|}}\partial_B\bigg(K^{AB}\sqrt{|\hat{g}|}\bigg)
\end{equation}
is an actual vector on phase space. In other words, under general coordinate transformations of phase space, it transforms as
\begin{equation}
\label{u_trans_rule}
    u'^M=\frac{\partial Z'^M}{\partial Z^A}u^A.
\end{equation}
In \citep{sorkin1986stochastic}, Sorkin uses the entropy density $s$ instead of $\log\sqrt{|\hat{g}|}$ to define $u$. To find the most general vector $u^A$ and tensor $K^{AB}$, we first specify the phase space for particles of mass $m$ in curved spacetime. In classical physics, phase space consists of three spatial coordinates and three momentum coordinates, with time as the evolution parameter. Here, however, we want to preserve general covariance, so following \citep{dowker2004quantum,philpott2009energy}, we use the proper time of the particles $\tau$ as the evolution parameter and take all the spacetime coordinates $x^\mu$ of a spacetime $\mathcal{M}$ as part of the phase space. Momentum coordinates can be taken to be either vectors or 1-forms living on spacetime. For convenience, in this section, we follow \citep{acuna2022introduction}, and we choose the latter option with lower indices. Therefore, the phase space can be parametrized by $(x^\mu,p_\nu)$, where $p_\nu$ is a one-form at point $x^\mu$. The set of all such points forms the cotangent bundle of the spacetime. However, we are only interested in the momenta that satisfy the on-shell condition; so we restrict ourselves to a 7-dimensional subspace of the cotangent bundle \citep{acuna2022introduction}:

\begin{equation}
\label{Gamma_m_plus}
    \Gamma_m^+=\Big\{(x,p)\in T^*\mathcal{M}\ \Big|\ g^{\mu\nu}(x)p_\mu p_\nu=-m^2,\ p^0>0\Big\}.
\end{equation}
This is our desired phase space $\mathcal{S}$.

To find vectors and tensors on this phase space, we need to first discuss coordinate transformations. The transformation rules \eqref{u_trans_rule} and \eqref{K_trans_rule} were written for arbitrary coordinate transformations of the phase space. However, a coordinate transformation on phase space that mixes the spacetime and momentum labels of the cotangent bundle arbitrarily is not physically meaningful. By construction, the above phase space has spacetime as its base manifold, separate from the momentum coordinates. We cannot use momentum coordinates to label spacetime points in a new coordinate system. Therefore, we do not require our equations to be form-invariant under every possible transformation. 

First, we require an \textit{admissible} coordinate transformation to map a spacetime point $x^\mu$ to $x'^\alpha(x)$, without any mixing with momentum coordinates. This induces a natural transformation on 1-forms via the push-forward on tangent spaces. On top of that, we can also allow for reparametrizations of the tangent (and cotangent) spaces of the spacetime. The only structure that has to be preserved is the linearity of the tangent space as a vector space. This means that we should include $x$-dependent Lorentz transformations of the tangent spaces. To formulate this idea, let us use another parametrization of the phase space \eqref{Gamma_m_plus} in terms of \textit{physical momentum} $q_a$ rather than spacetime momentum $p_\mu$. Remember that any physical measurement of the momentum of a particle is performed using an orthonormal tetrad field $e_a^{\ \mu}(x)$, $a=0,...,4$ that plays the role of a lab frame. The measured momentum is the projection of the spacetime momentum $p$ on this tetrad, given by $q_a=e_a^{\ \mu}(x)p_\mu=p.e_a$. After restricting to the mass shell, the physical momentum would live on the 3-dimensional hyperbolic space:

\begin{equation}
\label{H3_def}
    \mathbb{H}_3=\Big\{q\ \Big|\ \eta^{ab}q_aq_b=-m^2, q^0>0\Big\}.
\end{equation}

Note that $\mathbb{H}_3$ implicitly depends on $m$, but for the sake of simplicity, we do not include it in the notation. One can use $(x^\mu,q_a)$, $a=0,...,3$ as a coordinate system on $\Gamma_m^+$, but then it should be remembered that one of the momentum variables is redundant due to the mass-shell constraint. Instead, we simply restrict to their spatial components and use $(x^\mu,q_i)$, $i=1,2,3$, as the coordinates of the phase space. 

Now, a local ($x$-dependent) Lorentz transformation can be simply stated as a change of the orthonormal tetrad field $e_a\rightarrow\Tilde{e}_r$, $r=0,...,3$. The new momentum coordinates $\Tilde{q}_r$ would be related to the old ones by a Lorentz transformation at each spacetime point. And once again, we restrict to the spatial components of the momentum $\Tilde{q}_I$, $I=1,2,3$. In summary, the most general admissible coordinate transformation of the phase space is
\begin{equation}
\label{coord_trans}
    \big(x^\mu,q_i\big)\rightarrow\Big(x'^\alpha(x),\ \Tilde{q}_I(x,q)=\Tilde{e}_I(x).e^a(x)\ q_a\Big).
\end{equation}
It is easy to check that these transformations form a subgroup of the diffeomorphism group of the phase space.

In the spirit of general relativity, we require $u^A$ and $K^{AB}$ to transform like vectors and tensors, respectively, and to be form-invariant only under the set of admissible coordinate transformations. Consider first the diffusive part. Equation \eqref{jump_moment_eqs} can be rewritten as
\begin{equation}
    K^{\mu B} =\lim_{\delta \tau \to 0} \frac{1}{2 \delta\tau}\Big\langle \delta x^\mu \delta Z^B\Big\rangle.
\end{equation}
Here $B$ can be any of the spacetime or momentum indices. Now $dx^\mu$ is of order $d\tau$, and in the limit $d\tau\rightarrow 0$, the differential increment $dZ^B$ goes to zero. Therefore $K^{\mu B}=0$. This is essentially the same argument as in \citep{philpott2009energy}. One might wonder why the same argument does not conclude that any $K^{AB}$ has to be zero. The reason is that there are noise terms present in $dq_i$ that are not of the order $d\tau$. In other words, we are assuming that the stochastic noise terms do not directly act on $dx^\mu$. This assumption will be justified in the next section by demanding the particles not travel back in time.

The only non-zero components of $K^{AB}$ are therefore $K^{ij}$, where $i,j=1,2,3$ are the momentum indices. For these components, \eqref{K_trans_rule} reduces to \footnote{Note that since $q_a$ lives in the Minkowski spacetime, $q_i=q^i$ and $-q_0=q^0=E$.}
\begin{equation}
    K'^{IJ}=\frac{\partial \Tilde{q}^I}{\partial q^i}\frac{\partial \Tilde{q}^J}{\partial q^j}K^{ij}.
\end{equation}
This is just the statement that $K^{ij}$ should be a tensor on $\mathbb{H}_3$ under Lorentz transformations. As argued in \citep{philpott2009energy}, one obvious choice is the hyperbolic inverse metric on $\mathbb{H}_3$, $h^{ij}$, that is induced from the 4-dimensional Minkowski space where $\mathbb{H}_3$ is embedded. $\mathbb{H}_3$ is a maximally symmetric Riemannian manifold, so all other naturally defined geometric 2-tensors on it, like the Ricci or Einstein tensors, are proportional to its metric tensor. Therefore, $K_{ij}$ has to be proportional to $h_{ij}$. More explicitly, for our choice of parametrization, the metric tensor and its inverse are
\begin{align}
    h_{ij}&=\delta_{ij}-\frac{q_iq_j}{E^2},\\
    h^{ij}&=\delta^{ij}+\frac{q^iq^j}{m^2},
\end{align}
and we have $K^{ij}=\kappa h^{ij}$, where $\kappa$ is a free parameter independent of momentum. So far, we have concluded that the diffusion tensor has to have the following form:

\begin{equation}
\label{K_tensor}
    K=\begin{pmatrix}
  0 & 0 \\
  0 & K^{ij}=\kappa\Big(\delta^{ij}+\frac{1}{m^2}q^i q^j\Big)\\
\end{pmatrix}.
\end{equation}

For the true vector $u^A$, the choice is even more constrained, leaving no room for any additional free parameters. First, we look at the pseudo-vector $v^A$. Equation \eqref{jump_moment_eqs} gives
\begin{equation}
    v^\mu=\lim_{\delta \tau \to 0}\frac{1}{ \delta \tau}\Big\langle \delta x^\mu\Big\rangle.
\end{equation}
By definition, therefore, $v^\mu=p^\mu/m$. Being there no contribution from the diffusion tensor, $K^{\mu B}=0$, one finds $u^\mu = p^\mu/m$ for the spacetime part of $u$. This is also how Dowker et al. argued in \citep{philpott2009energy}. They continued by arguing that global Lorentz invariance enforces $u^i$, the momentum part of $u$, to be identically zero. This is what limits their derivation to only flat spacetime. Here, we shall see that $u^i$ is non-zero for curved spacetime.

As mentioned earlier, the vector $u$ encodes the dynamics of massive particles in the absence of diffusion processes, according to some least-action principle. For free particles, the action is proportional to the total proper time, and the dynamics is given by the geodesic equation. Alternatively, one can use the Polyakov version of the action of a point particle, which has a corresponding Hamiltonian that after subtraction of a constant term, is $\frac{1}{2m}p_\mu p_\nu g^{\mu\nu}$. The corresponding Hamiltonian vector field is nothing but $L/m$, where $L$ is the well-known Liouville operator:
\begin{equation}
    L=p^\mu\frac{\partial}{\partial x^\mu}+\Gamma^\sigma_{\mu\nu}p_\sigma p^\mu\frac{\partial}{\partial p_\nu}.
\end{equation}
Note that this is written in $(x^\mu,p_\nu)$ parametrization of the cotangent bundle; see Appendix~\ref{appendixCovLiouville} for its representation in $(x^\mu,q_i)$ coordinates of phase space. The integral curves of $L/m$ in the phase space are the trajectories of massive particles moving on the geodesics of spacetime. This is suggesting us to set
\begin{equation}
\label{u=L/m}
    u=\frac{1}{m}L.
\end{equation}
Note that spacetime components of both sides already agree, each being equal to $p^\mu/m$. Not surprisingly, one can prove that $L$ is indeed a vector under the transformations \eqref{coord_trans}, as detailed in Appendix \ref{appendixCovLiouville}. But is $L/m$ the only way to extend $\frac{1}{m}p^\mu\partial/\partial x^\mu$ to a true vector on the phase space? Imagine there is a second vector $u'$ on the phase space. Then $u''=u-u'$ is also a vector field but with spacetime components being zero. Similar to $K^{ij}$, now $u''^{\ i}$ must be a Lorentz invariant vector field on $\mathbb{H}_3$. If non-zero, each such vector would introduce preferred local directions in $\mathbb{H}_3$, thereby breaking its maximal symmetry. This is only possible in the presence of spacetime curvature, such that $u''^{\ i}$ would depend on curvature tensors. Our focus, however, is on the \textit{minimal coupling} of the diffusion to the curvature, forcing $u''^{\ i}$ to be identically zero. See Appendix~\ref{equiv_between_principles} for a non-minimal coupling example.

We have found the explicit covariant expressions of the terms in the Fokker-Planck equation \eqref{eq:Fokker-Planck}. All we have to do is to substitute $u$ and $K$ from \eqref{u=L/m} and \eqref{K_tensor}. So far, the proper time $\tau$ has served as the evolution parameter, so that $\rho$ is a function of $(\tau,x^\mu,p_\nu)$. In order to better compare with the usual Boltzmann equation and following \citep{philpott2009energy}, we integrate out the unobservable proper time and use a distribution function that only depends on $(x^\mu,p_\nu)$. This is done via
\begin{equation}
    \sqrt{|\hat{g}|}f(x,p)=\int_{-\infty}^{\infty} d\tau\ \rho.
\end{equation}
Recall that $\hat{g}$ is the determinant of the metric of the phase space. Note that this definition makes $f$ a scalar in the phase space. After picking a natural choice for this metric as well as performing a few steps on the Fokker-Planck equation, as outlined in Appendix~\ref{appendix_stochasticev}, one derives the following equation:\vspace{2.5mm}
\begin{equation}
\label{SW_Main_Equation}
L[f]=p^\mu\frac{\partial}{\partial x^\mu}f+\Gamma^\sigma_{\mu\nu}p_\sigma p^\mu\frac{\partial}{\partial p_\nu}f=m\kappa\nabla^2_{\mathbb{H}_3}f.\vspace{2.5mm}
\end{equation}
This is the \textit{covariant diffusion equation} in curved spacetime. It is the unique generally covariant extension to the Boltzmann equation for free, massive, spin 0 particles. It describes the distribution of Brownian particles that experience a frame-independent diffusion effect in curved spacetime. $\nabla^2_{\mathbb{H}_3}$ is the Laplacian operator on the hyperbolic space ${\mathbb{H}_3}$. $\kappa$ can be a spacetime scalar in its most general form, but at the zeroth order in a curvature expansion, it is a constant. It has units of $M^2/L$, therefore being a diffusion constant on the mass shell.  

Dudley's distribution function $f$ in \citep{dudley1966lorentz} differs from ours by a factor of $m^3$. Apart from re-scaling the parameters, his diffusion equation is the special case of the covariant diffusion equation \eqref{SW_Main_Equation} on flat spacetime and for a spatially homogeneous distribution $f$. Also, the cosmic-time Lorentz invariant diffusion equation in \citep{philpott2009energy}, is the special case of \eqref{SW_Main_Equation} in flat spacetime once we identify their $\rho_t/(\gamma\sqrt{g})$ and $k$ with our $f$ and $\kappa$, respectively.

A reader familiar with the Klein-Kramers equation might see some resemblance with the covariant diffusion equation. The Klein-Kramers equation describes the diffusion of particles in a classical non-relativistic phase space. However, it cannot be derived from the covariant diffusion equation in a non-relativistic limit. The Klein-Kramers has two free parameters—friction and diffusion constants—that can be related to the temperature of the background medium. There exists the distinguished rest frame of the background fluid. As a result, in the corresponding Langevin equation \citep{chandrasekhar1943stochastic} that describes an Ornstein–Uhlenbeck process, there is a drag term that prevents Brownian particles from attaining large velocities in that frame. The covariant diffusion equation, however, has only a single additional free parameter.  There is no preferred frame, and hence the Brownian particles do not feel any drag force. As a matter of fact, the effect acts in the opposite direction: The drift term causes the particles to run away from their rest frame at any given moment. To see this more clearly, we need to find the stochastic dynamics of individual particles. This is the subject of the next section.

\subsection{Covariant Brownian Motion: Langevin Equations}
\label{SDE_section}

Any diffusion process at the level of the distribution of particles uniquely corresponds to a set of Langevin or stochastic differential equations that describe the Brownian motion of individual particles. In this section, starting from first principles, we derive the most general conditions a stochastic extension to the timelike geodesic equation must satisfy. In the end, we specialize to the case of minimal coupling to curvature and derive the unique covariant Brownian motion in curved spacetime. The resulting equation has already appeared in the mathematics literature \citep{franchi2007relativistic,angst2020poisson} as the natural extension of Dudley's work \citep{dudley1966lorentz}. Similar equations have also been studied in the context of relativistic statistical mechanics, \citep{cai2023relativistic1,cai2023relativistic2}, with the difference that the systems of interest were in thermal equilibrium and a form of relativistic Einstein's relation was therefore imposed. Another suggestion for a stochastic contribution to the geodesic equation has been put forward in Moffat's stochastic gravity~\citep{Moffat_1997}. We believe that what follows sheds light on the uniqueness of the covariant Brownian motion, and gives physical intuition for the origins and the forms of the stochastic terms. 

For a massive particle, the least-action trajectory in curved spacetime is given by:
\begin{align}
    dx^{\mu} &= \frac{p^\mu}{m} d\tau \ , \label{dx=pdtau} \\
    dp^\mu &= -\frac{1}{m}\Gamma ^\mu_{\alpha \beta} p^{\alpha} p^\beta d\tau \ \label{eq:geodesic} ,
\end{align}
which is completely equivalent to the geodesic equation in a spacetime with metric $g_{\mu \nu}$ and associated Christoffel symbols $\Gamma^{\mu}_{\alpha \beta}$. This is supplied by the mass-shell constraint:
\begin{equation}
    \label{eq:mass_shell}
    g_{\mu \nu}p^\mu p^\nu = - m^2 \ ,
\end{equation}
which is trivially conserved by the geodesic motion. Introducing stochasticity into the equations of motion involves adding a random noise term to at least one of the equations. Under the condition of having almost surely continuous stochastic trajectories in spacetime and Markovian (i.e. memoryless) stochastic dynamics, the stochastic differential increment $dW^\mu$ is a Gaussian random variable with statistics:
\begin{equation}
    \Big\langle dW^\mu\Big\rangle = 0 \ , \qquad \Big\langle dW^\mu dW^\nu\Big\rangle = 2D^{\mu \nu}d\tau \ ,
\end{equation}
with $D^{\mu \nu}$ being the positive semi-definite covariance matrix of the noise. The mean can in general be non-zero, so in what follows, we explicitly subtract the mean from the noise to have the above statistics. Note that continuity and Markovianity are the effects of coarse-graining. The specific microscopic model might have some discrete jumps and forgetting time $\tau_f$. But when considering appropriate scales, the mentioned assumptions will be valid~\cite{philpott2009energy}. As we review in detail in Appendix \ref{app:equiv_itoFP}, $dW^\mu$ is of order $\mathcal{O}(\sqrt{d\tau})$, a statement made precise by It\^o's~lemma. 

The form of stochastic modification of the equations of motion is strongly constrained by the requirement of covariance and preservation of the particle on the mass shell. First, let us consider \eqref{dx=pdtau}. Adding noise to the evolution equation for $x^\mu$ inevitably leads to acausal trajectories in spacetime. To see this, consider the modified set of equations:
\begin{align}
    dx^{\mu} &= \frac{p^\mu}{m} d\tau + dW^\mu\ ,  \\
    dp^\mu &= -\frac{1}{m}\Gamma ^\mu_{\alpha \beta} p^{\alpha} p^\beta d\tau \ .
\end{align}
If we perform a boost into a frame where the particle is at rest, then $p^0 = m$, meaning that the equation of motion for the time coordinate reads:
\begin{equation}
    dx^0 = d\tau + dW^0 \ .
\end{equation}
No matter the size of the diffusion constant $\kappa$, which effectively controls the variance in $dW^\mu$, the particle will eventually experience $dx^0<0$ with probability 1, meaning that it will travel back in time. Further discussions on such stochastic processes can be found in \citep{kuipers2023stochastic,kuipers2023quantum}, where it is shown that such causality violations are suppressed for timescales larger than the particle's Compton wavelength.

To avoid causality violations entirely, the only option left is to add the stochastic term to \eqref{eq:geodesic}:
\begin{align}
    dx^{\mu} &= \frac{p^\mu}{m} d\tau   \ , \label{eq:pos_flat} \\
   Dp^\mu &= b^\mu d\tau + dW^\mu  \ .\label{eq:mom_flat} 
\end{align}
Here, $Dp^\mu$ is the covariant differential, which involves the Christoffel symbols. We have explicitly separated the mean of the noise as a drift 4-vector $b^\mu$. The physical origin of this extra drift seems mysterious at first sight. However, note that as the stochasticity is removed for $\kappa \to 0$, both the drift and the noise term should vanish due to $dW^\mu$ having zero mean and variance, and we should recover the usual geodesic equation. The necessity of adding the drift term becomes, nonetheless, evident when considering the conservation of mass shell. Using It\^o's lemma (i.e. recalling that $dW^\mu dW_\mu$ is of order $d\tau$):
\begin{equation}
\begin{split}
\label{dm^2=0}
    - d (m^2) &= d(p^\mu p_\mu) = 2 p_\mu Dp^\mu + dp^\mu dp_\mu \\
    & = 2 (p_\mu b^\mu+D^\mu_{\ \mu}) d\tau +2p_\mu dW^\mu ,
\end{split}
\end{equation}
which has to be zero for the stochastic term not to drive the system off the mass shell. In order for \eqref{dm^2=0} to vanish, however, both the $\mathcal{O}(d\tau)$ and the $\mathcal{O}(dW)\sim\mathcal{O}(d\tau^{1/2})$ terms need to vanish independently. Looking at the stochastic term first, this implies:
\begin{equation}
\label{pdW=0}
    p_\mu dW^\mu=0,    
\end{equation}
meaning that the noise term is orthogonal to the momentum 4-vector, i.e. tangent to the mass shell. We should also make sure that the second order contribution $dW^\mu dW_\mu$ allowed by It\^o's lemma preserves the mass shell condition. This means that also the other term present in \eqref{dm^2=0} has to be zero on the mass shell:
\begin{equation}
\label{p.b+trD=0}
    p_\mu b^\mu+D^\mu_{\ \mu}=0.
\end{equation}
Finally, note that the orthogonality condition \eqref{pdW=0} reduces to a condition on the covariance matrix by requiring the correlations between $p_\mu dW^\mu$ and $dW^\nu$ to vanish on the mass shell:
\begin{equation}
\label{p.D=0}
    p_\mu D^{\mu\nu}=0.
\end{equation}
In short, equations \eqref{p.b+trD=0} and \eqref{p.D=0} are the most general conditions that the drift and covariance matrix have to satisfy on the mass shell for the stochastic differential equation \eqref{eq:mom_flat} to preserve the mass shell condition. For the equations to be generally covariant, we only need $b^\mu$ and $D^{\mu\nu}$ to be a spacetime vector and tensor, respectively. And finally, to have a proper covariance matrix, $D^{\mu\nu}$ must be positive semi-definite.

Given these constraints, what is the most general form of $b^\mu$ and $D^{\mu\nu}$? In the absence of any other fields or additional geometric structures, the existing covariant objects that we have are the momentum $p^\mu$, the metric,  the covariant derivative, and all of the curvature tensors. We can combine them in any way to produce a vector and a tensor, and as long as we satisfy \eqref{p.b+trD=0} and \eqref{p.D=0}, we have an admissible drift and covariance matrix. As highlighted in the previous section, for the current paper, we are only interested in the zeroth order in a curvature expansion. A Brownian motion caused by quantum gravity is, in principle, also present in flat spacetime. In curved spacetime, many curvature terms become viable that introduce a non-minimal coupling of the Brownian motion to the curvature. We say more about this possibility in Appendix \ref{equiv_between_principles}. For the rest of the paper, we advocate for the case of minimally-coupled Brownian motion.

With this assumption, the only covariant objects available are $p^\mu$, $g^{\mu\nu}$, and $p^\mu p^\nu$. Therefore, the drift vector has to be proportional to $p^\mu$, and the covariance matrix has to be a linear combination of the metric and $p^\mu p^\nu$. After plugging in \eqref{p.b+trD=0} and \eqref{p.D=0}, the three constants can all be written in terms of one of them, which we call $\kappa$. The result is
\begin{align}
    b^\mu&=\frac{3\kappa}{m^2}p^\mu   \ , \nonumber \\
   D^{\mu\nu} &= \kappa \Big(g^{\mu\nu}+\frac{1}{m^2}p^\mu p^\nu\Big)  \ .\label{b_and_D} 
\end{align}

\begin{figure}[h!]
    \centering
    \begin{subfigure}[b]{0.49\textwidth}
        \centering
        \includegraphics[width=\textwidth,keepaspectratio=true, trim=7cm 2cm 3.8cm 3.8cm, clip]{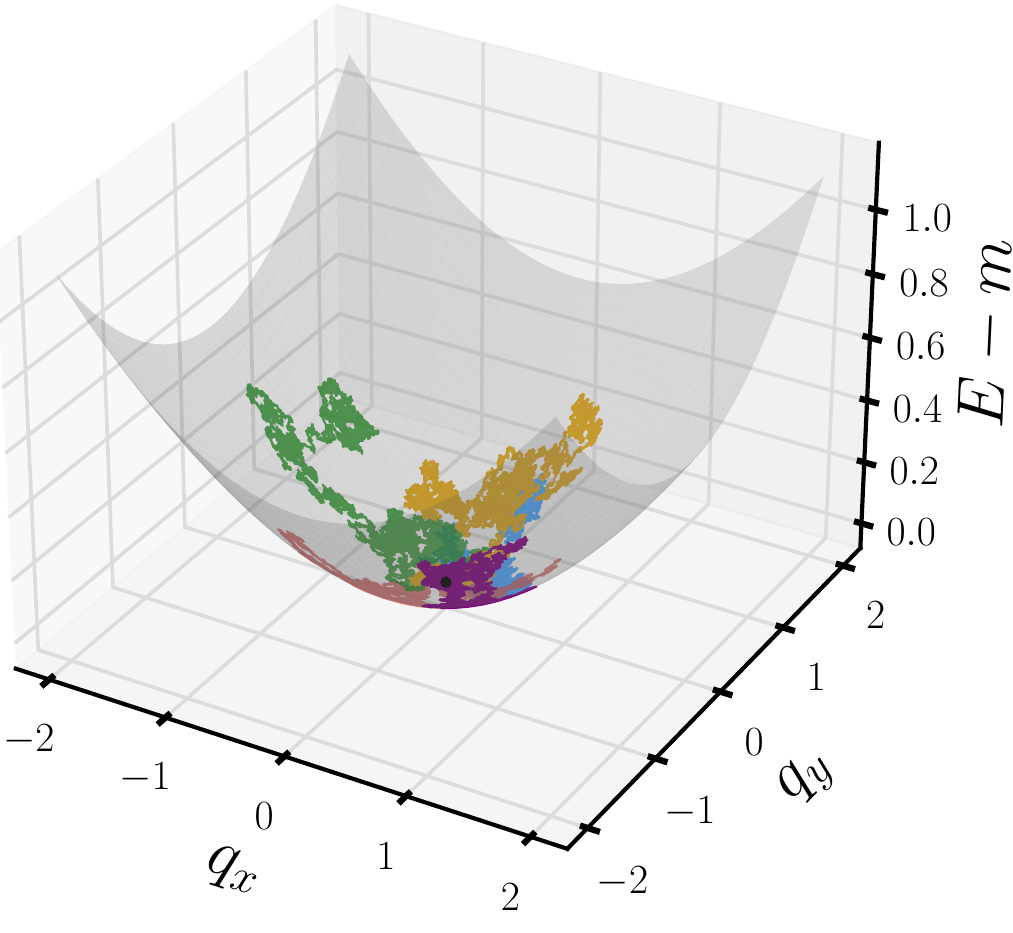}
        \caption{}       \label{qqE_diffusion}
    \end{subfigure}
    \hfill
    \begin{subfigure}[b]{0.49\textwidth}
        \centering
        \includegraphics[width=\textwidth,keepaspectratio=true, trim=7cm 2cm 3.8cm 3.8cm, clip]{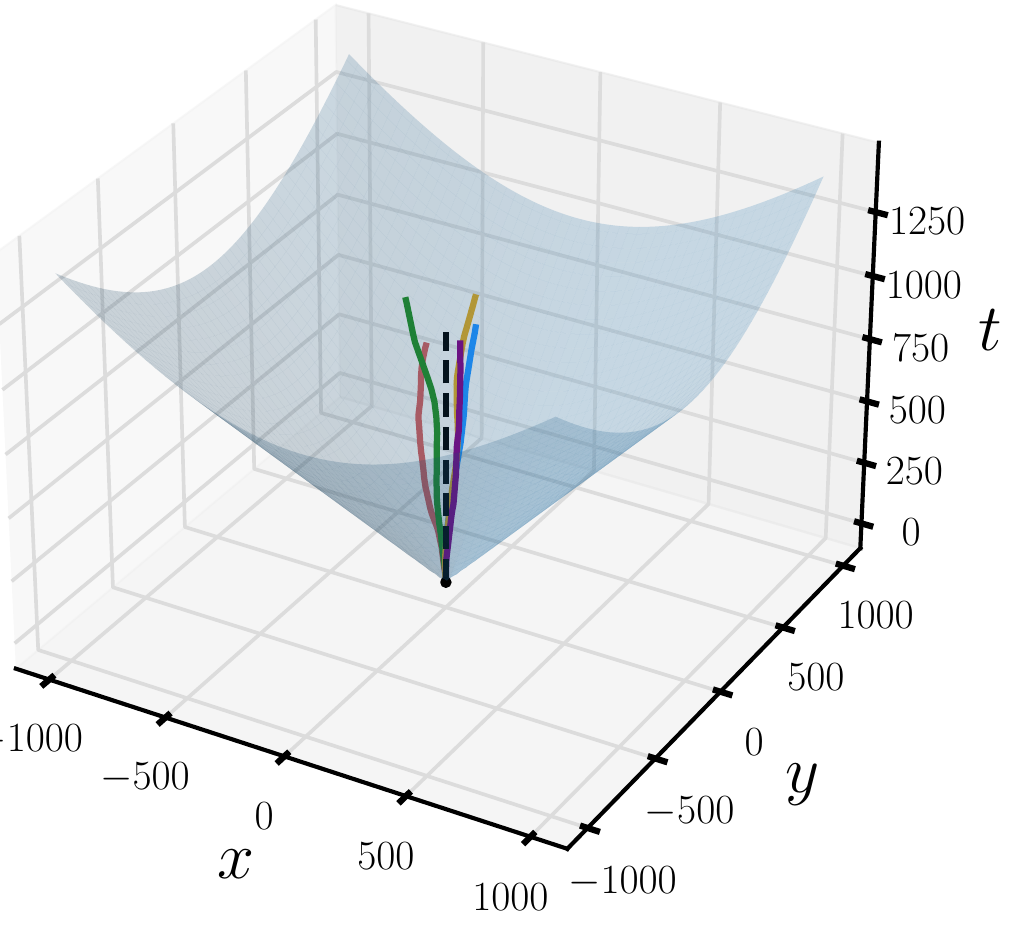}
        \caption{}\label{xyt_diffusion}
    \end{subfigure}
    \caption{\fontsize{9}{11}\selectfont Five realization of the covariant Brownian motion \eqref{stochastic_sw_eq} in 3d flat spacetime, with $m=1$, $\kappa=10^{-3}$, and total proper time $\tau=10^3$. Figure \ref{qqE_diffusion} shows the trajectories of the particles in the momentum space and on the mass shell. The particles are undergoing a random walk, and so the trajectories are not differentiable. Figure \ref{xyt_diffusion} shows the resulting trajectories in spacetime, where the shaded surface is the light cone. The particles have differentiable worldlines that deviate from geodesics.}
    \label{sde_realizations}
\end{figure}

We have derived that the most general form of the \textit{covariant Brownian motion} of massive particles minimally coupled to gravity and constrained to remain on the mass shell:\vspace{2.5mm}
\begin{align}
\label{stochastic_sw_eq}
    dx^{\mu} &= \frac{p^\mu}{m} d\tau \ ,\nonumber \\
    dp^\mu +\frac{1}{m}\Gamma ^\mu_{\alpha \beta} p^{\alpha} p^\beta d\tau&= \frac{3 \kappa}{m^2} p^\mu d\tau +  d W^\mu \ ,\\
    \Big\langle dW^\mu\Big\rangle = 0 \ , \quad \Big\langle dW^\mu &dW^\nu\Big\rangle = 2\kappa \Big(g^{\mu\nu}+\frac{1}{m^2}p^\mu p^\nu\Big)d\tau \ .\vspace{2.5mm}
\end{align}
It can be trivially extended to spacetimes of dimension other than 4, with the coefficient 3 in the drift term changing to $d-1$. Five different realizations of these equations in 3-dimensional flat spacetime are given in Figure~\ref{sde_realizations}. The particles perform a Brownian motion on the mass shell, and as a result of the change in their momentum, the spacetime trajectories deviate from a straight line.

It turns out that the parameter $\kappa$ is the same as the diffusion parameter we encountered in the previous section, and that these stochastic differential equations are completely equivalent to the covariant diffusion equation. Although we used seemingly different physical principles to arrive at the respective equations, there are parallels between the two arguments. First, with the identification 
\begin{equation}
\label{K=eeD}
    K^{ij}=e^i_{\ \mu}e^j_{\ \nu}D^{\mu\nu},   
\end{equation}
one can prove that the tensor transformation condition of $K$ on the phase space is equivalent to $D^{\mu\nu}$ being orthogonal to $p$ on the mass shell, equation \eqref{p.D=0}. Moreover, with the identification
\begin{equation}
\label{v=eb}
    v^i=e^i_{\ \mu}b^\mu,
\end{equation}
the vector transformation property of $u=(0,u^i)$ on the phase space is equivalent to the condition \eqref{p.b+trD=0}. In Appendix \ref{equiv_between_principles}, we prove these two statements and show that the covariant diffusion equation and the covariant Brownian motion are equivalent. 

\begin{figure}[h!]
    \centering
    \begin{subfigure}[b]{0.49\textwidth}
        \centering
        \includegraphics[width=\textwidth]{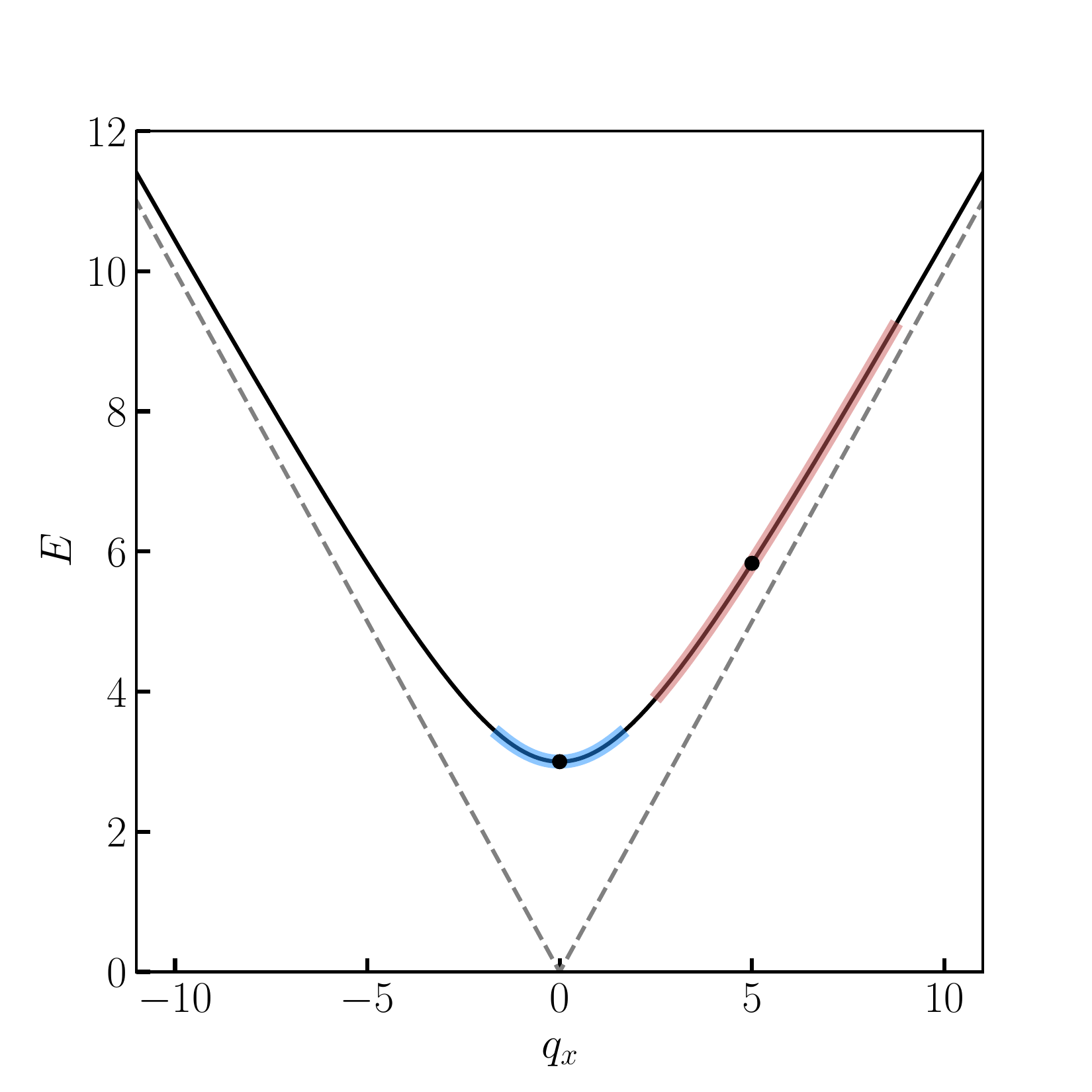}
        \caption{}       \label{2Dball}
    \end{subfigure}
    \hfill
    \begin{subfigure}[b]{0.49\textwidth}
        \centering
      \includegraphics[width=\textwidth,keepaspectratio=true, trim=7cm 2cm 3.7cm 3.8cm, clip]{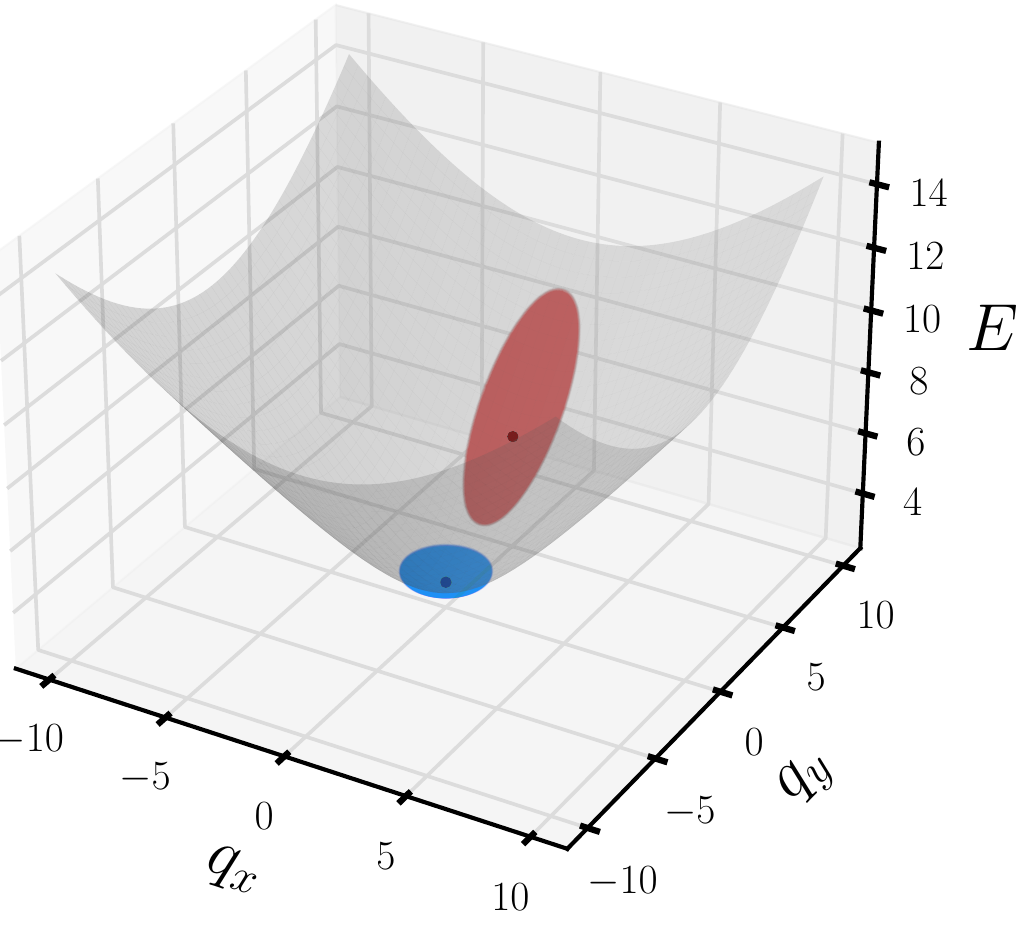}
        \caption{}\label{3dball}
    \end{subfigure}
    \caption{\fontsize{9}{11}\selectfont Balls of radius 1/2 in hyperbolic space. Figure \ref{2Dball} shows the ball of radius 1/2 centered on $q_x=0$ in blue and in red the region centered on $q_x=5$. The center of the respective regions is represented by a black dot. Note that the latter is asymmetric around its center, and in particular it extends to far higher momenta than naively expected. The same situation is shown in Figure \ref{3dball} in 2+1 dimension, with the blue region centered on $q_i=0$ and the red one on $(q_x=0,q_y=5)$.}
    \label{unitball}
\end{figure}

As the final note, we highlight that since $D^{\mu\nu}$ has to be positive semi-definite as a covariance matrix, $\kappa$ must be positive. Therefore, the drift term in \eqref{stochastic_sw_eq} acts like a force in the direction of the particle's momentum. We have seen that the existence of this drift term is necessary for covariance and the conservation of mass. Here we give a more intuitive account of that. Consider the hyperboloid of mass $m$, \eqref{H3_def}. As shown in Figure \ref{unitball}, in the rest frame of a given particle, the unit sphere around the origin $q^i=0$ is perfectly symmetric in all directions of the spatial momentum. When a particle starts a random walk from the origin, it has equal chances of going to any spatial momentum direction. However, for a particle with non-zero spatial momentum $|\Vec{q}|^2=q^iq^i$, the unit sphere is skewed toward larger $|\Vec{q}|^2$. So when doing a random walk on the mass shell, the particle has a bias toward larger energies. This is the geometric origin of the drift term.

\subsubsection{Massless case}
Before moving on to cosmology, for completeness, we briefly discuss covariant diffusion and Brownian motion for the case of massless particles. This further demonstrates the strength of the formulation we have developed.

At the level of trajectories, the stochastic correction to the massless geodesic equation as a function of an affine parameter $\lambda$ has to be of the general form:
\begin{align}
    \label{massless_SDE_ansatz}
    dx^\mu &= p^\mu d\lambda \\
    dp^\mu + \Gamma ^\mu_{\alpha \beta} p^{\alpha} p^\beta d\lambda&= b^\mu d\lambda +dW^\mu \ ,
\end{align}
where we have separated the mean of the noise as a drift vector $b^\mu(p)$ so that the Gaussian noise $d{W}^\mu$ has zero mean, and its covariance tensor $D^{\mu\nu}$ has to be determined. Preserving the light-cone condition $p^2=0$ leads to the same conditions \eqref{pdW=0}, \eqref{p.b+trD=0}, and \eqref{p.D=0} as the massive case. Once again, we are only interested in the minimal coupling case. However, there is an important distinction with the massive case: The same arguments as the previous section allow for two free parameters due to $p^2=0$. The drift vector and covariance tensor must be proportional to $p^\mu$ and $p^\mu p^\nu$, respectively, but possibly with two different proportionality constants. In other words, the conditions \eqref{p.b+trD=0} and \eqref{p.D=0} enforce $dW^\mu$ to be null and orthogonal to $p^\mu$. As a result, it has to be proportional to the momentum:
\begin{equation}
    dW^\mu = p^\mu dW_\lambda \ ,
\end{equation}
where $dW_\lambda$ is a 1-dimensional Wiener process in $\lambda$. We conclude that the covariant Brownian motion of massless particles is described by
\begin{align}
\label{massless_SDE_ansatz}
    dx^\mu &= p^\mu d\lambda \nonumber\\
    dp^\mu +\Gamma ^\mu_{\alpha \beta} p^{\alpha} p^\beta d\lambda &= \kappa_2 p^\mu d\lambda + p^\mu dW_\lambda \ ,\\
    \big\langle dW_\lambda\big\rangle = 0 \ , &\quad \big\langle dW_\lambda^2\big\rangle = 2\kappa_1d\lambda \ .
\end{align}
$\kappa_1$ and $\kappa_2$ are two free parameters that have units of $M/L$. The massless geodesic equation corresponds to the limit $\kappa_1,\kappa_2 \to 0$. 

Interestingly, in flat spacetime, the momentum equation
\begin{equation}
    dp^\mu =\kappa_2 p^\mu d\lambda + p^\mu dW_\lambda
\end{equation}
corresponds to a collection of geometric Brownian motions, one of the few stochastic processes that have a closed form solution. With the initial condition $p^\mu(0) = p^\mu_0$, this is given by:
\begin{equation}
    p^\mu(\lambda) = p^\mu_0 \ e^{\left(\kappa_2-\kappa_1\right)\lambda+W_\lambda} \ .
\end{equation}
From this, it is easy to show that the momentum of the diffusing particle has mean
\begin{equation}
    \langle p^\mu \rangle = e^{\kappa_2 \lambda} p^\mu_0 \ ,
\end{equation}
and variance:
\begin{equation}
    \text{Var}[p^\mu(\lambda)] = (p^\mu_0)^2 e^{2 \kappa_2 \lambda} \left(e^{2\kappa_1\lambda}-1\right).
\end{equation}
Unlike the massive case, the particle cannot change its orientation on the light cone, nor its direction. To see the former, note that all the components of the spatial momentum get rescaled equally in time. To understand the latter, note that such a factor, being an exponential, is always positive.

Finally, using the projection relations \eqref{K=eeD} and \eqref{v=eb} or the prescription in Appendix \ref{app:equiv_itoFP}, we can write down the covariant diffusion equation for massless particles corresponding the the above SDE:
\begin{equation}
    p^\mu\frac{\partial}{\partial x^\mu}f+\Gamma^\sigma_{\mu\nu}p_\sigma p^\mu\frac{\partial}{\partial p_\nu}f=\big(3\kappa_1-\kappa_2\big)\frac{1}{q}\frac{\partial}{\partial q}\big(q^2f\big)+\kappa_1\frac{1}{q}\frac{\partial}{\partial q}\Big(q^3\frac{\partial}{\partial q}f\Big).
\end{equation}

Geometrically, this reflects the fact that $u^i=\big(\kappa_2-3\kappa_1\big)q^i$ is a true vector and $K^{ij}=\kappa_1 q^iq^j$ is a tensor on the phase space of massless particles, as explained by Dowker et al. \citep{philpott2009energy}. As noted by the authors, there is currently no quantum gravity model of massless particles that explains the origin of the two free parameters, and the equation remains purely phenomenological. The above equation generalizes their Lorentz invariant equation to curved spacetime if we identify their $k_1$, $k_2$, and $\rho_t$ with $\kappa_1$, $\kappa_2-3\kappa_1$, and a multiple of $a^3q^2f$,~respectively. 

For a flat FRW universe with scale factor $a$ and conformal time $\eta$, the above equation~reads

\begin{equation}
    \frac{\partial f}{\partial\eta}-q\mathcal{H}\frac{\partial f}{\partial q}=a\big(3\kappa_1-\kappa_2\big)\frac{1}{q^2}\frac{\partial}{\partial q}\big(q^2f\big)+a\kappa_1\frac{1}{q^2}\frac{\partial}{\partial q}\Big(q^3\frac{\partial}{\partial q}f\Big),
\end{equation}
where $\mathcal{H}=a'/a$. This equation, among other effects, influences the free streaming of gravitational waves. Its impact on the power spectrum of primordial gravitational waves will be investigated in a future work.

\section{Cosmology}
\label{section_cosmo}
If baryons undergo covariant Brownian motion, they gain energy, which they subsequently lose via thermal radiation. This effect weakens the constraints on the diffusion parameter from cold baryonic systems \citep{kaloper2006low}. Therefore, dark matter particles—having very weak or no interaction with standard model particles—are better probes for covariant Brownian motion. Their coldness determines how much structure is formed in the late universe. So we expect tight constraints on the diffusion parameter from cosmological observations. 

When applied to dark matter particles, covariant Brownian motion gives rise to a model of stochastic dark matter. Beyond this, we remain agnostic about the particle nature of dark matter, such as its mass or field-theoretic description. Its primary governing equation is the covariant diffusion equation \eqref{SW_Main_Equation}, written in an FRW universe with metric
\begin{equation}
    ds^2=a(\eta)^2\Big(-\big(1+2\psi\big)d\eta^2+\big(1-2\phi\big)dx^idx^i\Big),
\end{equation}
where $a$ is the scale factor, $\eta$ is the conformal time, and $\phi$ and $\psi$ are gravitational potentials in Newtonian gauge.
To connect the covariant diffusion equation with the background and perturbation equations for stochastic dark matter, we integrate out the momentum variables. The energy-momentum tensor of dark matter is related to its phase space distribution via
\begin{equation}
    T_{\mu\nu}=\int\frac{d^3q}{E}\ p_\mu p_\nu f.
\end{equation}
Note that the integration measure is the well-known Lorentz-invariant volume element on the mass shell in terms of the physical momentum $q^i$, while $p_\mu$ in the integrand is the spacetime momentum~\citep{acuna2022introduction}. The background and perturbed energy density are governed by the following equations
\begin{align}
\label{bg_rho_eq}
    \bar{\rho}_{dm}'+3\mathcal{H}\big(\bar{\rho}_{dm}+\bar{P}_{dm}\big)&=a\Gamma m\bar{n}_{dm},\\
\label{pt_rho_eq}
    \delta\rho'+3\mathcal{H}(\delta\rho+\delta P)+ \big(\bar{\rho}+\bar{P}\big) \big(\theta-3\phi'\big)&=a\Gamma\Big( m\delta n+m\bar{n}\psi\Big),
\end{align}
where $m$ is the mass of dark matter, the prime denotes a derivative with respect to $\eta$, the barred variables are at background, and $P_{dm}$ and $n_{dm}$ are pressure and number density, respectively. $\theta$ is the divergence of the velocity field of dark matter. For simplicity, we drop the label $dm$ when there is no risk of confusion. Covariant diffusion has introduced an energy injection term into the continuity equations proportional to the number density. Also note that here we have introduced another constant, $\Gamma$, which has the units of $1$/time. It is related to the diffusion constant $\kappa$ and the mass of dark matter particles via
\begin{equation}
\label{Gamma_definition}
    \Gamma=\frac{3\kappa}{m^2}.
\end{equation}
One can think of $\Gamma$ as the \textit{diffusion rate} of dark matter particles. As far as background equations like \eqref{bg_rho_eq} are concerned, it has to be compared to $H_0$. If $\Gamma\ll H_0$, the effect of covariant Brownian motion on the background energy density would be negligible, and stochastic dark matter particles always remain non-relativistic. As we will see later, this condition is indeed necessary for sufficient structure formation. 

\begin{figure}[h!]
    \centering
    \begin{subfigure}[b]{0.49\textwidth}
        \centering
        \includegraphics[width=\textwidth]{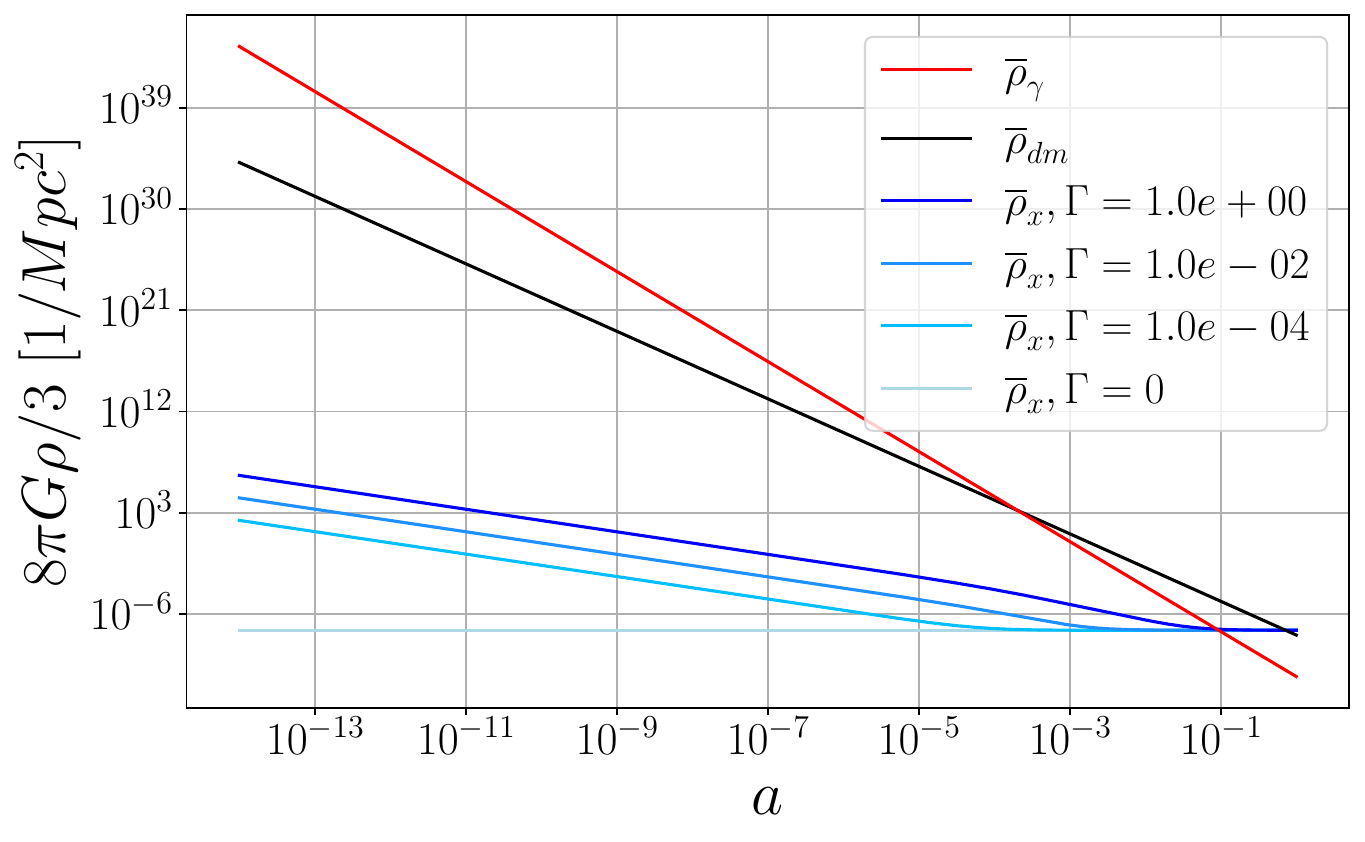}
        \caption{}       \label{rho_plot}
    \end{subfigure}
    \hfill
    \begin{subfigure}[b]{0.5\textwidth}
        \centering
        \includegraphics[width=\textwidth]{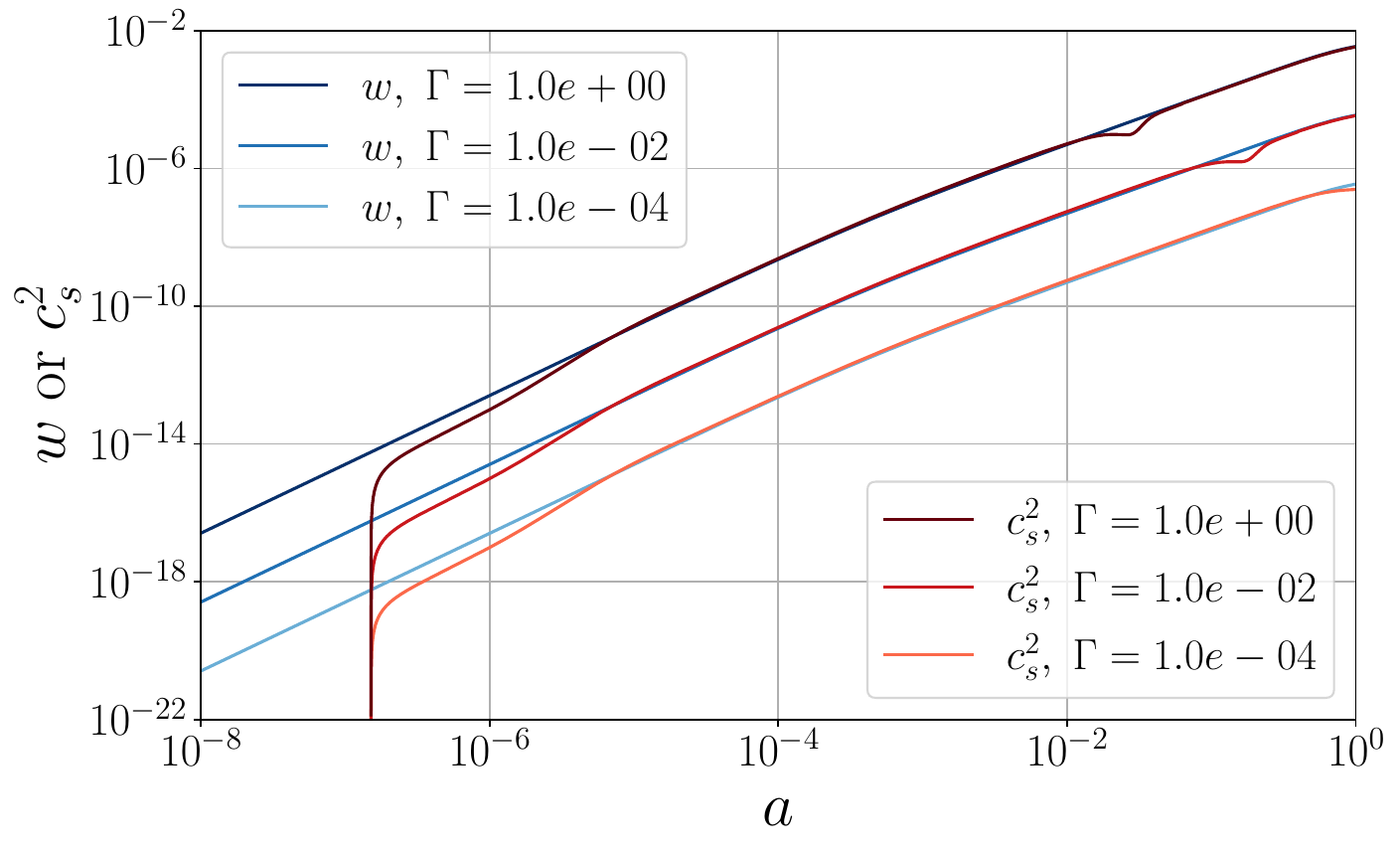}
        \caption{}\label{w_plot}
    \end{subfigure}
    \caption{\fontsize{9}{11}\selectfont Figure \ref{rho_plot} shows the log-log plot of the background energy density of photons, dark matter, and dark energy for various choices of $\Gamma$ versus scale factor. The units of $\Gamma$ are $\text{km}/\text{s}/\text{Mpc}$. One can see that even for values of $\Gamma$ as large as $1\text{km}/\text{s}/\text{Mpc}$, the deviation of dark energy density from a constant $\Lambda$ is substantial only when its contribution to the total energy density is negligible. Figure \ref{w_plot} shows the evolution of the equation of state and sound speed squared of stochastic dark matter at $k=1/$Mpc. $w_{dm}$ follows a power law behavior at all times according to \eqref{w_dm_approx}. The numerical integration of the perturbation equations has started later than the background with $c_s^2=0$ as the initial condition. $c_s^2$ quickly catches up with its superhorizon attractor solution \eqref{cs2=2/9gT/a2}.}
    \label{bg_plots}
\end{figure}

When solving \eqref{bg_rho_eq}, things would simplify greatly if we could set the dark matter pressure density to zero, as for cold dark matter. This is related to whether we can identify $m\bar{n}_{dm}$ with $\bar{\rho}_{dm}$ to have a closed system of equations without referencing the number density or mass of dark matter particles. However, we expect heating to be important in general. Indeed, a non-relativistic expansion of $E$ in powers of $q/m$ shows that
\begin{equation}
\label{mn=rho-3/2P}
    m\bar{n}=\bar{\rho}-\frac{3}{2}\bar{P},
\end{equation}
The number density is conserved for stochastic dark matter, while the energy density is not. Therefore, pressure density cannot be conserved. Even starting with a perfectly cold distribution, covariant Brownian motion causes dark matter particles to heat up over time, leading to a growing equation of state $w=\bar{P}/\bar{\rho}$ and sound speed $c_s^2=\delta P/\delta\rho$ as is evident in the numerical solutions presented in Figure \ref{w_plot}. 

Another possible simplification would be to use a fluid approximation to model stochastic dark matter. While tempting, this approach—where higher moments of velocity are ignored when integrating out the momentum variables—leads to spurious growing acoustic oscillations in the dark matter overdensity $\delta=\delta\rho/\bar{\rho}$, see the dashed curve in Figure \ref{k=.1overdensity}. In the absence of collisions between stochastic dark matter particles, we do not expect any acoustic oscillations. In essence, the fluid quantities alone cannot capture the irreversibility inherent in Brownian motion.

Therefore, the full Boltzmann hierarchy must be used to solve the perturbation equations~\citep{ma1995cosmological}. However, the traditional hierarchy is not well-suited to stochastic dark matter. It depends explicitly on the particle mass $m$, which we do not make any assumptions about. Also in Boltzmann codes, the background distribution function is assumed to be of the form $f_0(\eta,q)=f_0(aq)$
indicating that particle momentum merely redshifts.  In contrast, as discussed in section \ref{background_f0_solution},  the background distribution function for stochastic dark matter behaves like a heat kernel in the non-relativistic approximation, leading not only to redshifting but also to broadening over time in momentum space. This is why we need the “generalized Boltzmann hierarchy" that introduces both multipole expansions and velocity expansions of the distribution function.

\subsection{Generalized Boltzmann Hierarchy}
\label{GBH_section}
To numerically solve for the background and perturbation equations of stochastic dark matter, we modify \texttt{CLASS} \citep{blas2011cosmic,lesgourgues2011cosmic} to implement the generalized Boltzmann hierarchy. This was introduced by Nascimento in \citep{de2021generalized} as an alternative approach to solving for massive neutrinos in Boltzmann codes. It has the advantage of being much easier to implement in \texttt{CLASS}, and while Nascimento skipped the full implementation in his paper, he found that the generalized Boltzmann hierarchy gives comparable results to the \texttt{ncdm} component in \texttt{CLASS} in the high-precision setting with sub-percent accuracy. For us, it also has the advantage of being explicitly independent of the mass of dark matter particles so that we can only work with the diffusion rate $\Gamma$. 

\begin{figure}[h!]
    \centering
    \begin{subfigure}[b]{0.49\textwidth}
        \centering
        \includegraphics[width=\textwidth]{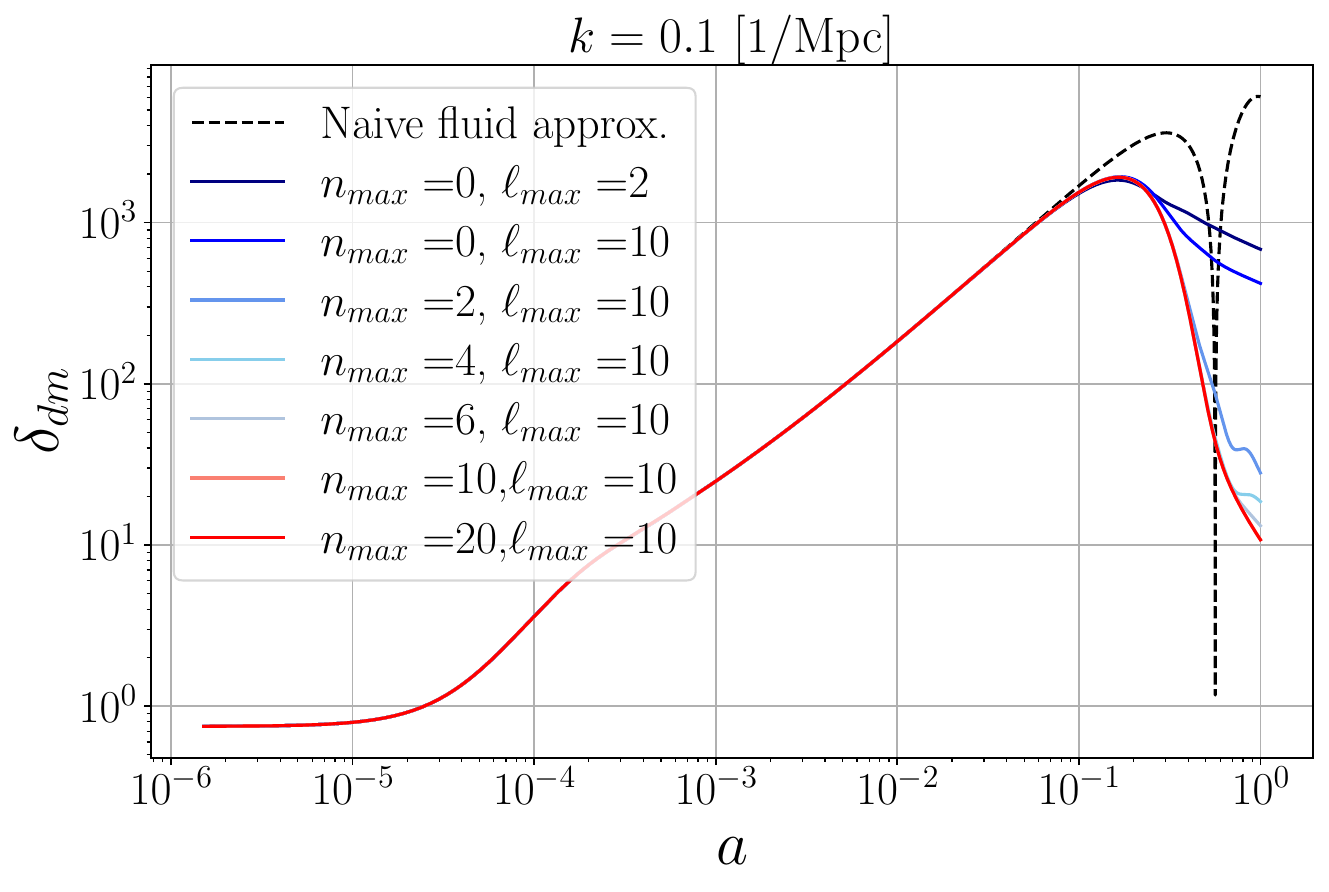}
        \caption{}       \label{k=.1overdensity}
    \end{subfigure}
    \hfill
    \begin{subfigure}[b]{0.5\textwidth}
        \centering
        \includegraphics[width=\textwidth]{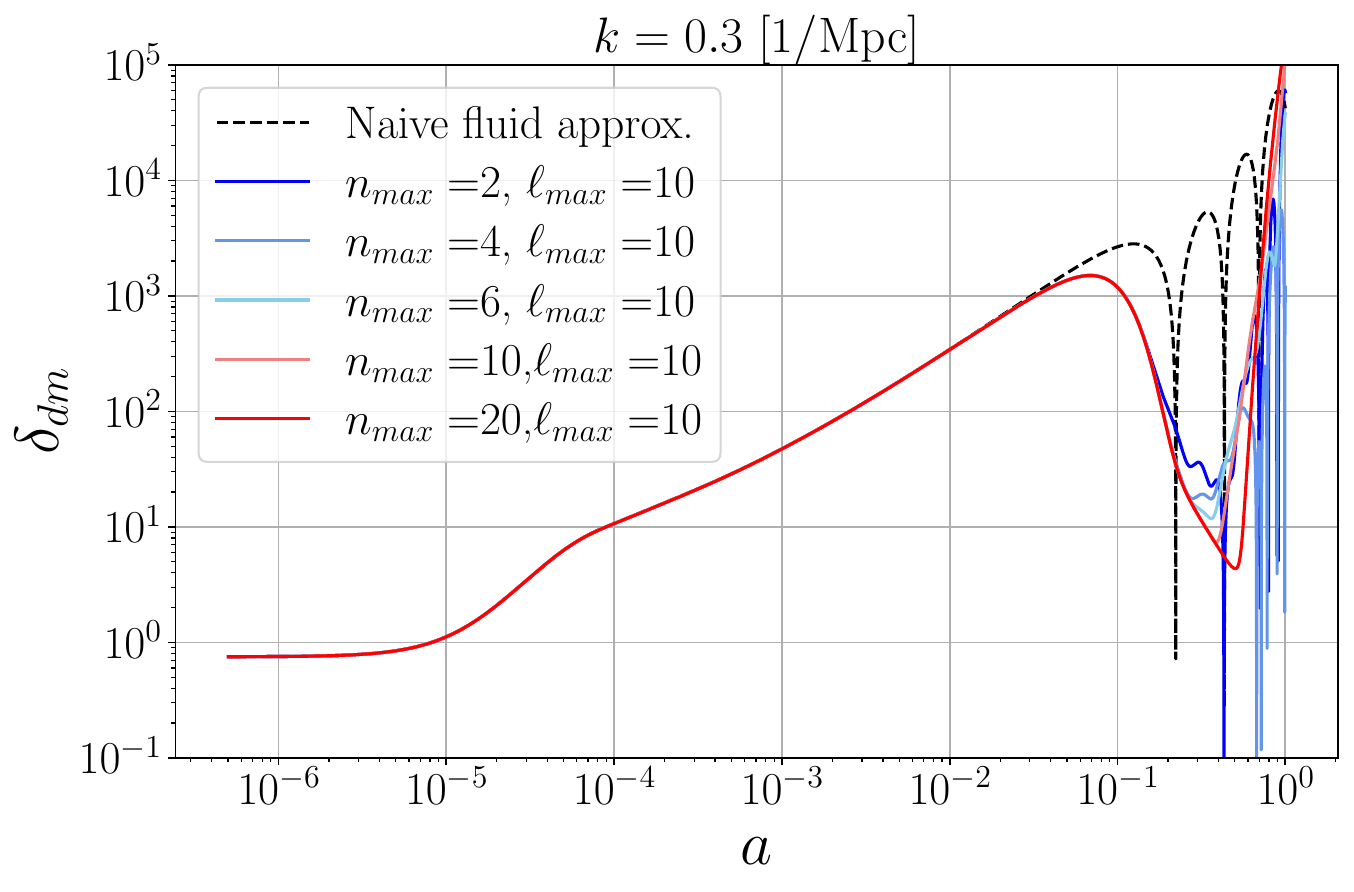}
        \caption{}\label{k=.3overdensity}
    \end{subfigure}
    \caption{\fontsize{9}{11}\selectfont Evolution of the stochastic dark matter overdensity versus scale factor for $\Gamma=0.1\text{km}/\text{s}/\text{Mpc}$ and two different wavenumbers $k$, normalized by the initial photons overdensity. A naive fluid approximation that only keeps the fluid quantities when integrating the momentum variables gives spurious growing oscillations. It is only after taking enough angular and velocity moments of the generalized Boltzmann hierarchy that the solution converges to a decaying power law. Even then, the system of equations becomes unstable at large $ky$, when the free streaming length of the particles becomes large. This is depicted in Figure \ref{k=.3overdensity}. Semi-analytic solutions are needed to tame these numerical instabilities.}
    \label{delta_dm_vs_a_fig}
\end{figure}

The main idea is that to have a systematic non-relativistic expansion of the usual Boltzmann hierarchy, one has to multiply the distribution function by $P_\ell(\mu)(q/E)^{2n+\ell}$ where $\mu$ is the cosine of the angular component of the momentum, with $P_\ell$ the Legendre polynomial, and then integrate the momentum variables. The resulting mode is the $(n,\ell)$ moment of the distribution function which we show by $f_{n,\ell}$. Its evolution equation is found by suitably integrating the covariant diffusion equation, which couples it to the adjacent moments. The lowest moments in $n$ and $l$ represent fluid variables, which are coupled to higher moments. In practice for numerical solutions, this generalized Boltzmann hierarchy has to be terminated at some finite $n_{\text{max}}$ and $l_{\text{max}}$, in such a way that including even higher moments would not affect the fluid quantities within numerical precision.  With trial and error, we find that for our case, $(n_{\text{max}},l_{\text{max}})=(20,10)$ is enough to achieve convergence for fluid quantities, as shown in Figure \ref{k=.1overdensity}.

The generalized Boltzmann hierarchy provides a straightforward and efficient way of numerically solving for the background and perturbation quantities. However, as already noted in \citep{de2021generalized}, it has the downside of breaking down at large $ky(\eta)$, where $k$ is the wavenumber of perturbations and $y$ is the typical free streaming length at conformal time $\eta$. This can be intuitively understood as follows: When the free streaming length of particles becomes comparable to a given length scale, then the perturbations at that length scale become increasingly sensitive to the high-velocity tail of the distribution function, and higher moments in $n$ are needed to achieve convergence. In Figure \ref{k=.3overdensity}, one can see that choosing a higher $n_{\text{max}}$ integrates further in time, but eventually, the moments start to diverge. Therefore, our numerical solutions using the generalized Boltzmann hierarchy have to be supplemented by a semi-analytic understanding of the behavior of each of the distribution moments at large $ky(\eta)$. This will be the purpose of the next section.

\subsection{Semi-Analytic Solutions}
\label{section_semi_analytic}
The generalized Boltzmann hierarchy is an infinite system of coupled ODEs and is challenging to solve analytically per se. However, all the equations originate from the single covariant diffusion equation \eqref{SW_Main_Equation}. In the context of the Boltzmann equation for photons, the line-of-sight integration provides an integral solution for the CMB temperature fluctuations. The solution includes the Sachs-Wolfe (SW) term, which accounts for the initial temperature fluctuations and gravitational potential. There is also the integrated Sachs-Wolfe (ISW) term, which incorporates a time integral of the varying gravitational potential. The ISW term specifically captures the gravitational redshift of the photons as they climb out of potential wells from the last scattering surface to the present time. 

We will follow a similar approach in order to solve for the dark matter fluctuations. Analogous to CMB anisotropies, the covariant diffusion equation admits an integral solution comprising SW and ISW terms. There are additional heat kernel integrals to account for diffusion. In general, the time integral cannot be performed explicitly. Nonetheless, in the regime of strong diffusion, when the generalized Boltzmann hierarchy becomes unstable, we can use saddle-point approximations to infer the power law behavior of all the moments of the distribution. Before that, let us discuss the background distribution first.

\subsubsection{Background}\label{background_f0_solution}

To start, let us look at the background covariant diffusion equation in FRW spacetime:
\begin{equation}
\label{swerves_for_f0}
    \frac{\partial f_0}{\partial\eta}-q\mathcal{H}\frac{\partial f_0}{\partial q}=a\frac{\Gamma m}{3}\frac{1}{q^2}\frac{\partial}{\partial q}\big(Eq^2\partial_q f_0\big).
\end{equation}
Finding a fully relativistic solution to this equation is a difficult task. If there were an extra factor of $E$ on the right, then this equation could have been put exactly in the form of a heat equation on the hyperbolic space, whose heat kernel is known analytically \citep{grigor1998heat}. In fact, this can be done for the diffusion equation in the proper time of the particles, rather than the cosmic or conformal time \citep{dowker2004quantum}. However, as we mentioned earlier, the proper time of diffusive particles is not observable, so we need approximations for solving \eqref{swerves_for_f0}. It will be seen later that we are only interested in the case where the dark matter particles remain non-relativistic up to the present time. Therefore, we assume that $E\simeq m$, which makes the right-hand side proportional to the flat Laplacian. 
Switching to comoving momenta $\Tilde{q}=aq$, the equation becomes a simple diffusion equation with the new time parameter 
\begin{equation}
\label{time_variable}
    d\mathcal{T}=a^3d\eta.
\end{equation}
Therefore, in the non-relativistic approximation, the solution can be written as 
\begin{equation}
\label{background_dist_solution}
    f_0(\mathcal{T},\Tilde{q})=\frac{n_0}{\big(\frac{4}{3}\pi\Gamma m^2 \mathcal{T}\big)^{3/2}}e^{-\frac{\Tilde{q}^2}{\frac{4}{3}\Gamma m^2 \mathcal{T}}},
\end{equation}
where $n_0$ is the present number density. We have assumed that the initial distribution function is a delta function at zero momentum. In other words, throughout the paper, we assume the initial condition for stochastic dark matter is that of cold dark matter. This is of course, an idealization; for any specific particle model of dark matter, the production mechanism gives an initial distribution function \citep{dodelson1994sterile,boyanovsky2008clustering,boyanovsky2011small} that should be convolved with the above heat kernel to give the correct distribution at later times. For now, we remain agnostic about the detailed particle model of dark matter and use CDM as the initial condition. 

As noted in~\citep{dowker2004quantum}, the distribution~\eqref{background_dist_solution} can be seen as a Maxwell-Boltzmann distribution with a time-dependent temperature
\begin{equation}
    k_BT=\frac{2}{3}\Gamma m\frac{\mathcal{T}}{a^2}.
\end{equation}
This is one way of seeing that covariant Brownian motion spontaneously warms up the stochastic dark matter particles. Note that in both radiation domination and matter domination, $\mathcal{T}\sim a^3\eta\sim a^2t$, where $t$ is the cosmic time. Therefore, the temperature grows linearly with the cosmic time and is independent of the choice of $a_0$, as expected. Also, remember that this equation is only valid as long as the particles remain non-relativistic, so $k_BT\ll m$. This translates to $\Gamma\mathcal{T}/a^2\ll1$, or $a\Gamma\eta\ll1$. So we can only trust these non-relativistic approximations if $\Gamma\ll H_0$.

Having the solution to the background distribution at hand, we can directly infer the behavior of several background quantities:
\begin{align}
    \Bar{P}_{dm}&=\frac{2}{3}\frac{\Gamma\mathcal{T}}{a^2}m\Bar{n}_{dm},\nonumber\\
    \Bar{\rho}_{dm}&=\Big(1+\frac{\Gamma\mathcal{T}}{a^2}\Big)m\Bar{n}_{dm},\nonumber\\
    \label{w_dm_approx}
    w_{dm}&=\frac{2}{3}\frac{\frac{\Gamma\mathcal{T}}{a^2}}{1+\frac{\Gamma\mathcal{T}}{a^2}}\simeq \frac{2}{3}\frac{\Gamma\mathcal{T}}{a^2}.
\end{align}
The number density is conserved and scales as $1/a^3$. So for $\Gamma\ll H_0$, we conclude that the deviation of $\Bar{\rho}_{dm}$ from $1/a^3$ scaling must be very small, and the equation of state, although growing with time, remains negligible up to the present time. One might be worried that the condition $\Gamma\ll H_0$ is too restrictive, not allowing for any observable signatures. This is only true at the background level. In the next section, we will see that the picture is very different for perturbations. While background is only sensitive to the overall transfer of energy to dark matter particles, perturbations are sensitive to the diffusion length. From equation~\eqref{background_dist_solution}, the typical velocity squared of stochastic dark matter particles at some time $\mathcal{T}$ is 
\begin{equation}
\label{typical_velocity}
    v^2\sim \frac{1}{a^2}\Gamma \mathcal{T}\sim a\Gamma\eta,
\end{equation}
which is just the variance of a random walk on the mass shell. The typical “diffusion length" is given by
\begin{equation}
    \lambda_D\sim vt\sim av\eta.
\end{equation}
The free streaming length $y$ is just the comoving diffusion length, i.e. $y=\lambda_D/a$. A typical stochastic dark matter particle could travel a distance of about $1$Mpc during the age of the universe, with an equation of state consistently less than $w_{dm} \sim10^{-6}$. Such a large diffusion length has a significant impact on the small-scale perturbations, as we will see in the subsequent sections.

\subsubsection{Perturbations}
\label{pert_semianalytic_solution}

We expand the distribution function up to first order as
\begin{equation}
    f=f_0+F.
\end{equation}

We are mostly interested in solving for $F$ when the diffusion becomes strong. For the length scales of interest and reasonable values of the diffusion rate, this happens well after radiation-matter equality. We checked numerically that even in the presence of diffusion, the total anisotropic stress is negligible and we can use the approximation $\phi=\psi$. Moreover, to be able to give an analytic solution, we have to go the non-relativistic limit, where $q\ll m$ and $ E\simeq m$. This is justified since, as we will see later, we will get a bound of $w\lesssim4.\times10^{-6}$ from CMB, which allows velocities of order $v\lesssim 2.\times10^{-3}c$ (consistent with other reports \citep{mueller2005cosmological,kumar2014observational}). In the non-relativistic approximation, the hyperbolic Laplacian in the covariant diffusion term becomes a simple flat Laplacian in momentum space. Under these approximations, if we transform to the comoving momentum variable $\Tilde{q}=aq$ again, and use $\mathcal{T}$ defined in~\eqref{time_variable} as the time parameter, the evolution equation for $F$ becomes
\begin{equation}
\label{simplified_F_eq}
    \frac{\partial F}{\partial\mathcal{T}}+\frac{1}{ma^4}ik\Tilde{q}\mu F-\frac{\partial f_0}{\partial \Tilde{q}}\frac{m}{a^2}ik\mu\phi=\frac{\Gamma m^2}{3}\delta_{ij}\frac{\partial^2}{\partial\Tilde{q}_i\partial\Tilde{q}_j}F,
\end{equation}
where $\mu=\hat{k}.\hat{q}$. Without the diffusion term on the right-hand side, one could solve the Boltzmann equation using integration along the line of sight. This would lead to a Sachs-Wolfe term accounting for the initial condition and an ISW term accounting for the integral effect of the gravitational source along the line of sight. Although the diffusion term makes things more complicated, we can still solve this equation. The crucial observation is that~\eqref{simplified_F_eq} resembles a diffusion equation with a time-dependent linear potential and a non-homogeneous gravitational source term. Alternatively, by analytically continuing the time variable $\mathcal{T}$ to the imaginary axis, this would be a Schrodinger equation with a time-dependent linear potential and a non-homogeneous source term. Forgetting about the non-homogeneous term for a moment, this quantum mechanical problem has been studied and solved in~\citep{feng2001complete}. Following the same steps, we can solve our equation~\eqref{simplified_F_eq} without the source term $\phi$, and then use Duhamel's principle \citep{kadivar2018techniques} to solve for the whole equation including the non-homogeneous source term. We leave the details for Appendix \ref{Appendix_Analytic_sol}. The final solution is
\begin{align}
\label{integral_solution}
&F(\mathcal{T}, k, Q) = \int d^3Q' \left( e^{-\frac{1}{3}\Gamma k^2\mathcal{G}(\mathcal{T},\mathcal{T}_i)} \frac{e^{-\frac{|Q-Q'|^2}{\frac{4}{3}\Gamma m^2 (\mathcal{T}-\mathcal{T}_i)}}}{\left(\frac{4}{3}\pi\Gamma m^2 (\mathcal{T}-\mathcal{T}_i)\right)^{3/2}} e^{-i\Vec{k}.\Vec{y}(Q',\mathcal{T}_i)+i\Vec{k}.\Vec{y}(Q,\mathcal{T})} F(\mathcal{T}_i, k, Q') \right)\nonumber\\ &
+ \int_{\mathcal{T}_i}^{\mathcal{T}} d\mathcal{T}' \int d^3Q' \left( e^{-\frac{1}{3}\Gamma k^2\mathcal{G}(\mathcal{T},\mathcal{T}')} \frac{e^{-\frac{|Q-Q'|^2}{\frac{4}{3}\Gamma m^2 (\mathcal{T}-\mathcal{T}')}}}{\left(\frac{4}{3}\pi\Gamma m^2 (\mathcal{T}-\mathcal{T}')\right)^{3/2}} e^{-i\Vec{k}.\Vec{y}(Q',\mathcal{T}')+i\Vec{k}.\Vec{y}(Q,\mathcal{T})} \frac{m}{a(\mathcal{T}')^2}i\Vec{k}.\Hat{q}' \frac{\partial f_0}{\partial \Tilde{q}} \phi(k,\mathcal{T}')\right).
\end{align}
$\mathcal{T}_i$ is the initial time when the initial conditions are set, $y$ is the free streaming length, $Q$ is some redefined momentum parameter, and $\mathcal{G}(\mathcal{T},\mathcal{T}')$ is some form of a time difference between $\mathcal{T}'$ and $\mathcal{T}$ (see Appendix \ref{Appendix_Analytic_sol} for details):
\begin{equation}
    \mathcal{G}(\mathcal{T},\mathcal{T}_i)\sim \frac{\mathcal{T}^3}{a^8}-\frac{\mathcal{T}_i^3}{a_i^8}\sim a\eta^3-a_i\eta_i^3,
\end{equation}
The important factor that we have to pay attention to is the first exponential in the two integrals. Assuming that $T_i\ll T$, the first SW term in~\eqref{integral_solution} gets exponentially suppressed by $\exp\big(-\frac{1}{3}\Gamma k^2\mathcal{G}(\mathcal{T},\mathcal{T}_i)\big)$ whenever
\begin{equation}
    a\Gamma k^2\eta^3\gg1.
\end{equation}
As for the second integral, only the late-time part of the integral ($\mathcal{T}'\sim\mathcal{T}$) would give a non-negligible contribution, in contrast to the early-time which would be exponentially suppressed by the same exponential factor. Therefore, the strength of the diffusion at some scale $k$ is determined by the factor $a\Gamma k^2\eta^3$. We say that diffusion has become relevant at this scale if this number is comparable to 1.

There is yet another more intuitive way of understanding the same result. As we found out in the previous section, at the background level, the strength of the diffusion is controlled by $\Gamma/H$. If this is small, then the effect of diffusion can be ignored in the background. However, perturbations are sensitive to the diffusion length. If the growing diffusion length becomes comparable to the length scale of interest, the clustering of dark matter particles starts to get washed away. From the previous section, we know that the typical diffusion length is $\lambda_D\sim av\eta$. To see its relevance, this should be compared to the physical length scale of interest, $\lambda=a/k$. We have
\begin{equation}
\label{ky^2}
    \big(ky\big)^2=\Big(\frac{\lambda_D}{\lambda}\Big)^2\sim a\Gamma k^2\eta^3.
\end{equation}
Therefore, $a\Gamma k^2\eta^3$, the parameter that controls the exponential suppression of initial conditions and early ISW, is the square of the diffusion length divided by the length scale of interest. When this number becomes very large, the diffusion is so strong that it erases the effect of initial conditions. Hence, only the very late behavior of the gravitational potential determines the distribution of stochastic dark matter particles. This aligns with our understanding of diffusion. For stochastic dark matter particles alone, entropy increases, and the dynamics is irreversible, effectively erasing the initial data.

\begin{figure}[h!]
    \centering
    \begin{subfigure}[b]{0.49\textwidth}
        \centering
        \includegraphics[width=\textwidth]{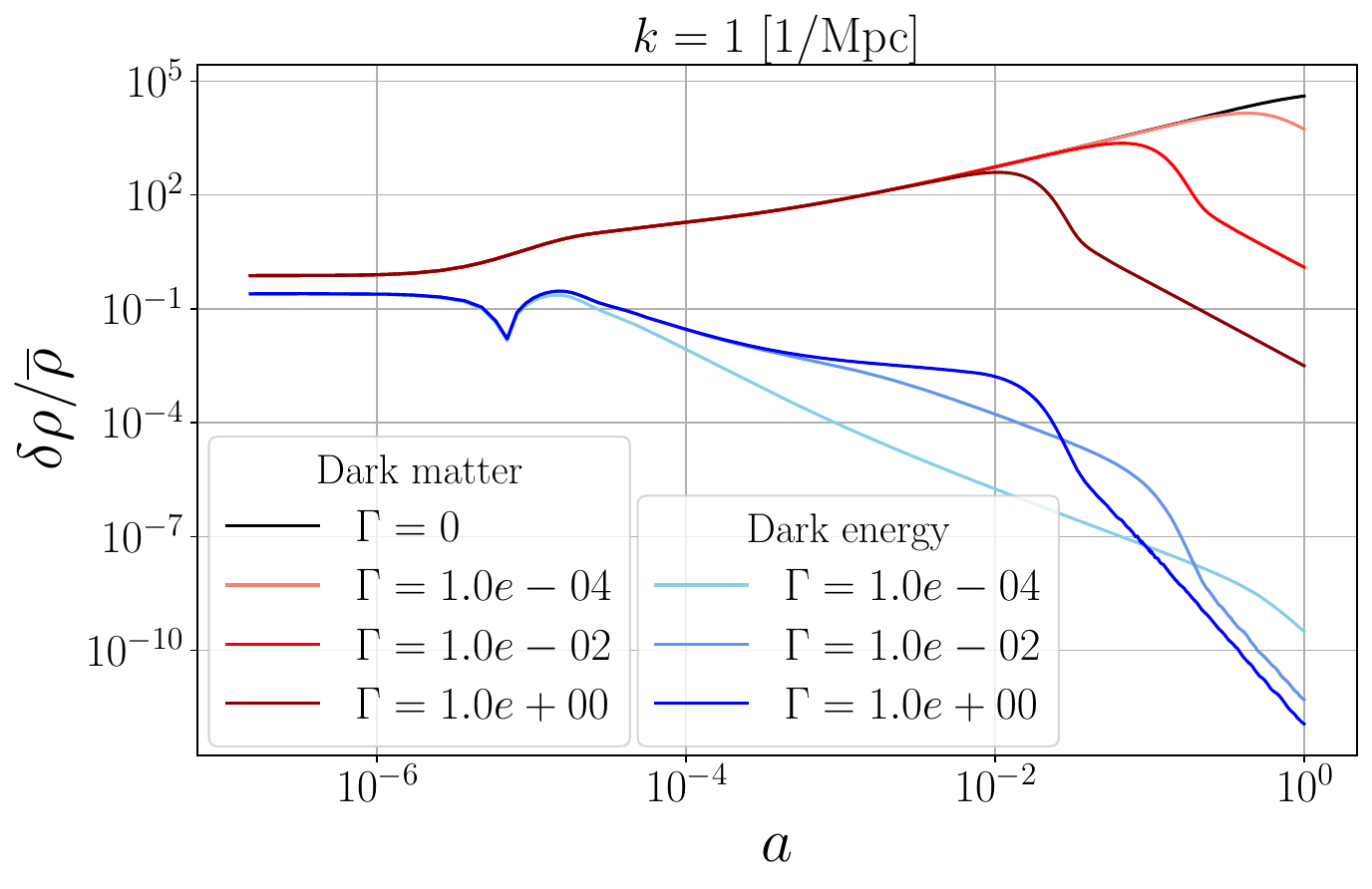}
        \caption{}       \label{full_delta_dm_x}
    \end{subfigure}
    \hfill
    \begin{subfigure}[b]{0.49\textwidth}
        \centering
        \includegraphics[width=\textwidth]{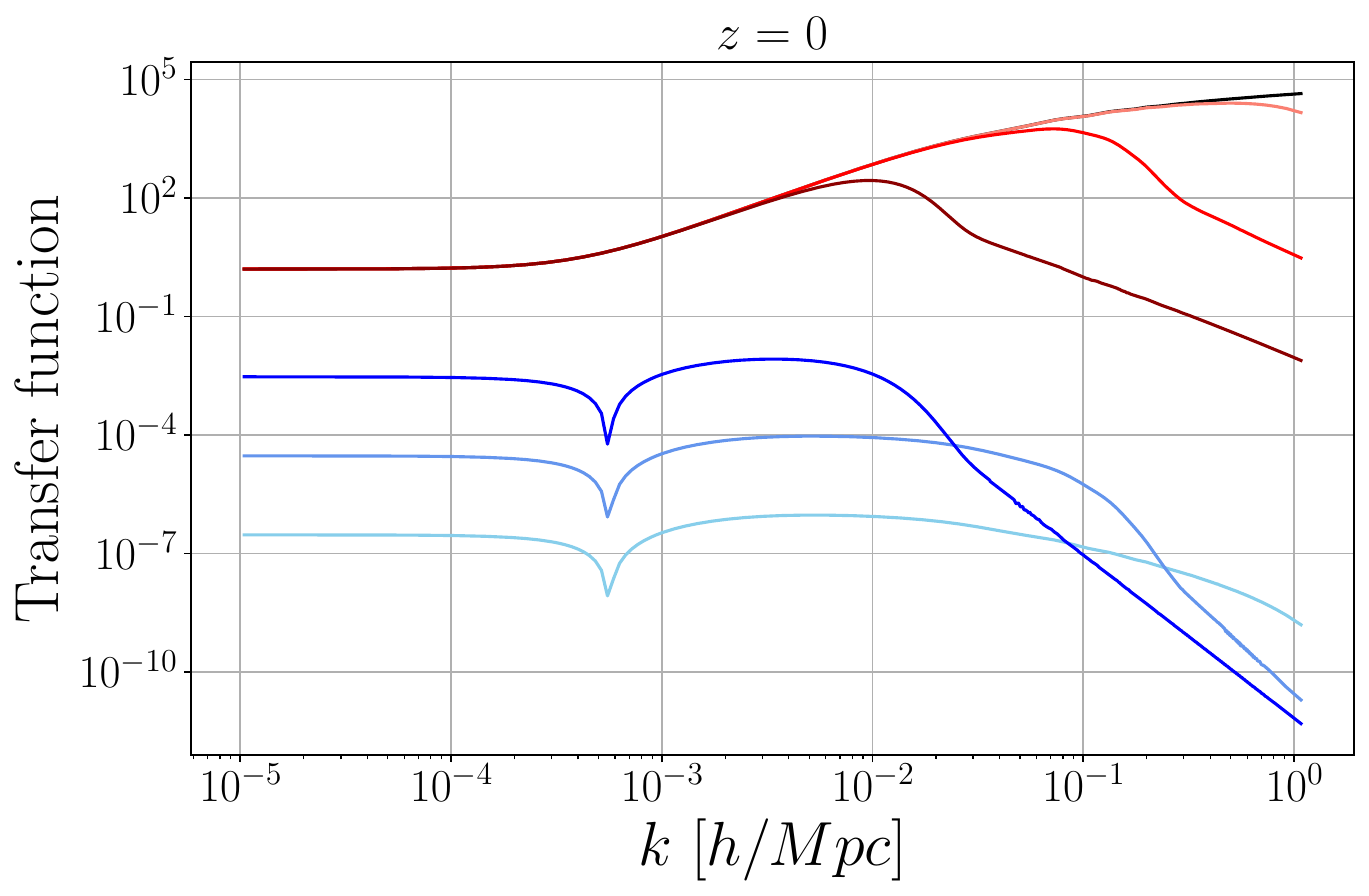}
        \caption{}\label{transfer_func_fig}
    \end{subfigure}
    \caption{\fontsize{9}{11}\selectfont Figure \ref{full_delta_dm_x} shows the stochastic dark matter and dark energy overdensity versus scale factor at $k=1/$Mpc, normalized by the initial photons overdensity. These are numerical solutions obtained by the generalized Boltzmann hierarchy supplemented with the asymptotic solution \eqref{delta_dm_asymptotic} at large $a\Gamma k^2\eta^3$. The dark~energy overdensity is at all times much smaller than stochastic dark matter overdensity, except at very early times when the background dark energy density is negligible. Figure \ref{transfer_func_fig} shows the transfer function of the stochastic dark matter and dark energy versus $k$ at the present time. The perturbations of dark energy are negligible compared to the stochastic dark matter across all scales. The units of $\Gamma$ are $\text{km}/\text{s}/\text{Mpc}$ and the two plots share the same legend. }
    \label{asymptotic_solutions_implemented_fig}
\end{figure}

Relation \eqref{ky^2} indicates that the regime $a\Gamma k^2\eta^3\gg1$ is also the regime where the generalized Boltzmann hierarchy becomes unstable, according to section \ref{GBH_section}. Therefore, if we could find the analytic behavior of~\eqref{integral_solution} in this regime, we would have everything we need for the full solution to dark matter perturbations. Fortunately, this is feasible. In this strong diffusion regime, we can ignore the SW term. For the ISW term, notice that the integrand consists only of either Gaussian or plane wave functions of $Q'$. So, the momentum integral can be carried out analytically. For the remaining time integral, we can use the steepest descent method or numerical techniques to capture only the very late-time contribution of the integral; see Appendix~\ref{Appendix_Analytic_sol}. For example, the dark matter density contrast is given by the following integral
\begin{equation}
    \delta(\eta,k)=-k^2\int_{\eta_i}^{\eta}d\eta'\ \eta'\big(1-\frac{\eta'}{\eta}\big)\phi(\eta',k)e^{-\frac{1}{315}a\Gamma k^2\eta^3\big(1-6(\frac{\eta'}{\eta})^5+5(\frac{\eta'}{\eta})^6\big)}.
\end{equation}
This equation shows that, more precisely, we need
\begin{equation}
    a\Gamma k^2\eta^3\gg315
\end{equation}
to be in the strong diffusion regime. The numerical factor comes from the specifics of the matter domination era. Now, the integrand is exponentially larger at the right-end of the integration interval. The saddle-point approximation gives
\begin{equation}
\label{delta_dm_asymptotic}
    \delta(\eta,k)\approx\frac{-21}{2a\Gamma\eta}\phi(\eta,k).
\end{equation}
A more accurate numerical approximation gives $23$ in the numerator instead; see Appendix~\ref{Appendix_Analytic_sol}. Notice how $\delta$ depends only on the present gravitational potential and not on its past history. More importantly, this shows that when diffusion takes over, the density contrast of stochastic dark matter begins to decrease following a power law. In other words, the clustering becomes washed away, and the dark matter distribution tends toward homogeneity. This asymptotic solution matches well with the numerical results from the generalized Boltzmann hierarchy just before the onset of numerical instability. Thus, by combining the two, we get the full solution for the dark matter overdensity and transfer function, as depicted in Figure \ref{asymptotic_solutions_implemented_fig}. The transfer function further shows the bottom-up smoothing of structures, with more suppression at smaller scales.

Similar expressions to~\eqref{delta_dm_asymptotic} can be found for the pressure density, velocity, and anisotropic stress of stochastic dark matter:
\begin{equation}
    \frac{\delta P}{\Bar{\rho}}\approx-\phi\ ,\quad (1+w)\theta\approx-48.6\frac{1}{a\Gamma\eta^2}\phi\ , \quad (1+w)\sigma\approx-\frac{12.6}{\sqrt{a\Gamma k^2\eta^3}}\phi\ .
\end{equation}
Once again, a more accurate numerical integration gives $\delta P /\rho\approx-1.06\phi$, see Appendix~\ref{Appendix_Analytic_sol}.

Finally, let us recall the approximations we used to derive these asymptotic solutions. We assumed that the particles are non-relativistic, and we ignored the total anisotropic stress to write $\phi=\psi$. And last, we ignored the effect of dark energy when writing the functions $\mathcal{T}$ and $\mathcal{G}$ in terms of the conformal time $\eta$. Since we only switch from numerical integration to the asymptotic solutions at very late times and small length scales, and also since the allowed diffusion rates by the data are small, $\Gamma/H_0<10^{-4}$, we expect the error introduced by this approximation in the cosmological observables to be negligible.

\subsection{Dark Energy Model}

\label{DEmaintext}
When we add the covariant diffusion term to the Boltzmann equation for stochastic dark matter, the energy-momentum tensor is no longer conserved since \citep{acuna2022introduction}:
\begin{equation}
\label{dT=intLf}
    \nabla_\mu T^\mu_{\ \nu}= \int \frac{d^3 q}{E}\ p_\nu L[f].
\end{equation}
$L[f]$ being equal to $m\kappa\nabla^2_{\mathbb{H}_3}f$, the right-hand side is no longer zero. However, the Einstein equations require the total energy-momentum tensor to be conserved. Thus, it is necessary to identify the source of the energy and momentum transfer to the stochastic dark matter.

From a fundamental point of view, covariant Brownian motion happens because the very notion of geometry at the Planck scale is ambiguous. The stochastic term in the geodesic equation accounts for our ignorance of the effect of such a discrete or fuzzy geometry on the motion of a point particle. Therefore, the back-reaction of this stochastic motion of a particle on the dynamics of spacetime is of equal importance. Ideally, this should be effectively described by a stochastic term in the Einstein equations. For a single particle moving in spacetime, the energy-momentum tensor has a stochastic contribution from $p^\mu$ \eqref{stochastic_sw_eq}. To balance this, there must be a stochastic term on the geometric side of the Einstein equations. Finding such a correction is beyond the scope of this paper and is an important avenue for future research.

Here, we adopt a phenomenological approach to this problem. Although a single particle exhibits stochastic behavior, their collective has a deterministic behavior. Hence, the correction terms to the geometry behave deterministically as well. We propose that this can be effectively described by an imperfect dark energy fluid. In this sense, dark energy should be thought of as part of the geometry rather than a matter species. In other words, an imperfect dark energy fluid represents the mechanism responsible for the back-reaction of covariant Brownian motion on geometry. It transfers energy and momentum to stochastic dark matter particles, conserving the total energy-momentum tensor. Consequently, it has to obey
\begin{equation}
\label{balance eq}
    \nabla_\mu T^{\mu\nu}_x=-\nabla_\mu T^{\mu\nu},
\end{equation}
where $T^{\mu\nu}_x$ is the energy-momentum tensor of dark energy.

At this point, we have to choose a kinematic ansatz for $T^{\mu\nu}_x$. The simplest choice would be to promote the cosmological constant to a spacetime scalar $\Lambda(x)$. In fact, in unimodular gravity, the diffeomorphism invariance is relaxed in such a way that it naturally allows for non-conservation in the matter energy-momentum tensor such that $\nabla_\nu\Lambda=8\pi G\nabla_\mu T^{\mu}_{\ \nu}$ \citep{velten2021conserve}. Such theories have even been motivated by discreteness at the Planck scale \citep{perez2016dark,perez2019dark}; see \citep{corral2020diffusion,perez2021resolving,cedeno2021revisiting,fabris2022nonconservative,de2021interacting,amadei2022planckian,leon2022inflation} for cosmological applications and inflationary scenarios. However, this does not apply to our case because \eqref{dT=intLf} cannot be written as the gradient of a scalar. 

The next simplest ansatz is to consider a perfect fluid dark energy. If one chooses the dark energy equation of state to be $w_x=-1$, then the perturbed velocity divergence of dark energy, $\theta_x$, would not appear in any dynamical equation as it is always multiplied by $1+w_x$. Then the spatial part of the equation \eqref{balance eq} enforces a constraint equation on the pressure perturbation of dark energy, which translates into a constraint on its total sound speed squared, $c_x^2$. Such a constraint equation fixes $c_x^2$ in terms of the energy density of dark energy and velocity divergence of dark matter, $\theta_{dm}$. As a result, $c_x^2$ can in general become negative, creating gravitational instability in the perturbation equations. So one has to allow for $w_x\neq-1$. In this case, it has been shown that interacting dark energy-dark matter theories could have early-time instabilities~\citep{valiviita2008large}. To avoid the instabilities in the case that energy flows from dark energy to dark matter, one has to ensure $w_x<-1$; i.e. one needs phantom dark energy \citep{gavela2009dark}. To avoid physical problems associated with phantom dark energy models \citep{ludwick2017viability}, we discard this option as well.

Finally, we get to imperfect fluid models for dark energy. In \citep{madsen1988scalar,pimentel1989energy,faraoni2018imperfect,sawicki2013consistent,zimdahl2019matter} it is shown that many modified gravities and scalar dark energy models can be expressed in GR using an imperfect fluid energy-momentum tensor. In its most general form, it contains heat flux and anisotropic stress terms in addition to the perfect fluid quantities. For our current purpose, since \eqref{balance eq} has four equations, we only need four dynamical degrees of freedom. Therefore we use the following ansatz
\begin{equation}
\label{DETmunu}
    T_x^{\mu\nu}=P_xg^{\mu\nu}+(P_x+\rho_x)U^\mu U^\nu+\mathcal{Q}_x^\mu U^\nu+\mathcal{Q}_x^\nu U^\mu.
\end{equation}
$\mathcal{Q}^\mu_x$ is the heat flux vector that satisfies $\mathcal{Q}_x^\mu U_\mu=0$, so it has three independent degrees of freedom. We set $w_x=-1$, and to avoid gravitational instability, we impose $c_x^2=1$ manually. Unlike perfect fluids, setting $w_x=-1$ does not enforce any constraint equation on $c_x^2$ due to the presence of $\mathcal{Q}_x^\mu$. We still have three extra degrees of freedom in $U^\mu$. To fix them, we make the extra assumption that the imperfect dark energy fluid is tightly coupled to the stochastic dark matter so that they share the same velocity vector $U^\mu$ even at the perturbations level. The background and perturbation equations derived from \eqref{balance eq} can be found in Appendix~\ref{DEappendix}.

We highlight again that this is a purely phenomenological way of dealing with the back-reaction of the covariant Brownian motion on geometry, and the correct formulation will most probably correct our assumptions. For the present, we believe this is suitable enough for our purposes since the main physics that we are confident about and want to investigate is the diffusion of dark matter. As the dark energy sector is more speculative, we want it to have minimal deviations from a constant $\Lambda$. In other words, we want the background energy density of dark energy to closely resemble a cosmological constant and the clustering of dark energy to be negligible compared to dark matter. In Figure \ref{rho_plot}, one can see that if $\Gamma\ll H_0$, the deviation from a cosmological constant behavior is indeed small. Also, Figure \ref{asymptotic_solutions_implemented_fig} shows that the dark energy overdensity is small compared to dark matter, therefore having a negligible impact on structure formation.

\begin{figure}[h!]
    \centering
    \begin{subfigure}[b]{0.48\textwidth}
        \centering
        \includegraphics[width=\textwidth]{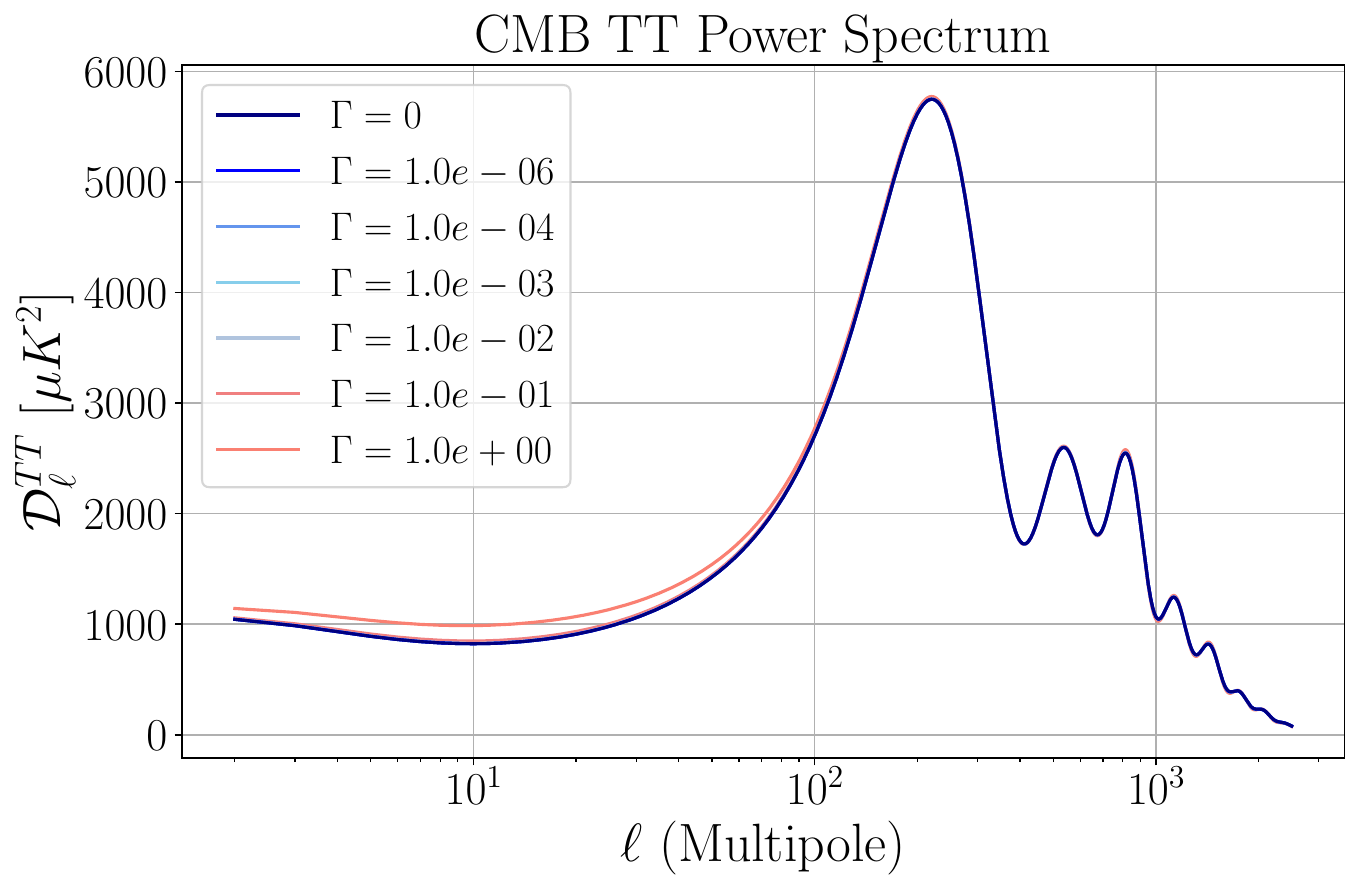}
        \caption{}       \label{C_l}
    \end{subfigure}
    \hfill
    \begin{subfigure}[b]{0.51\textwidth}
        \centering
        \includegraphics[width=\textwidth]{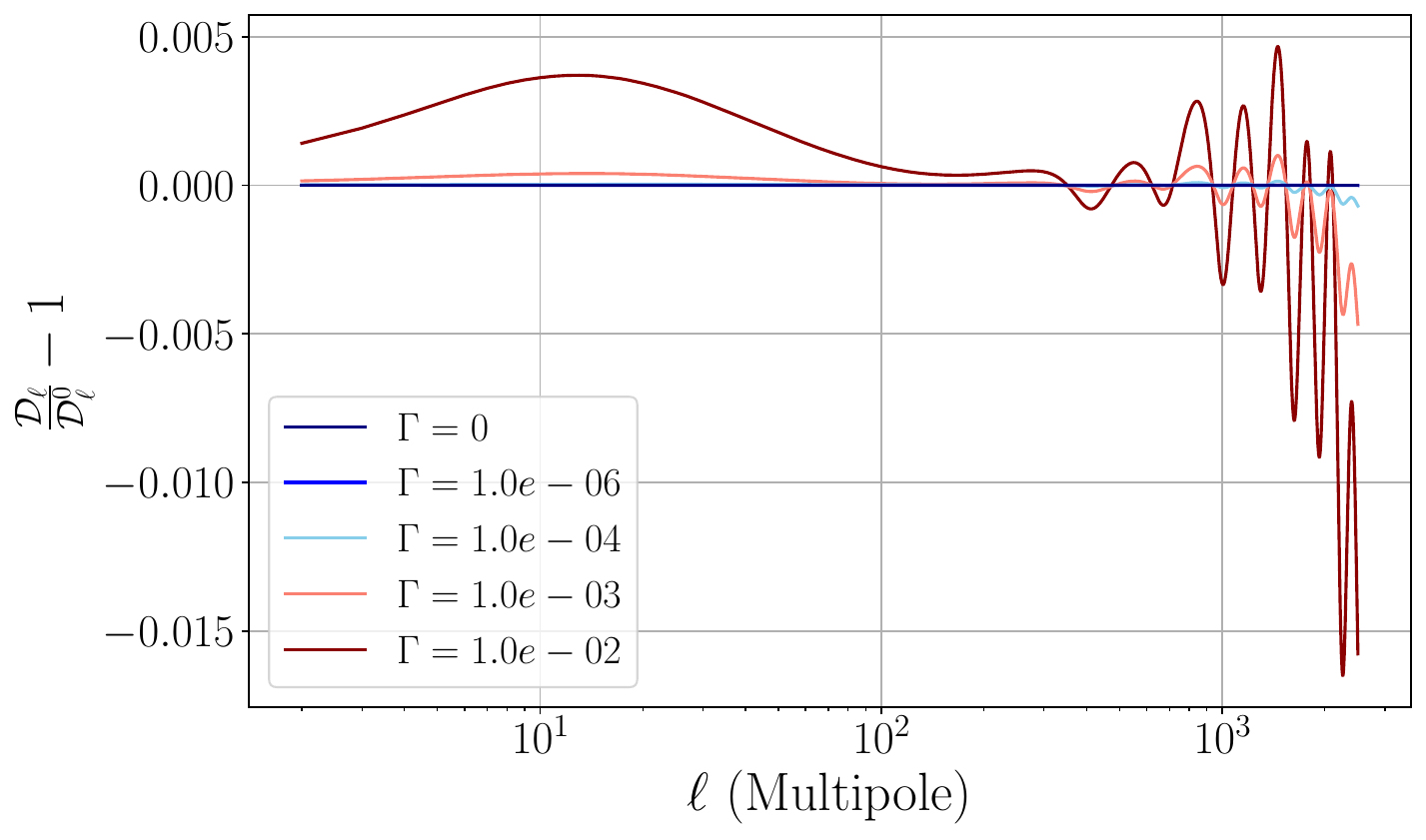}
        \caption{}\label{delta_C_l}
    \end{subfigure}
    \caption{\fontsize{9}{11}\selectfont The CMB TT power spectrum for increasing values of $\Gamma$ in $\text{km}/\text{s}/\text{Mpc}$. Figure \ref{delta_C_l} shows the fractional change in the power spectrum compared to $\Lambda$CDM, $\mathcal{D}_\ell^{0}$. At low $\ell$ there is an enhancement due to an increased late ISW contribution. At large $\ell$, the diffusion has a de-lensing effect.}
    \label{CMB_TT_power_spec}
\end{figure}

\subsection{Effect on Observables}

Figure \ref{C_l} shows the CMB TT power spectrum for various $\Gamma$ values. Apart from an enhancement of $C_\ell$ at $\ell<100$ for $\Gamma$ values around $1\text{km}/\text{s}/\text{Mpc}$, other changes are less visually apparent. Figure \ref{delta_C_l} better shows the fine details by plotting the fractional change in $C_\ell$ compared to $\Lambda$CDM ($\Gamma=0$). Here, we can better see the change at small angular scales. This pattern resembles the decaying dark matter model discussed in \citep{poulin2016fresh}. As detailed in Appendix~\ref{Appendix_Analytic_sol} and Figure \ref{phi_evolution}, stochastic dark matter induces a decay in the gravitational potential at small scales even during matter domination. A time-dependent gravitational potential enhances the late ISW effect, hence increasing the power at small $\ell$. On the other hand, due to the suppression of structures at small scales, the amount of lensing is reduced, which results in sharper peaks but an overall decrease in the power at large $\ell$. 

\begin{figure}[h!]
    \centering
    \begin{subfigure}[b]{0.48\textwidth}
        \centering
        \includegraphics[width=\textwidth]{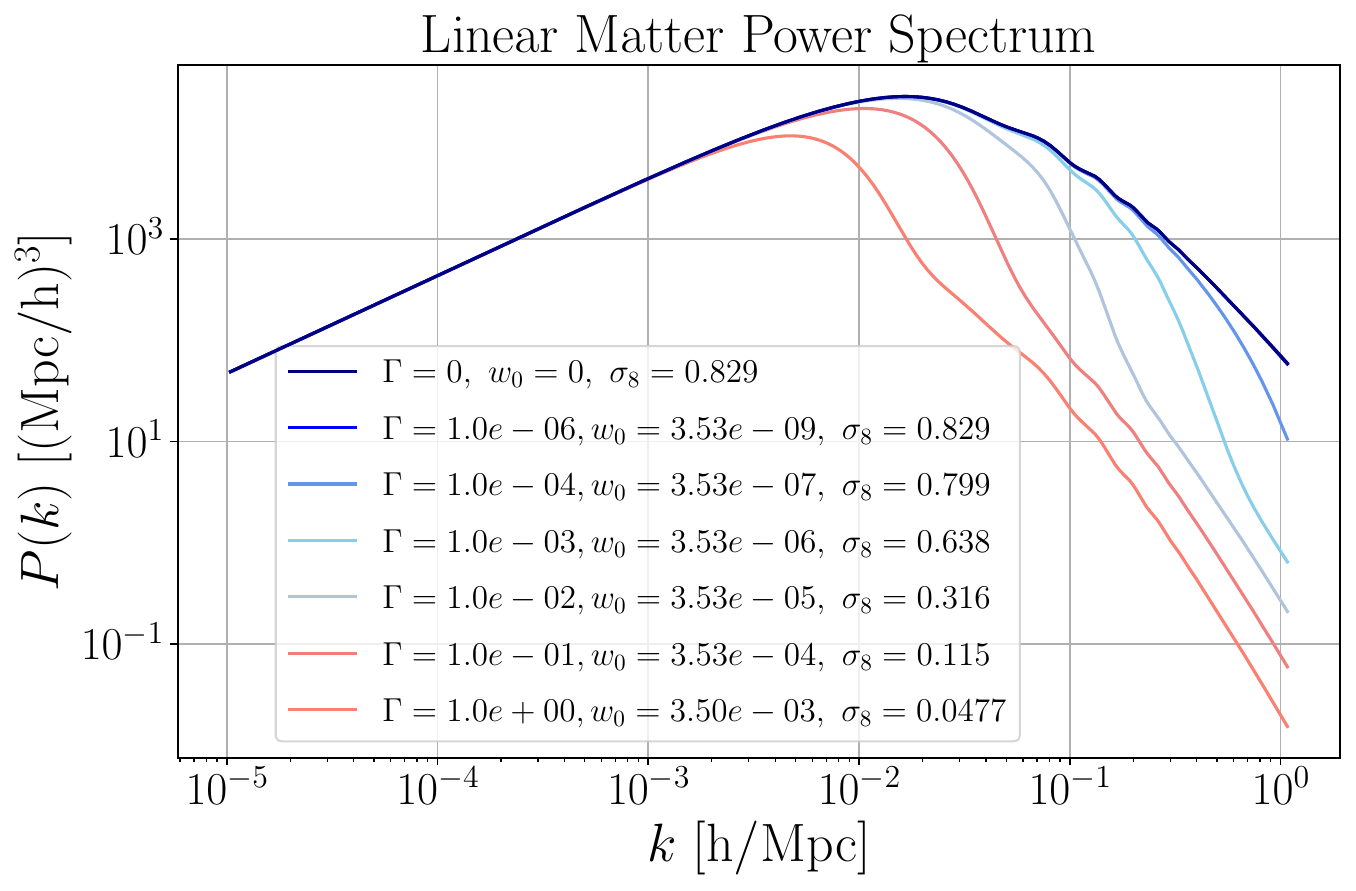}
        \caption{}       \label{P_k}
    \end{subfigure}
    \hfill
    \begin{subfigure}[b]{0.51\textwidth}
        \centering
        \includegraphics[width=\textwidth]{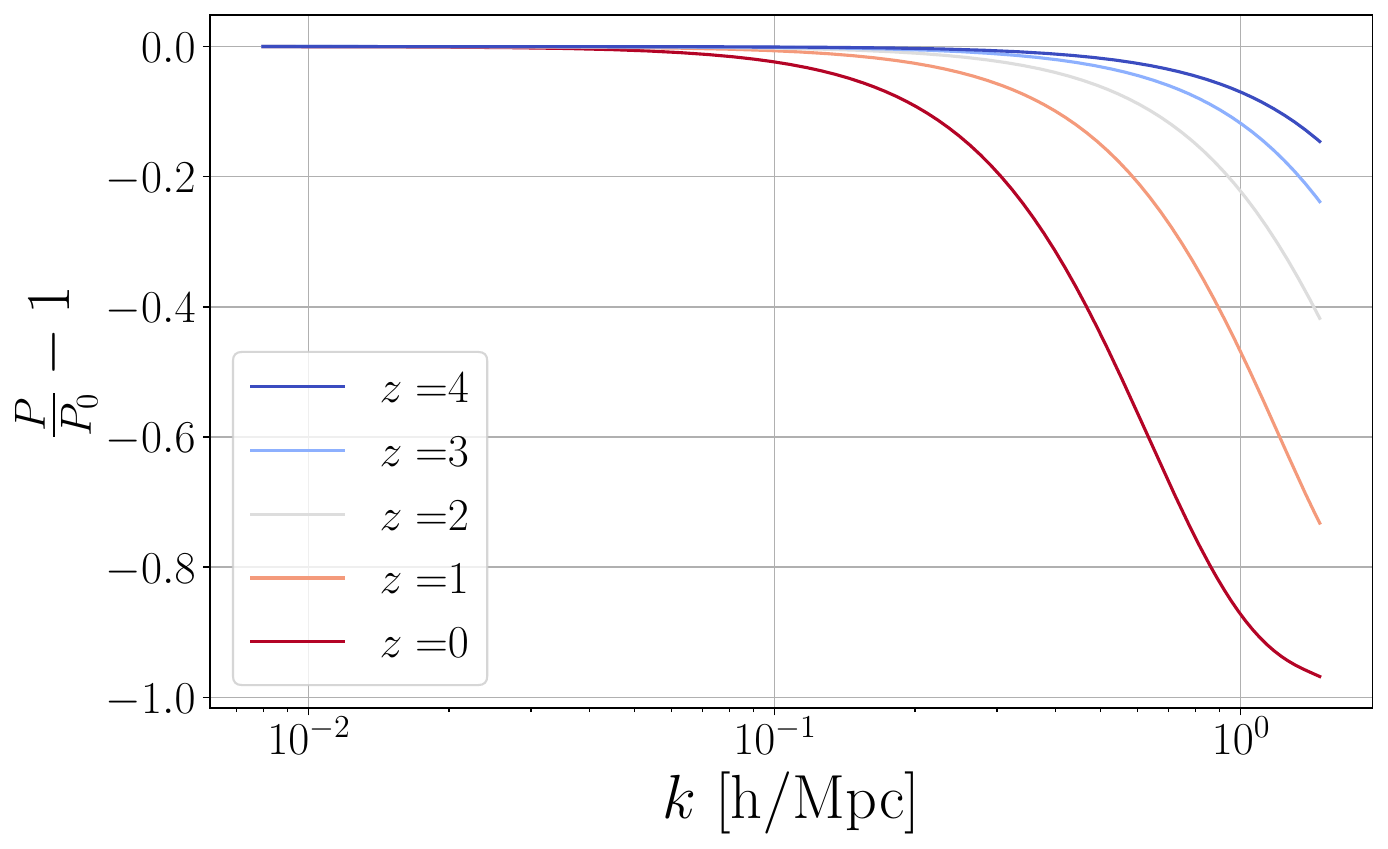}
        \caption{}\label{delta_P_k}
    \end{subfigure}
    \caption{\fontsize{9}{11}\selectfont Suppression of the matter power spectrum in stochastic dark matter scenario. Figure \ref{P_k} shows the power law suppression of the matter power spectrum at $z=0$ for increasing values of $\Gamma$ in $\text{km}/\text{s}/\text{Mpc}$, which is accompanied by increasing values of $w_{dm}$ at present and decreasing values of $\sigma_8$. Figure \ref{delta_P_k} shows the fractional change in the matter power spectrum compared to the power spectrum of $\Lambda$CDM, $P_0$, for $\Gamma=1.44\times 10^{-4}\text{km}/\text{s}/\text{Mpc}$ at various redshifts.}
    \label{Matter_power_spec}
\end{figure}

Figure \ref{P_k} shows the effect of increasing values of $\Gamma$ on the matter power spectrum. We can clearly see the power law suppression at small scales, which is accompanied by increasing values of the present-day equation of state of dark matter and decreasing values of $\sigma_8$—the expected amplitude of matter fluctuations at the scale of $8h^{-1}$Mpc. Figure \ref{delta_P_k} shows the relative suppression of the matter power spectrum compared to $\Lambda$CDM at different redshifts for a fixed diffusion rate. The suppression of power at small scales is redshift-dependent, with a stronger suppression at lower redshifts. These are the distinguishing features of stochastic dark matter. For example, the interacting dark energy-dark matter model studied in \citep{matteo2021dark}, with a net flow of energy from dark energy to dark matter, was shown to suppress the matter power spectrum and relieve the $S_8$ tension. However, the power spectrum is modified at all scales $k$, there is a shift in the location of the peak, and the suppression plateaus at very large $k$ relative to $\Lambda$CDM. For massive neutrinos and many warm dark matter or interacting dark matter models, the interesting new physics happens at sufficiently early times so that the relative matter power spectrum is weakly dependent on redshift at late times \citep{agarwal2011effect,boyanovsky2011small,beltran2021probing,he2023s8}. The closest models in the literature to stochastic dark matter are decaying dark matter models that feature similar redshift-dependent suppression of power at small scales to offer resolution for the $S_8$ tension \citep{abellan2022implications}. However, they also modify the matter power spectrum at large scales, and the suppression plateaus at very large $k$ \citep{wang2012effects,aoyama2014evolution}. Unlike such scenarios, for stochastic dark matter, the new physics is happening only at small scales and at late times. This feature helps the model to satisfy CMB lensing constraints which are mainly sensitive to matter overdensity at $2\lesssim z\lesssim5$, while suppressing the matter power spectrum strong enough at $z\lesssim1$ to decrease $S_8$ and make it consistent with the weak lensing measurements.

Figure \ref{s8vsGamma} shows the dependence of $\sigma_8$ on $\Gamma$ while keeping the other cosmological parameters fixed. One can see that for $\Gamma<10^{-5}\text{km}/\text{s}/\text{Mpc}$, $\sigma_8$ remains virtually unchanged, while for $\Gamma>10^{-3}\text{km}/\text{s}/\text{Mpc}$, the decay in structure formation is too strong. This gives a clear picture of the order of magnitude of $\Gamma$ if it is to satisfy the cosmological constraints.

\begin{figure}[h!]
    \centering
    \begin{subfigure}[b]{0.48\textwidth}
        \centering
        \includegraphics[width=\textwidth]{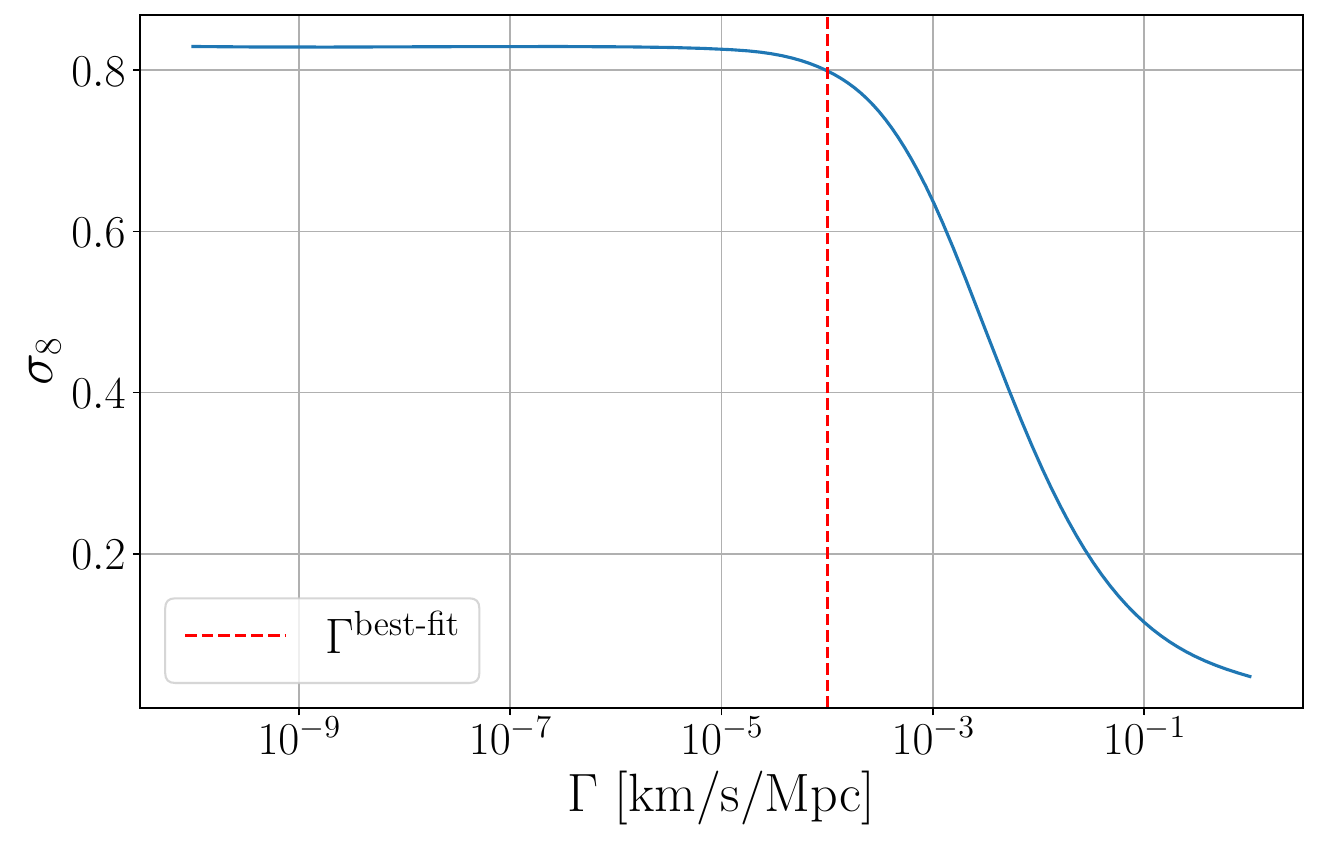}
        \caption{}       \label{s8vsGamma}
    \end{subfigure}
    \hfill
    \begin{subfigure}[b]{0.49\textwidth}
        \centering
        \includegraphics[width=\textwidth]{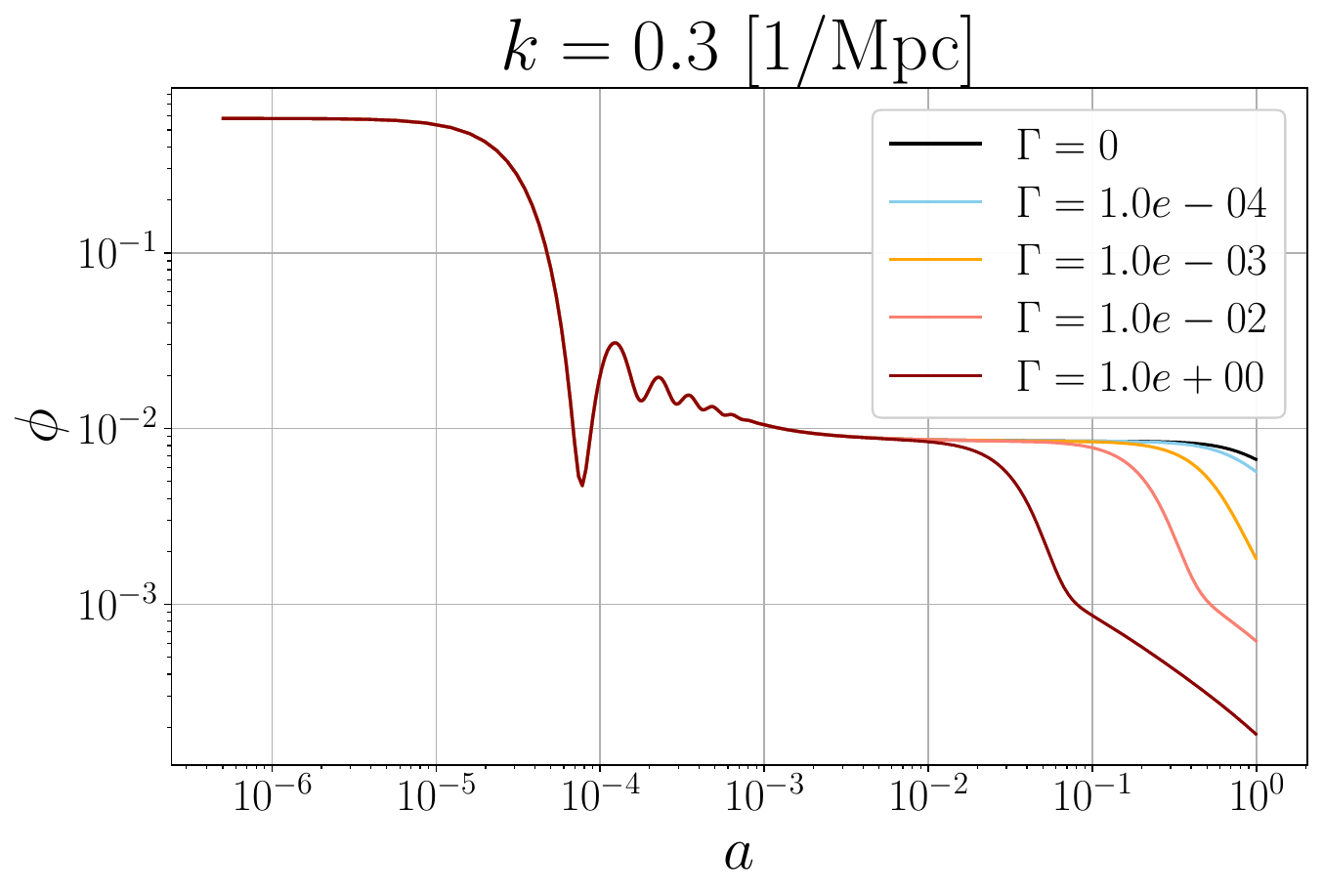}
        \caption{}\label{phi_evolution}
    \end{subfigure}
    \caption{\fontsize{9}{11}\selectfont Figure \ref{s8vsGamma} shows the dependence of $\sigma_8$ on $\Gamma$. Increasing the diffusion rate decreases the clustering of matter. To have enough structure formation, we need to constrain the diffusion rate to $\Gamma\lesssim 10^{-3}\text{km}/\text{s}/\text{Mpc}$. Figure \ref{phi_evolution} shows the gravitational potential $\phi$, normalized by the initial photons overdensity, versus scale factor at $k=0.3/\text{Mpc}$. Unlike CDM, for which $\phi$ is constant during matter domination and only starts to decay during dark matter-dark energy transition epoch, here the decay starts earlier.}
    \label{sigma8vsG}
\end{figure}

\subsection{Data Analysis}
\label{mcmc_section}
The stochastic dark matter scenario has only a single additional free parameter $\Gamma$ to the standard six free parameters of $\Lambda$CDM: $\{H_0,$ $\Omega_{dm}h^2,$ $\Omega_{b}h^2,$ $n_s,$ $ln(10^{10}A_s),$ $\tau_{reio}\}$. To constrain these parameters from cosmological measurements, we perform Markov Chain Monte Carlo (MCMC) using \texttt{MontePython} \citep{Brinckmann:2018cvx,Audren:2012wb}. We employ flat priors for all parameters, including a flat prior of $[0,1.0]\text{km}/\text{s}/\text{Mpc}$ for $\Gamma$. 

As explained in \citep{abdalla2022cosmology}, any model that claims a resolution for the $S_8$ tension should be tested against a variety of different datasets. Our datasets include CMB high-$\ell$ TT, TE, EE, low-$\ell$ TT, EE, and lensing power spectra \citep{aghanim2020planck,aghanim2020plancklike}, as well as expansion and growth histories data from SDSS DR7 Main Galaxy Sample (MGS) \citep{howlett2015clustering}, BOSS DR12 LOWZ and LRG samples \citep{alam2017clustering}, eBOSS DR16 LRG and Quasar (QSO) samples \citep{bautista2021completed,gil2020completed,hou2021completed,neveux2020completed,alam2021completed}, eBOSS DR16 Ly-$\alpha$ \citep{des2020completed}, and uncalibrated Pantheon+ SNIa catalog \cite{brout2022pantheon+}. Together, we call these our Baseline.  We also include weak lensing measurements as the main drivers of the $S_8$ tension in $\Lambda$CDM. The latest cosmic shear measurement by the DES collaboration reports $S_8=0.759^{+0.025}_{-0.023}$~\citep{amon2022dark,secco2022dark}. Following similar analyses of the $S_8$ tension for power-suppressing models \citep{beltran2021probing,abellan2022implications,he2023s8}, we only use this measured value of $S_8$ as a prior for our analysis. This requires model independence of the $S_8$ measurements. Studies by both the KiDS and DES collaborations \citep{troster2021kids,abbott2023dark} indicate robustness of $S_8$ value for various models, including a massive neutrino model $\nu$CDM and a massive sterile neutrino model $N_\text{eff}-m_\text{eff}$ which show power suppression at small scales. However, to the best of our knowledge, none of these tested models show significant modification to the late-time structure growth like the stochastic dark matter. So when using $S_8$ measurements as priors, this caveat should be borne in mind. A more complete analysis of stochastic dark matter in light of weak lensing measurements, with accurate modeling of non-linearities or with proper scale cuts is beyond the scope of the present paper, and we leave it to future work. We have also tested that using the latest cosmic shear measurement by KiDS \citep{li2023kids}, $S_8=0.776^{+0.029}_{-0.027}$, does not change our conclusions.

\begin{figure}[h!]
    \centering
    \begin{subfigure}[b]{0.49\textwidth}
        \centering
        \includegraphics[width=\textwidth]{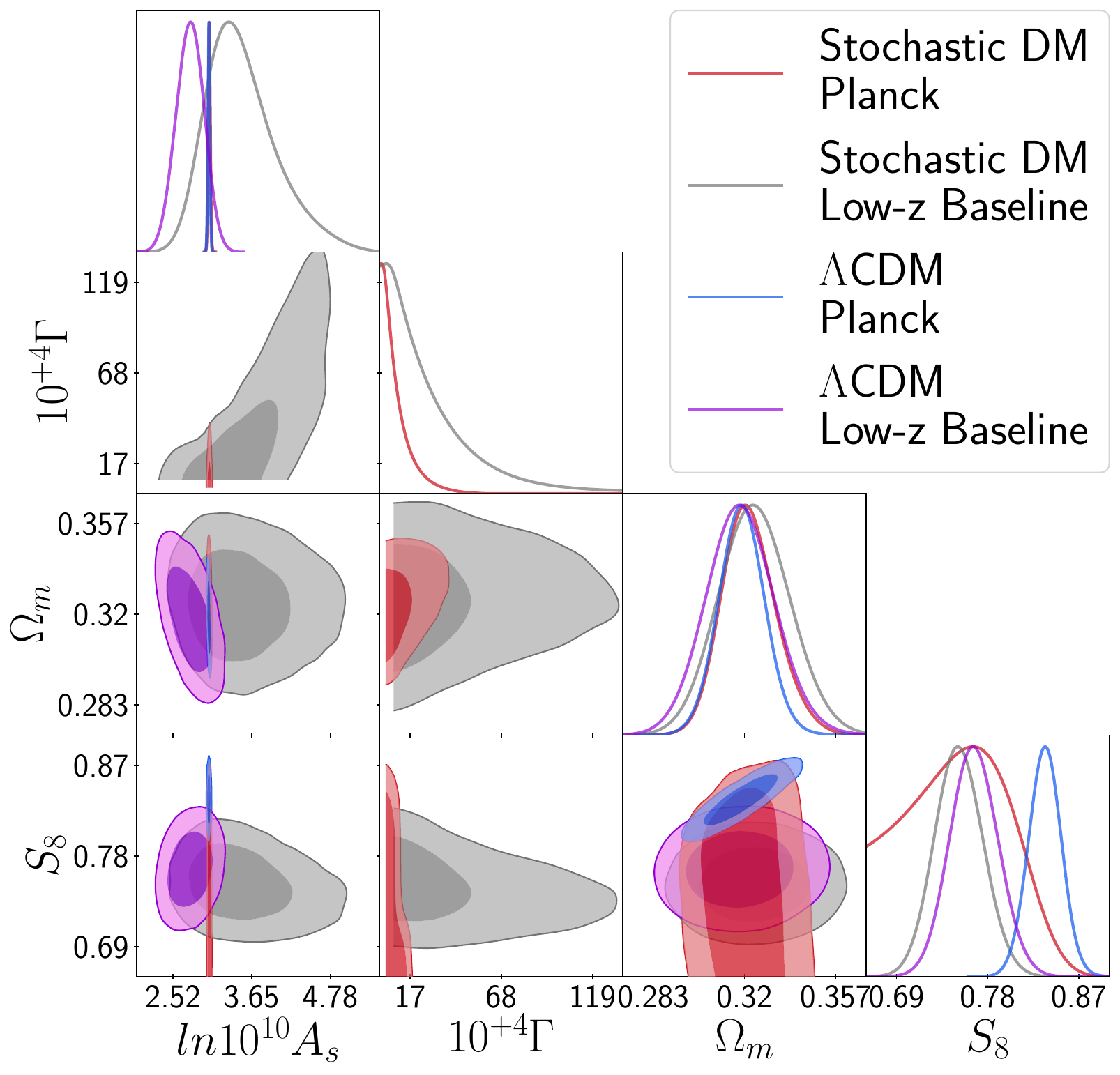}
        \caption{}       \label{mini_triangle_lowz_vs_planck}
    \end{subfigure}
    \hfill
    \begin{subfigure}[b]{0.49\textwidth}
        \centering
        \includegraphics[width=\textwidth]{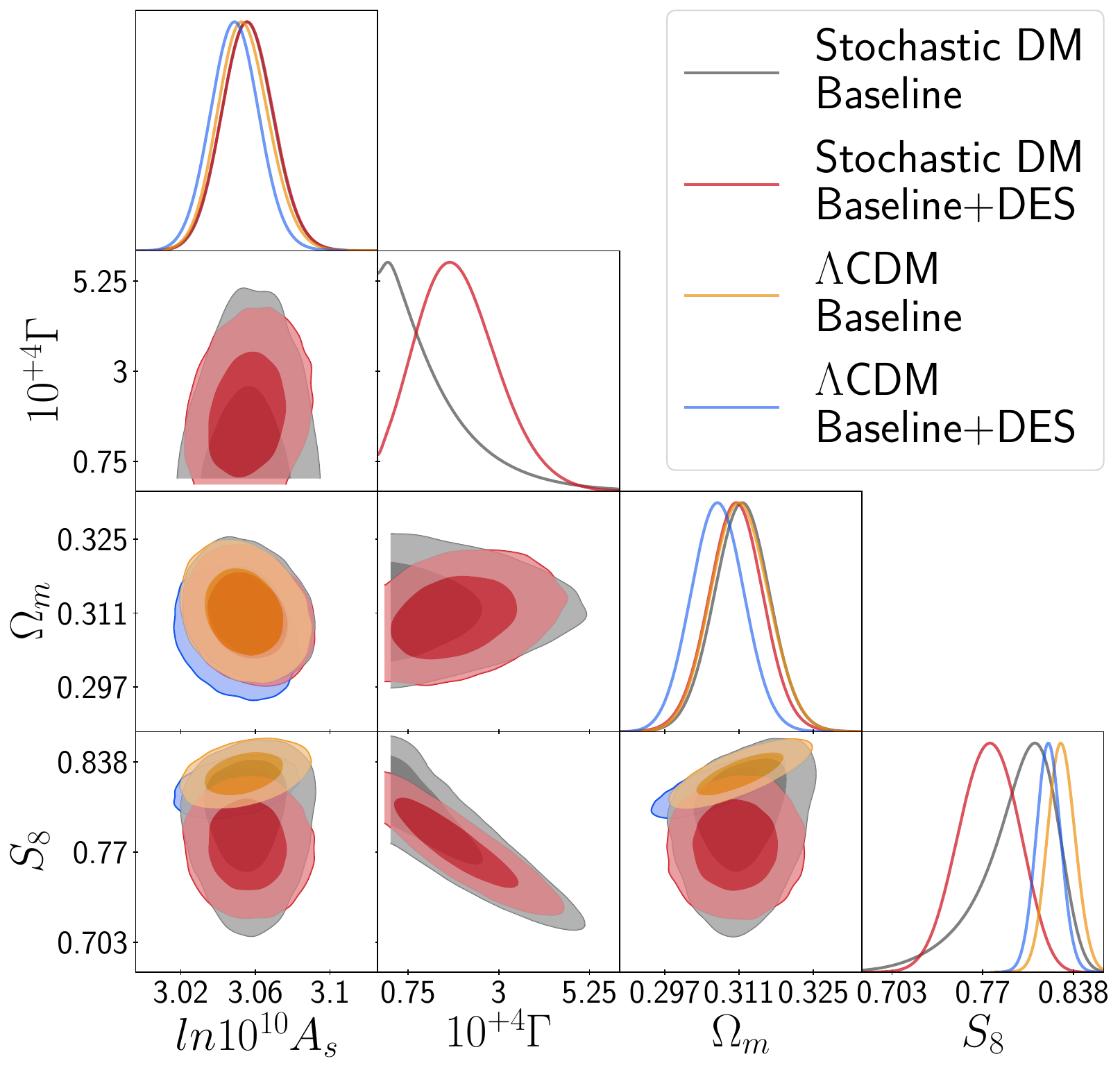}
        \caption{}\label{mini_triangle_Baseline_DES}
    \end{subfigure}
    \caption{\fontsize{9}{11}\selectfont 68\% and 95\% confidence-level posterior distributions of the amplitude of primordial fluctuations, the diffusion rate of stochastic dark matter, $\Omega_m$, and $S_8$ for stochastic dark matter and $\Lambda$CDM from various combinations of datasets. Figure \ref{mini_triangle_lowz_vs_planck} emphasizes the discrepancy in $S_8$ between Planck and low-z measurements in $\Lambda$CDM, while showing consistency in the case of stochastic dark matter. Figure \ref{mini_triangle_Baseline_DES} shows the $1.7\sigma$ preference for non-zero $\Gamma$ when all of the datasets are taken into account together with the $S_8$ prior from DES.}
    \label{mini_triangle}
\end{figure}

We want to assess how stochastic dark matter affects the $S_8$ tension and how it compares to $\Lambda$CDM. First, we only keep those datasets that, when combined with Pantheon+ and an $S_8$ prior, result in a posterior average $S_8<0.785$ for $\Lambda$CDM. These are SDSS DR7 MGS,  eBOSS DR16 LRG, eBOSS DR16 Ly-$\alpha$, uncalibrated Pantheon+ SNIa, and DES, which we call the “Low-$z$ Baseline" data. Figure \ref{mini_triangle_lowz_vs_planck} compares the posterior plots of $\Lambda$CDM and stochastic dark matter for Planck2018 with Low-$z$ Baseline. For $\Lambda$CDM, there is a $2.4\sigma$ tension in the respective $S_8$ posteriors. In contrast, the stochastic dark matter model effectively resolves this tension. The $S_8$ posterior from Planck2018 under this model has an extended lower tail, such that it is consistent with the low-$z$ constraint. It is important to note that this is not just through having a large parameter space and enlarged error bars. The diffusion rate $\Gamma$ specifically targets $S_8$, while maintaining the posterior precision of other parameters at almost the same level as $\Lambda$CDM. This is shown in Appendix \ref{full_posterior_appendix} by comparing the full posteriors of $\Lambda$CDM and stochastic dark matter for Planck2018. In particular, stochastic dark matter does not affect the $H_0$ tension. This is not surprising because the allowed range for the diffusion rate satisfies $\Gamma\ll H_0$, and hence the background is unaffected by the model.

We have seen that in the stochastic dark matter model, the Planck data has a very low constraining power on $S_8$, making it consistent with the Low-z Baseline data. The consistency allows us to combine these datasets. Another approach to evaluate the tension or consistency in $S_8$ is to compare the best-fit $\chi^2$ values for a given model when combining all datasets, both with and without the $S_8$ prior. The tension can then be estimated by calculating the square root of the $\chi^2$ difference between these two cases~\citep{abellan2022implications}. Table~\ref{chi_squared_table} gives the minimum $\chi^2$ for various combinations of datasets. After including the $S_8$ prior, the minimum $\chi^2$ for $\Lambda$CDM increases by $\sim6.7$, while for stochastic dark matter, it only increases by $\sim1.2$. Again, this shows a reduction in the $S_8$ tension from~$2.6\sigma$ in $\Lambda$CDM to~$1.1\sigma$ in stochastic dark matter. 

\begin{table}[h]
\centering
\caption{\fontsize{10}{11}\selectfont Best-fit \(\chi^2\) of stochastic dark matter and $\Lambda$CDM from different datasets. \(\Delta \chi^2\) denotes the best-fit $\chi^2$ of stochastic dark matter minus that of $\Lambda$CDM.}
\label{chi_squared_table}
\begin{tabular}{lcccc}
    \hline
    \textbf{} & \textbf{Planck} & \textbf{Low-z Baseline} & \textbf{Baseline} & \textbf{Baseline+DES} \\
    \hline
    \(\text{Stochastic DM }\chi^2\) & 1192.44 & 1418.16 & 4215.26 & 4216.5 \\
    \(\text{$\Lambda$CDM }\chi^2\) & 1191.76 & 1419.30 & 4212.98 & 4219.7 \\
    \(\Delta \chi^2\) & +0.68 & \scalebox{2.5}[1]{-}1.14 & +2.28 & \scalebox{2.5}[1]{-}3.2 \\
    \hline
\end{tabular}
\end{table}

Finally, we use the Baseline and Baseline+DES to get the strongest constraints on $\Gamma$ and $S_8$ for stochastic dark matter. Figure \ref{mini_triangle_Baseline_DES} shows the posterior plots of this analysis. First, in the case of $\Lambda$CDM, adding the DES datapoint has the effect of slightly shifting $\Omega_m$ to lower values to obtain a smaller $S_8$. For stochastic dark matter, however, $\Omega_m$ remains virtually unchanged after adding the DES datapoint. Instead, by tuning $\Gamma$, $S_8$ becomes consistent with the DES measurement. Although the posteriors are consistent with the diffusion rate being zero, here we see a preference for non-zero $\Gamma$ that ranges between $1.2\sigma$ and $1.7\sigma$, depending on whether we use the $S_8$ prior from KiDS or DES. We highlight that this non-zero preference is entirely driven by the $S_8$ prior from weak lensing measurements. While the Baseline still prefers the $\Lambda$CDM model, as shown in Table \ref{chi_squared_table}, stochastic dark matter provides a better best-fit $\chi^2$ with Baseline+DES by about $3.2$. Precise model comparison, however, remains beyond the scope of this paper and is deferred to future studies.

\begin{table}[h]
\label{Full_parameter_constraints_table}
    \centering
    \caption{\fontsize{10}{11}\selectfont Set of best-fit parameters and posterior averages and standard deviations from the analysis of Baseline and Baseline+DES. At each entry, the number on top is the best-fit parameter.}
    \begin{tabular}{lcccc}
        \hline
        \multirow{2}{*}{\textbf{Parameter}} & \multicolumn{2}{c}{\textbf{Stochastic DM}} & \multicolumn{2}{c}{\textbf{$\Lambda$CDM}} \\
        \cline{2-3} \cline{4-5}
        & \textbf{Baseline} & \textbf{Baseline+DES} & \textbf{Baseline} & \textbf{Baseline+DES} \\
        \hline
        \multirow{2}{*}{$100 \Omega_{b}h^2$} & 2.244 & 2.245 & 2.235 & 2.240 \\
        & $2.239^{+0.013}_{-0.014}$ & $2.240^{+0.013}_{-0.014}$ & $2.240^{+0.013}_{-0.014}$ & $2.245^{+0.013}_{-0.013}$ \\
        \hline
        \multirow{2}{*}{$\Omega_{dm}h^2$} & 0.1190 & 0.1195 & 0.1200 & 0.1188 \\
        & $0.1198^{+0.0009}_{-0.0009}$ & $0.1196^{+0.0009}_{-0.0009}$ & $0.1197^{+0.0009}_{-0.0009}$ & $0.1190^{+0.0008}_{-0.0008}$ \\
        \hline
        \multirow{2}{*}{$\ln(10^{10}A_s)$} & 3.060 & 3.057 & 3.052 & 3.056 \\
        & $3.055^{+0.014}_{-0.015}$ & $3.055^{+0.014}_{-0.015}$ & $3.053^{+0.014}_{-0.015}$ & $3.047^{+0.013}_{-0.015}$ \\
        \hline
        \multirow{2}{*}{$n_s$} & 0.9680 & 0.9676 & 0.9671 & 0.9687 \\
        & $0.9661^{+0.0037}_{-0.0036}$ & $0.9664^{+0.0036}_{-0.0036}$ & $0.9662^{+0.0036}_{-0.0036}$ & $0.9674^{+0.0036}_{-0.0037}$ \\
        \hline
        \multirow{2}{*}{$\tau_{reio}$} & 0.0630 & 0.0606 & 0.0574 & 0.0618 \\
        & $0.0591^{+0.0067}_{-0.0078}$ & $0.0594^{+0.0070}_{-0.0077}$ & $0.0583^{+0.0068}_{-0.0076}$ & $0.0567^{+0.0066}_{-0.0074}$ \\
        \hline
        \multirow{2}{*}{$10^{+4}\Gamma$} & 0.157 & 1.44 & — & — \\
        & $1.29^{+0.24}_{-1.29}$ & $2.00^{+0.84}_{-1.18}$ & —& — \\
        \hline
        \multirow{2}{*}{$H_0$} & 67.82 & 67.70 & 67.40 & 67.88 \\
        & $67.52^{+0.39}_{-0.40}$ & $67.60^{+0.39}_{-0.39}$ & $67.54^{+0.39}_{-0.40}$ & $67.84^{+0.38}_{-0.38}$ \\
        \hline
        \multirow{2}{*}{$\Omega_m$} & 0.3076 & 0.3096 & 0.3133 & 0.3064 \\
        & $0.3119^{+0.0053}_{-0.0054}$ & $0.3107^{+0.0052}_{-0.0053}$ & $0.3115^{+0.0052}_{-0.0055}$ & $0.3074^{+0.0050}_{-0.0050}$ \\
        \hline
        \multirow{2}{*}{$S_8$} & 0.820 & 0.786 & 0.832 & 0.821 \\
        & $0.795^{+0.036}_{-0.017}$ & $0.774^{+0.023}_{-0.022}$ & $0.829^{+0.010}_{-0.010}$ & $0.819^{+0.009}_{-0.010}$ \\
        \hline
    \end{tabular}
    \label{tab:parameter_constraints}
\end{table}

\subsection{Discussion}

The best-fit diffusion rate from Baseline+DES is
\begin{equation}
\label{Gamma_bf}
    \Gamma=1.44\times 10^{-4}\text{km}/\text{s}/\text{Mpc}.
\end{equation}
This is about $6$ orders of magnitude smaller than $H_0$, implying a “diffusion time" $\Gamma^{-1}$ that is a million times the age of the universe. At first glance, this could suggest a fine-tuning problem from a quantum gravity perspective, potentially making \eqref{Gamma_bf} too small to be explained even by Planckian effects. 

However, $\Gamma^{-1}$ should not be considered the primary time scale for evaluating the strength of quantum gravity effects. Rather, it is the time scale over which the covariant Brownian motion affects the background energy density on cosmological scales. In principle, different quantum gravity theories would provide distinct mechanisms for diffusion. Within each model, $\Gamma$ has to be related to some more fundamental parameters. Thus, the interpretation of the above value is highly model-dependent.

As a specific example, we consider causal sets. In swerves models, there is a more fundamental time scale, the non-local forgetting time $\tau_f$. In \citep{philpott2010particle}, Philpott used numerical simulations in two-dimensional Minkowski spacetime to show that the diffusion constant $\kappa$ can be related to the forgetting time via
\begin{equation}
\label{kappa_tauf}
    \kappa\approx 2m^2\frac{t_{pl}^4}{\tau_f^5},
\end{equation}
where $t_{pl}$ is the Planck time. Combining with the definition of $\Gamma$ in \eqref{Gamma_definition}, we can relate $\Gamma$ to the forgetting time:
\begin{equation}
\label{tau_f_Gamma}
    \Gamma\approx 6\frac{t_{pl}^4}{\tau_f^5}\quad\rightarrow\quad \tau_f\sim\Big(\Gamma^{-1}t_{pl}^4\Big)^{1/5}.
\end{equation}
Strictly speaking, this relation is confirmed only in flat two-dimensional spacetime. However, assuming that the ratio of powers is similar in more general cases, \eqref{Gamma_bf} gives an estimate for the forgetting time of stochastic dark matter particles
\begin{equation}
\label{10^13t_pl}
    \tau_f\sim 10^{13}t_{pl}.
\end{equation}
This is a time scale much larger than the Planck time, but still four orders of magnitude below typical time scales in the Standard Model of particle physics and significantly smaller than any cosmological scale. Another non-local scale that appears in the discussion of d'Alembertian on causal sets has been estimated to be 20 orders of magnitude larger than the Planck length~\citep{sorkin2007does}. Had these non-local scales been of the order of cosmological scales, we would have had a fine-tuning problem. Further implications of the above forgetting time require more sophisticated models of matter propagation on a causal set, potentially relating the forgetting time to the Compton wavelength of the particle or its wave-packet properties \citep{dowker2013introduction}.

How does our best-fit $\Gamma$ compare to previous bounds on the diffusion parameter? Since we are agnostic about the mass of dark matter particles, we cannot translate our result into a bound on $\kappa$. Instead, the bounds on $\kappa$ presented in \citep{dowker2004quantum,kaloper2006low}, which are based on cold baryonic systems, can be translated into bounds on their diffusion rate $\Gamma_b$. These studies suggest an upper bound within the range
\begin{equation}
\label{Gamma<1kmsMpc}
    \Gamma_b\lesssim [0.1, 10] \text{km}/\text{s}/\text{Mpc}.
\end{equation}
Therefore, in terms of the diffusion rate, our cosmological bounds on dark matter are much tighter. Nevertheless, due to the larger power of the Planck time in \eqref{tau_f_Gamma}, they yield approximately the same lower bound \eqref{10^13t_pl} on the forgetting time of baryons. The authors in \citep{kaloper2006low} also estimated an upper bound on the diffusion constant $\kappa$ of massive neutrinos from existing bounds on hot dark matter. Although their bound on $\kappa$ is much tighter than that for baryonic systems, the small mass of neutrinos results in a weaker bound on the diffusion rate than \eqref{Gamma<1kmsMpc}.

\section{Conclusions and Future Directions}

Inspired by swerves models that describe the deviation of massive particles from geodesics in discrete spacetimes of causal set theory, we built on previous work on the Lorentz-invariant diffusion equation in flat spacetime. We laid the groundwork needed to formulate all possible covariant stochastic processes in curved spacetime. By focusing on the minimal coupling case, we derived the unique covariant Brownian motion.

We applied the covariant Brownian motion framework to dark matter particles, developing a model of stochastic dark matter. We found that the effect of Brownian motion on perturbations becomes significant much earlier than on the background. It counteracts the dark matter clustering and dissipates overdensities caused by gravitational collapse, leading to a bottom-up suppression of the matter power spectrum. This allows stochastic dark matter to meet cosmological constraints on large scales and at higher redshifts while resolving the $S_8$ tension.

A key question is the behavior of stochastic dark matter at sub-megaparsec scales, and whether the model can satisfy Milky Way satellite measurements \citep{kim2018missing,nadler2021constraints} and Ly-$\alpha$ forest flux power spectrum data \cite{villasenor2023new,hooper2022one}, which impose strong constraints on warm dark matter models. Since matter power suppression is strongest at $z\lesssim1$, Milky Way satellite measurements are expected to place tighter constraints on stochastic dark matter. Addressing this requires fully exploring the non-linear regime through N-body simulations for stochastic dark matter, an important avenue for future research. This will also provide more accurate theoretical computations to compare with CMB lensing and weak-lensing measurements. Moreover, it would determine whether stochastic dark matter can resolve small-scale problems of CDM \citep{de2010core,bullock2017small,pontzen2014cold,salucci2019distribution,zavala2019dark}.

We can further test the covariant Brownian motion by applying it to other matter particles, including baryons and electrons. The bounds \eqref{Gamma<1kmsMpc} already rule out significant diffusion in baryons in the early universe that could affect CMB spectral distortions. Nonetheless, these bounds allow for spontaneous heating of baryons in the late universe. Our formulation of the covariant diffusion equation in curved spacetime and its semi-analytic treatment in FRW spacetime pave the way for a systematic treatment of baryons. The diffusion rate of baryons can then be constrained through various methods, including its impact on the matter power spectrum, Ly-$\alpha$ bounds on the temperature of the intergalactic medium \citep{liu2021lyman}, and the heating of gas-rich dwarf galaxies \citep{wadekar2022strong}. We expect the resulting bounds on the diffusion rate of baryons to improve \ref{Gamma<1kmsMpc}, making them comparable to those we have found for dark matter.\\

There are several open questions on the theoretical side. Our motivation for considering covariant Brownian motion stemmed from discrete quantum gravity. Another motivation concerns the effect of vacuum fluctuations of a scalar or electromagnetic field on a point particle coupled to that field \citep{gour1999will,johnson2002stochastic,bessa2009brownian}. As noted in \citep{gour1999will}, this process should be Lorentz invariant in the vacuum. The effective approach we developed in section \ref{SDE_section} can also inform all possible covariant stochastic processes in the presence of an additional field.  Future work should explore the connection between our effective approach and existing Langevin equations. Another independent motivation for studying particle diffusion could arise from spontaneous collapse models \citep{bedingham2013single}. In relativistic models \citep{tumulka2006relativistic,bedingham2011relativistic}, we expect that a suitable Markovian, point-particle limit of these dynamics will recover the covariant Brownian motion. Furthermore, attempts to reconcile quantum field theory with general relativity without quantizing the gravitational degrees of freedom \citep{PhysRevX.13.041040} inevitably lead to collapse within the quantum system and a stochastic modification to Einstein's equations. It is plausible, therefore, that alongside effects due to the stochasticity of spacetime itself \cite{oppenheim2024emergencephantomcolddark}, an accurate description of the phenomenology of the dark sector in such theories necessarily requires taking the covariant Brownian motion of particles into account.

Returning to our original motivation from discrete spacetime, a natural generalization is to consider particles of higher spin. Lorentz invariant fluctuations of photon polarization have been studied in \citep{Contaldi_2010}. Our work prepares the ground for generalizing this to curved spacetime and finding the corresponding Langevin equations on the Bloch sphere. Moving on from the point particle approximation, it is crucial to find a field-theoretic basis for covariant Brownian motion through a stochastic correction to the Klein-Gordon equation. A key question is whether Lorentz invariance imposes similar strict restrictions on stochastic corrections within the classical covariant phase space of scalar fields \citep{crnkovic1987covariant}. We are interested in recovering covariant Brownian motion from the particle approximation of such a Brownian Klein-Gordon equation. An example of a causal but non-local scalar field theory on a causal set exists, which, in the continuum limit, yields non-local corrections to the Klein-Gordon equation \citep{belenchia2015nonlocal,sorkin2007does,nasiri2023synge}. It remains an open question whether this non-locality effectively manifests as a stochastic effect and whether it can recover covariant Brownian motion.

Finally, there is the question of back-reaction. In this work, for cosmological purposes, we used a phenomenological imperfect dark energy fluid to balance total energy-momentum conservation. Some assumptions were made to ensure the dark energy sector is well-behaved and stable, but these must ultimately be justified by a self-consistent treatment of covariant Brownian motion within the framework of Einstein equations. The Langevin-Einstein equations in \citep{hu2008stochastic} offer a potential path forward if we demand the noise tensor to balance energy-momentum non-conservation.

\section*{Acknowledgements} It is a pleasure to thank Fay Dowker and Niayesh Afshordi for many detailed discussions and thoughtful insights, and Rafael Sorkin for his valuable feedback. We also thank Jessie Muir, Caio Nascimento, Ghazal Geshnizjani, Neal Dalal, Benjamin Wallisch, John Moffat, and Farnik Nikakhtar for helpful comments and discussions. This work was funded in part by STFC grant ST/W006537/1. EA is supported by the
STFC Consolidated Grant ST/W507519/1. AN is
funded by the President’s PhD Scholarship from Imperial College London and the Canada
First Research Excellence Fund through the Arthur B. McDonald Canadian Astroparticle
Physics Research Institute. AN also appreciates the hospitality received as a visitor at
the Perimeter Institute for Theoretical Physics during the completion of this work. Research at Perimeter Institute is supported by the Government of Canada through Industry Canada and by the
Province of Ontario through the Ministry of Economic Development and Innovation. EP is supported by the Cosmoparticle Initiative at University College London.

\begin{appendices}
\section{General Covariance of Fokker-Planck Equation} \label{appendix_stochasticev}
In this appendix, we give a brief summary of Sorkin's arguments in \citep{sorkin1986stochastic} for deriving the Fokker-Planck equation from first physical principles. An alternative approach is the standard textbook derivation in stochastic processes, but the following method offers useful insights into the transformation properties of the drift and diffusion coefficients.

Given any region $\Omega$ of the phase space,
\begin{equation}
    \int_\Omega dZ\ \rho(Z,\tau)
\end{equation}
is equal to the number of particles that are in $\Omega$ when their proper time is $\tau$ \footnote{In Sorkin's notation in \citep{sorkin1986stochastic}, $\rho$ is a probability density, so its integral over the whole phase space is 1. This differs from our notation by a normalization factor equal to the total number of particles.}. Since $\rho$ is not integrated with any volume measure, it acts as a density. In other words, under an arbitrary coordinate transformation $Z^A\rightarrow Z'^M(Z)$ of the phase space, $\rho$ changes as 
\begin{equation}
\label{rho_trans_rule}
    \rho'(Z')=\Big|\frac{\partial Z}{\partial Z'}\Big|\rho(Z),
\end{equation}
where $|\frac{\partial Z'}{\partial Z}|$ is the determinant of the Jacobian matrix of the transformation, and primed quantities are in the new coordinate system. 

The basic ansatz for the evolution equation of $\rho$ is 
\begin{equation}
\label{Contin}
    \frac{\partial \rho}{\partial\tau}=K^{AB}\partial_A\partial_B\rho+F^A\partial_A\rho-C\rho,    
\end{equation}
where $K^{AB}(X)$ is a yet-unknown symmetric matrix in the phase space. $F^A(X)$ and $C(X)$ are unknown functions as well.

This is the most general second-order equation that can be written for $\rho$. One might question the generality of this ansatz by considering the inclusion of higher-order derivatives of $\rho$. The first-order derivative in $\tau$ is justified by the need for an evolution equation for $\rho$. In Appendix A in \cite{sorkin1986stochastic}, Sorkin proves that higher phase space derivatives on the right-hand side of this equation will lead to a negative $\rho$ for some particular choices of the initial distribution at $\tau=0$. Hence such terms cannot be present unless one keeps an infinite series of these terms. The same positivity requirement from $\rho$ at all $(Z,\tau)$ forces $K^{AB}$ to be a positive semi-definite matrix.

Now the assumption of conservation of particle number can be written as
\begin{equation}
    \frac{d}{d\tau}\int_\mathcal{S} dZ\ \rho(Z,\tau)=0 \quad\Rightarrow\quad \int_\mathcal{S} dZ\ \frac{\partial\rho}{\partial\tau}(Z,\tau)=0.
\end{equation}
Using \eqref{Contin} and integrating by parts leads to
\begin{equation}
    \int_\mathcal{S} dZ\ (\partial_A\partial_B K^{AB}-\partial_A F^A-C)\rho=0.
\end{equation}
Since this holds for any distribution $\rho$, one concludes that
\begin{equation}
    C=\partial_A\partial_B K^{AB}-\partial_A F^A.
\end{equation}
Defining $v^A=2\partial_B K^{AB}-F^A$ transforms the ansatz into
\begin{equation}
\label{Cont2}
    \frac{\partial\rho}{\partial\tau}=-\partial_A J^A,
\end{equation}
where
\begin{equation}
\label{jdef}
    J^A= v^A\rho-\partial_B(K^{AB}\rho),
\end{equation}
which leads to \eqref{eq:Fokker-Planck}.

Given that $\rho$ is a scalar density, the transformation properties of the other quantities can be derived. Equation~\eqref{Cont2} says that $J^A$ must be a vector density; i.e. it transforms like
\begin{equation}
\label{vecdens}
    J'^M(Z')=\Big|\frac{\partial Z}{\partial Z'}\Big|\frac{\partial Z'^M}{\partial Z^A}J^A(Z).
\end{equation}
This transformation rule for $J$ is valid for \textit{any} distribution $\rho$. Expanding \eqref{vecdens} using the definition \eqref{jdef} together with the transformation rule \eqref{rho_trans_rule}, one finds
\begin{equation}
\label{Ktrans}
    K'^{MN}=\frac{\partial Z'^M}{\partial Z^A}\frac{\partial Z'^N}{\partial Z^B}K^{AB},
\end{equation}
while $v$ mixes with $K$ when transformed;
\begin{equation}
    v'^M=\frac{\partial Z'^M}{\partial Z^A}v^A+\frac{\partial^2 Z'^M}{\partial Z^A\partial Z^B}K^{AB}.
\end{equation}

\subsection{Detailed Derivation of Covariant Diffusion Equation} \label{AppendixC}

Here, we outline the steps in deriving the covariant diffusion equation from the Fokker-Planck equation, which were omitted in section \ref{SorkinStochastic}.

Having found the most general form of $u^A$ and $K^{AB}$ (equations \eqref{u=L/m} and \eqref{K_tensor}) on the phase space $\Gamma_m^+$, we substitute them back into the Fokker-Planck equation to find

\begin{equation}
    \frac{\partial\rho}{\partial\tau}=-\partial_A \Big(\frac{1}{m}L^A\rho\Big)+\kappa\frac{\partial}{\partial q^i}\bigg(\sqrt{|\hat{g}|}h^{ij}\frac{\partial}{\partial q^j}\Big(\frac{\rho}{\sqrt{|\hat{g}|}}\Big)\bigg).
\end{equation}
Now, we have to integrate out the proper time since it is not observable. Using the definition
\begin{equation}
    \sqrt{|\hat{g}|}f(x,p)=\int_{-\infty}^{\infty} d\tau\ \rho,
\end{equation}
one finds
\begin{equation}
\label{redPl}
    \frac{1}{\sqrt{|\hat{g}|}}\partial_A \Big(\frac{1}{m}L^A\sqrt{|\hat{g}|}f\Big)=\kappa\frac{1}{\sqrt{|\hat{g}|}}\frac{\partial}{\partial q^i}\Big(\sqrt{|\hat{g}|}h^{ij}\frac{\partial}{\partial q^j}f\Big).
\end{equation}

To proceed further, we need to be explicit about the choice of the metric $\hat{g}$ on the phase space $\Gamma_m^+$. The natural choice is the Sasaki metric on the cotangent bundle \citep{acuna2022introduction}, for which the volume measure, when restricted to $\Gamma_m^+$, is $d^4xd^3q\sqrt{-g}\sqrt{h}$. As a result, the metric determinant separates into spacetime and momentum contributions, i.e.
\begin{equation}
    \sqrt{|\hat{g}|}=\sqrt{-g}\ \sqrt{h}.
\end{equation}
Let $\hat{\nabla}$ be the covariant derivative corresponding to the Sasaki metric. The left-hand side of \eqref{redPl} is now equal to 
\begin{equation}
    \frac{1}{m}\hat{\nabla}_A(fL^A)=\frac{1}{m}f\hat{\nabla}_AL^A+\frac{1}{m}L[f].
\end{equation}
Liouville's theorem states that $\hat{\nabla}_AL^A=div(L)=0$ \citep{acuna2022introduction}. Therefore, \eqref{redPl} can be written as
\begin{equation}
    L[f]=m\kappa\frac{1}{\sqrt{h}}\frac{\partial}{\partial q^i}\Big(\sqrt{h}h^{ij}\frac{\partial}{\partial q^j}f\Big).
\end{equation}
$h_{ij}$ being the hyperbolic metric on $\mathbb{H}_3$, the right-hand side of this equation gives us the Laplacian operator on $\mathbb{H}_3$. We have found the desired covariant diffusion equation.

\subsection{Covariance of Liouville Operator}\label{appendixCovLiouville}

On the cotangent bundle of the spacetime, the Liouville operator is defined as
\begin{equation}
    L=p^\mu\frac{\partial}{\partial x^\mu}+\Gamma^\sigma_{\mu\nu}p_\sigma p^\mu\frac{\partial}{\partial p_\nu}.
\end{equation}
The spacetime components give the definition of momentum, and its momentum components yield the geodesic motion. In other words, for a particle of mass $m$ with no diffusion, the Liouville operator gives the total derivative of its phase space motion, $dZ/d\tau=L/m$. Here we want to show that $L$ is indeed a true vector under the set of admissible coordinate transformations \eqref{coord_trans}. First, we change the coordinates of the cotangent bundle from $(x^\mu,p_\mu)$ to $(x^\mu,q_a=e_a^{\ \mu}p_\mu)$, where $a=0,...,3$. The coordinate vector fields are written in the new coordinates as
\begin{align}
    \frac{\partial}{\partial x^\mu}&\rightarrow\frac{\partial}{\partial x^\mu}+p_\nu e_{a\ ,\mu}^{\ \nu}\frac{\partial}{\partial q_a},\nonumber\\
    \frac{\partial}{\partial p_\mu}&\rightarrow e_a^{\ \mu}\frac{\partial}{\partial q_a}.
\end{align}
Therefore,
\begin{equation}
    L=p^\mu\frac{\partial}{\partial x^\mu}+p.\nabla_pe_a\frac{\partial}{\partial q_a}.
\end{equation}
Now, we demonstrate that this is indeed a true vector. In components form, $L$ is expressed as 
\begin{equation}
    L=\Big(L^\mu=p^\mu,\ L_a=p.\nabla_p e_a\Big)
\end{equation}
In a new coordinate system $(x'^\alpha,\Tilde{q}_r)$, this would be 
\begin{equation}
    L'=\Big(L'^\alpha=p'^\alpha,\ \Tilde{L}_r=p.\nabla_p \Tilde{e}_r\Big),
\end{equation}
and we have to show that 
\begin{align}
\label{tbs1}
    L'^\alpha=\frac{\partial x'^\alpha}{\partial x^\mu}L^\mu+\frac{\partial x'^\alpha}{\partial q_a}L_a\\
    \label{tbs2}
    \Tilde{L}_r=\frac{\partial \Tilde{q}_r}{\partial x^\mu}L^\mu+\frac{\partial \Tilde{q}_r}{\partial q_a}L_a.
\end{align}
The Jacobian for the coordinate transformation is
\begin{equation}
\setlength{\arraycolsep}{10pt} 
\begin{pmatrix}
  \frac{\partial x'^\alpha}{\partial x^\mu} &  \frac{\partial x'^\alpha}{\partial q_a}=0 \\
  \frac{\partial \Tilde{q}_r}{\partial x^\mu}=\partial_\mu\big(\Tilde{e}_r(x).e^a(x)\big)q_a & \frac{\partial \Tilde{q}_r}{\partial q_a}=\Tilde{e}_r(x).e^a(x)\\
\end{pmatrix}
\end{equation}
Since $\partial x'/\partial q=0$, \eqref{tbs1} reduces to 
\begin{equation}
\label{ptrans}
p'^\alpha=\frac{\partial x'^\alpha}{\partial x^\mu}p^\mu,
\end{equation}
which is trivially valid due to $p^\mu$ being a covariant vector on the spacetime. For equation \eqref{tbs2}, the right-hand side equals
\begin{align}
    p^\mu\partial_\mu(\Tilde{e}_r.e^a)q_a+(\Tilde{e}_r.e^a)p.\nabla_p e_a&=p.\big(p^\mu\partial_\mu(\Tilde{e}_r.e^a)e_a+(\Tilde{e}_r.e^a)\nabla_p e_a\big)\nonumber\\ &=p.\nabla_p\big((\Tilde{e}_r.e^a) e_a\big)=p.\nabla_p \Tilde{e}_r,
\end{align}
and therefore, \eqref{tbs2} is also true. We have shown that $L$ is a true vector field under coordinate transformations of the cotangent bundle of the form $(x^\mu,q_a)\rightarrow(x'^\alpha,\Tilde{q}_r)$, where $a,r=0,...,3$. Now note that since $L/m$ is the Hamiltonian vector field corresponding to the Hamiltonian $H=\frac{1}{2m}p_\mu p_\nu g^{\mu\nu}$, it is tangent to the surfaces of constant $H$. In particular, $L$ is also tangent to the phase space $\Gamma_m^+$ and can be thought of as a vector field on the phase space, covariant under admissible coordinate transformations \eqref{coord_trans}. 
Since $L$ is tangent to $\Gamma_m^+$, restricting it to this submanifold is done just by dropping $\partial/\partial q_0$. In other words, when restricted to $(y^\mu,q_i)$, $i=1,2,3$, $L$ is written as 
\begin{equation}
\label{L_in_p_q_i}
    L=p^\mu\frac{\partial}{\partial y^\mu}+p.\nabla_pe_i\frac{\partial}{\partial q_i}.
\end{equation}

\section{Unraveling the Covariant Diffusion Equation}
\label{app:unravelling}
\subsection{Equivalence Between Fokker-Planck Equation and It\^o SDEs}
\label{app:equiv_itoFP}
The evolution of a stochastic system can be described equivalently in two ways. On the one hand, if one is interested purely in the ensemble properties of the system, the natural description is in terms of the evolution of the probability distribution on the space of the dynamical degrees of freedom ${Z^A}$. In the case of a Markovian, almost surely continuous evolution, the evolution equation for $\rho$ is of the form of a Fokker-Planck equation:
\begin{equation}
\label{eq:non_cov_FPE}
    \partial_\tau \rho = - \partial_A \Big( v^A \rho \Big) + \partial_A \partial_B \Big( K^{AB} \rho \Big) \ ,
\end{equation}

Following Sorkin's argument in \citep{sorkin1986stochastic}, it is possible to gain insight into the meaning of $K^{AB}$ and $v^A$ by using a particular substitution for $\rho$. Take the special case where the particles are prepared at a particular point $Z_0^A$ in the phase space at their zero proper time:
\begin{equation}
\label{delta}
    \rho(Z,\tau=0)=N\delta(Z-Z_0).
\end{equation}
Here $N$ is the total number of particles. Then he computes the average of quantities $Z^A$ and $Z^AZ^B$, and differentiates with respect to $\tau$ at $\tau=0$:
\begin{equation}
    \frac{d \langle Z^A\rangle}{d\tau}|_{\tau=0}=\frac{d}{d\tau}\int_{\mathcal{S}}dX \ \rho(\tau,X)Z^A=\int_{\mathcal{S}}dX \ \frac{\partial\rho}{\partial\tau}Z^A=-\int_{\mathcal{S}}dX \ \partial_B(J^B)Z^A
\end{equation}
Integration by parts and use of \eqref{delta} reveals that
\begin{equation}
\label{dxdta}
    \frac{d \langle Z^A\rangle}{d\tau}\Big|_{\tau=0}=v^A(Z_0).
\end{equation}
In a similar fashion,
\begin{equation}
\label{dxdxdta}
    \frac{d}{d\tau} \Big\langle (Z^A-Z_0^A)(Z^B-Z_0^B)\Big\rangle\Big|_{\tau=0}=2K^{AB}(Z_0).
\end{equation}
Since $Z_0$ was an arbitrary point, the last two equations are valid everywhere in the phase space independent of the choice of the distribution function.
These are essentially the same equations as \eqref{jump_moment_eqs} written in a slightly different form:
\begin{align}
\label{jump_moments_appendix}
   v^A &= \lim_{\delta \tau\to 0}\frac{1}{\delta\tau}\Big\langle \delta Z^A\Big\rangle \notag \\
   K^{AB} &= \frac{1}{2\delta\tau}\Big\langle \delta Z^A \delta Z^B\Big\rangle
\end{align}

If one is interested in a sample realization of the stochastic dynamics, the Fokker-Planck equation is limiting. It is then more convenient to \textit{unravel} the dynamics in terms of individual stochastic trajectories $Z^A(\tau)$. Indeed, there exists an equivalent description of the stochastic process in terms of It\^o SDEs. Here one should determine the differential increments $dZ^A$ along a trajectory. Separating the mean, this can be described by the following SDE:
\begin{equation}
\label{general_sde}
    dZ^A=v^Ad\tau+dW^A.
\end{equation}
Furthermore, to satisfy \eqref{jump_moments_appendix}, the stochastic process $W^A(\tau)$ should satisfy the following statistics
\begin{equation}
\label{noise_correlations}
    \Big\langle dW^A \Big\rangle=0\ , \qquad \Big\langle dW^AdW^B\Big\rangle = 2K^{AB}d\tau \  \ 
\end{equation}
with $dW^A$ being defined as the infinitesimal limit of the increment in $W^A$. We assume sufficient coarse graining so that the dynamics is Markovian, with each increment consisting of numerous independent random kicks. So motivated by the central limit theorem, we take the increments to follow a Gaussian distribution with the above statistics. It is useful to think of $dW_\tau$ as order $\sqrt{d\tau}$, with It\^o's lemma making this precise. In practice, this means that the increments satisfy It\^o's rules:
\begin{equation}
    dW^A dW^B =  2K^{AB} d\tau\ ,\quad dW^A d\tau=0.
\end{equation} 
 
A common, though imprecise, way these equations appear in the physics literature is in the Langevin form:
\begin{equation}
    \partial_\tau Z^A = v^A(Z,\tau) +  \xi^{A}(\tau),
\end{equation}
where $\xi^{A}$, formally the distributional 
time derivative of the stochastic process $W^A(\tau)$, is a collection of uncorrelated Gaussian white noise fields with the following stochastic properties:
\begin{equation}
    \Big\langle\xi^A(\tau)\Big\rangle = 0 \ , \qquad \Big\langle\xi^A(\tau) \xi^B(\tau')\Big\rangle = 2K^{A B} \delta(\tau-\tau') \ .
\end{equation}

Therefore, moving from the Fokker-Planck description to the SDE description is straightforward. After determining the true vector $u^A$ and the true tensor $K^{AB}$, we identify the drift pseudo-vector $v^A$ from \eqref{u_vector}. Then, we can easily write down the corresponding SDE \eqref{general_sde}, together with the correlations of the noise term \eqref{noise_correlations}. In the next appendix, we apply this to the covariant diffusion equation. 

\subsection{From Covariant Diffusion Equation to Stochastic Trajectories}
\label{app:SDE_toSwerves}
Up to zeroth order in a curvature expansion, we have already identified the matrix $K^{AB}$; see \eqref{K_tensor}. The spacetime components of $v^A$ have also been found to be proportional to the momentum. For the momentum components of $v$, we write
\begin{equation}
\begin{split}
    v^{i} &= u^i + \frac{1}{\sqrt{h}} \frac{\partial}{\partial q^j} \left(\sqrt{h} h^{ij} \right) \\
    &= p.\nabla_p e^i +\kappa E\frac{\partial}{\partial q^j}\Big(\frac{\delta^{ij}}{E}+\frac{q^iq^j}{m^2E}\Big) \\
    &= p.\nabla_p e^i +\frac{3\kappa}{m^2}q^i \\
\end{split}
\end{equation}

where in the second line, we have used the representation of the Liouville operator in the $(x^\mu,q_i)$ coordinates \eqref{L_in_p_q_i}. We can already see the similarity between $3\kappa q^i/m^2$ and the bias 4-vector $b^\mu$ in \eqref{b_and_D}. 
With $v^i$ and $K^{ij}$ determined, we can now write the following SDE
\begin{align}
\label{sde_eq_for_dx_dqi}
    dx^\mu&=\frac{1}{m}p^\mu d\tau,\\
    \label{sde_for_dqi_only}
    dq^i&=p.\nabla_pe^i d\tau +\frac{3\kappa}{m^2}q^id\tau+dW^i,
\end{align}
where the noise term satisfies
\begin{equation}
    \Big\langle dW^i \Big\rangle=0\ , \qquad \Big\langle dW^idW^j\Big\rangle = 2\kappa\Big(\delta^{ij}+\frac{1}{m^2}q^iq^j\Big)d\tau
\end{equation}
To make this look manifestly covariant, we next find the differential increment of energy~$E=\sqrt{q^2+m^2}$ using It\^o's Lemma:
\begin{align}
    dE&=\frac{\partial E}{\partial q^i}dq^i+\frac{1}{2}\frac{\partial^2E}{\partial q^i\partial q^j}dq^idq^j\\
    &=\frac{1}{E}q^ip.\nabla_pe_i d\tau+\frac{3\kappa q^2}{Em^2}d\tau+\frac{1}{E}q_idW^i+\frac{3\kappa}{E}d\tau\\
    &=p.\nabla_pe^0d\tau+\frac{3\kappa}{m^2}Ed\tau+dW^0,
\end{align}
where in the last line we have defined
\begin{equation}
\label{dw0_def}
    dW^0\coloneqq \frac{1}{E}q_i dW^i,
\end{equation}
and used the fact that $q^ap.\nabla_pe_a=0$. This holds whenever $q_a=p.e_a$ and the $e_a$ form a complete set of vectors.

Therefore, we can now extend \eqref{sde_for_dqi_only} to include $dq^0$ as well:
\begin{equation}
    dq^a=p.\nabla_pe^a d\tau +\frac{3\kappa}{m^2}q^ad\tau+dW^a,
\end{equation}
such that the noise has the following statistics
\begin{equation}
    \Big\langle dW^a \Big\rangle=0\ , \qquad \Big\langle dW^adW^b\Big\rangle = 2\kappa\Big(\eta^{ab}+\frac{1}{m^2}q^aq^b\Big)d\tau.
\end{equation}
Note that by definition \eqref{dw0_def} satisfies
$q_adW^a=0$. Finally, transforming from $q^a$ coordinates to $p^\mu$ is straightforward by contracting with the tetrad field. If one defines
\begin{equation}
    dW^\mu\coloneqq e_a^{\ \mu}dW^a,
\end{equation}
one can easily check that the above SDE gives the covariant Brownian motion \eqref{stochastic_sw_eq}. We have shown that the covariant diffusion equation at the level of the distribution function is equivalent to the Langevin equation of covariant Brownian motion.

\subsection{Relation Between Spacetime Covariance and Phase Space Covariance}
\label{equiv_between_principles}
Our derivation of the covariant diffusion equation \eqref{SW_Main_Equation} and the corresponding Langevin equation \eqref{stochastic_sw_eq} followed from different lines of reasoning. For the former, we required the Fokker-Planck equation to respect general covariance under admissible coordinate transformations of the phase space $\Gamma_m^+$. For the Langevin equation, however, we required manifest general covariance of the correction to the geodesic equation, together with conservation of mass. In the previous appendix, we showed that the final results are two equivalent descriptions of the same phenomenon. It is illuminating to also find interconnections between the two sets of first principles we used to derive the equations. This is the purpose of this appendix.

First of all, notice that in section \ref{app:equiv_itoFP}, we showed that the matrix tensor $K^{AB}$ is the covariance matrix of the differentials of the motion. On the other hand, in section \ref{SDE_section}, we had $D^{\mu\nu}$ as the covariance matrix of the noise terms in the geodesic equation. To connect the two, we only have to remember that our phase space coordinate system uses a projection on some given tetrad field and only keeps the spatial components of the physical momentum as independent variables. Therefore,
\begin{equation}
    K^{ij}=e^i_{\ \mu}e^j_{\ \nu}D^{\mu\nu}.
\end{equation}
Similarly, in the Fokker-Planck equation, we had a pseudo-vector $v^A$ that we found to be the drift of the differentials of motion. On the other hand, we had the drift 4-vector $b^\mu$ in our discussion of the SDEs. Again, the two are related by a projection on the tetrad field:
\begin{equation}
    v^i=e^i_{\ \mu}b^\mu,
\end{equation}
and this can be used to find the true vector $u$ using \eqref{u_vector}. Now we prove the following two statements:

\vspace{4mm}
\hrule\vspace{10pt} 
\textbf{Statement 1.}
The matrix
\begin{equation}
    K=\begin{pmatrix}
  0 & 0 \\
  0 & K^{ij}\\
\end{pmatrix}
\end{equation}
is a tensor on the phase space $\Gamma_m^+$ if and only if $D^{\mu\nu}$ is a symmetric spacetime tensor that satisfies
\begin{equation}
\label{appendix_p.d=0}
    p_\mu D^{\mu\nu}=0,
\end{equation}
whenever $p^2=-m^2$.
\vspace{2mm}
\hrule\vspace{10pt}

The proof is straightforward. Remember that the set of admissible coordinate transformations of the phase space are
\begin{equation}
    \big(x^\mu,q_i\big)\rightarrow\Big(x'^\alpha(x),\ \Tilde{q}_I(x,q)=\Tilde{e}_I(x).e^a(x)\ q_a\Big),
\end{equation}
which have the following Jacobian in the momentum coordinates
\begin{equation}
    \frac{\partial \Tilde{q}^I}{\partial q^i}=\Tilde{e}^I.e_i+\frac{q_i}{q^0}\Tilde{e}^I.e_0\ .
\end{equation}
The tensor condition on $K$ reduces to
\begin{equation}
\label{eeD=JacobianeeD}
    \Tilde{e}^I_{\ \alpha}\Tilde{e}^J_{\ \beta}D^{\alpha\beta}=\frac{\partial \Tilde{q}^I}{\partial q^i}\frac{\partial \Tilde{q}^J}{\partial q^j}e^i_{\ \mu}e^j_{\ \nu}D^{\mu\nu}.
\end{equation}
With some algebra and using the completeness of the vector fields $e_a$, one can show that
\begin{equation}
\label{momentum_Jac}
    \frac{\partial \Tilde{q}^I}{\partial q^i}e^i_{\ \mu}=\Tilde{e}^I_{\ \mu}+\frac{1}{q^0}\Tilde{e}^I.e_0\ p_\mu\ .
\end{equation}
Therefore, the right-hand side of \eqref{eeD=JacobianeeD} equals the left-hand side, plus three additional terms involving the contraction of $D^{\mu\nu}$ with either $p_\mu$, $p_\nu$, or both. Since this equation has to be valid for any choice of the tetrad fields, one concludes that $p.D=0$. Note that this only needs to be true on the mass shell.

The same argument works for vectors as well. Given a vector field on the spacetime, its projection on the tetrad gives a vector field on the phase space if and only if it is orthogonal to $p$. This will be used in the proof of the next statement.
\vspace{4mm}
\hrule\vspace{10pt}
\textbf{Statement 2.} Assume that $p_\nu D^{\mu\nu}=0$ on the mass shell. Then $u^A$ given by
\begin{equation}
    u^A=\Big(u^\mu=0,u^i\Big)
\end{equation}
is a vector field on the phase space $\Gamma_m^+$ if and only if $b^{\mu}$ is a vector field on the spacetime that satisfies
\begin{equation}
\label{appendix_p.b+D=0}
    p_\mu b^{\mu}+D^{\mu}_{\ \mu}=0,
\end{equation}
whenever $p^2=-m^2$.
\vspace{2mm}
\hrule\vspace{10pt}
For the proof, first let us find $u^i$ in terms of $b$ and $D$ using its definition \eqref{u_vector}:
\begin{equation}
    u^i=e^i_{\ \mu}b^\mu-\frac{1}{\sqrt{h}}\frac{\partial}{\partial q^j}\Big(\sqrt{h}D^{\mu\nu}\Big)e^i_{\ \mu}e^j_{\ \nu}.
\end{equation}
Next we switch from the $\partial/\partial q_i$ basis to $\partial/\partial p_\mu$ using
\begin{equation}
    \frac{\partial}{\partial q^i}=\Big(e_{i\ \mu}+\frac{q_i}{E}e_{0\ \mu}\Big) \frac{\partial}{\partial p_\mu}.
\end{equation}
Using the fact that both $p_\nu D^{\mu\nu}$ and its tangential derivative along the mass shell vanish on the mass shell, one can write $u$ as the projection of a spacetime vector on the tetrad:
\begin{align}
\label{B_vector_appendix}
    u^i&=e^i_{\ \mu}B^\mu,\\
    B^\mu&=b^\mu-\Big(g^{\nu\sigma}+\frac{1}{m^2}p^\nu p^\sigma\Big)\frac{\partial}{\partial p^\sigma}D^\mu_{\ \nu}.
\end{align}

Similar to our proof for the previous statement, since the spacetime components of $u$ are assumed to be zero, the vector condition on $u$ reduces to $B$ being orthogonal to $p$. In order to show this, first note that the following vector field is tangent to the mass shell:
\begin{equation}
    \Big(g^{\nu\sigma}+\frac{1}{m^2}p^\nu p^\sigma\Big)\frac{\partial}{\partial p^\sigma},
\end{equation}
which implies that when acted on $p_\mu D^\mu_{\ \nu}$ gives zero. Therefore, on the mass shell,
\begin{equation}
    p_\mu B^{\mu}=p_\mu b^\mu+D^\mu_{\ \mu}.
\end{equation}
So $u$ is a vector field if and only if the right-hand side is zero on the mass shell. The statement follows.

We highlight that the covariance of the Liouville operator cannot be inferred from this statement, as its spacetime components are non-zero. It requires an independent proof, which we provided in Appendix \eqref{appendixCovLiouville}. Furthermore, using \eqref{B_vector_appendix}, it can be easily shown that the $u$ vector corresponding to \eqref{b_and_D} is identically zero.\\

It is useful to see what curvature corrections to the Brownian motion might look like. Notice that any non-zero vector field on the spacetime introduces a preferred direction in each tangent space so that local Lorentz invariance no longer enforces $u^i$ to be zero. In the presence of curvature, $\nabla^\mu R$ is such a vector field. The following choice satisfies the conditions \eqref{appendix_p.d=0} and \eqref{appendix_p.b+D=0} on the mass shell:
\begin{align}
\label{R2correction}
    D^{\mu\nu}&= \kappa'\Big(g^{\mu\alpha}+\frac{1}{m^2}p^\mu p^\alpha\Big) \Big(g^{\nu\beta}+\frac{1}{m^2}p^\nu p^\beta\Big)R_{;\alpha} R_{;\beta}\ ,\\
    b^\mu&=\frac{1}{m^2}p^\mu D^\nu_{\ \nu}=\frac{\kappa'}{m^2}\Big(R_{;\alpha}R^{;\alpha}+\frac{1}{m^2}\big(p^\alpha R_{;\alpha}\big)^2\Big)p^\mu\ .
\end{align}
It is possible to use $R_{\alpha\beta}$ instead of $R_{;\alpha}R_{;\beta}$ in \eqref{R2correction} to have a first order curvature factor. However, the positive semi-definiteness of the covariance matrix would then require the Ricci tensor to be positive semi-definite over spacelike vectors, which is not true in general. The resulting vector field $u$ for the above choice of the drift and diffusion is
\begin{equation}
\label{R2corrected_u}
    u=-\frac{4}{m^2}p^\alpha R_{;\alpha}\Big(R_{;\mu}+\frac{1}{m^2}p^\beta R_{;\beta}p_\mu\Big)\frac{\partial}{\partial p_\mu},
\end{equation}
which is tangent to the mass shell. It would have been difficult to come up with this vector field starting from the vector condition on $u^i$. The two statements in this appendix pave the way for constructing all possible covariant Brownian motions coupled to the curvature.

\section{Generalized Boltzmann Hierarchy Equations}
The generalized Boltzmann hierarchy provides a systematic way of doing non-relativistic approximation, without the need to assume that the stochastic dark matter is a fluid. It provides a hierarchy of equations both at the background and perturbations level, in which the fluid quantities constitute only the lowest moments of the distribution function. In this appendix, we will see how the moments are defined and what their evolution equations are for the case of stochastic dark matter that is governed by the covariant diffusion equation.

At the background level, the pressure moments are defined as follows \citep{de2021generalized}:
\begin{equation}
\Bar{P}_n = \frac{4\pi}{3} \int_0^\infty dq\ q^2E\Big(\frac{q}{E}\Big)^{2n} f_0  = w_n \Bar{\rho}_{dm}.
\end{equation}
Therefore, the zeroth moment gives the energy density $\Bar{P}_0=\Bar{\rho}_{dm}/3$, and $w_0=1/3$. The first moment is the pressure itself $\Bar{P}_1=\Bar{P}_{dm}$, and $w_1=w_{dm}$ is the equation of state of dark matter. These two quantities are coupled to higher-order moments via their evolution equation. In practice, as we are only interested in the regime $\Gamma\ll H_0$ of the diffusion parameter, even the pressure has a negligible impact on the background dynamics. Nevertheless, writing down the coupled hierarchy equations sets the stage for the analogous calculations for perturbations.

Multiplying the background covariant diffusion equation \eqref{swerves_for_f0} by $q^{2n+2}/E^{2n-1}$ and integrating $dq$ gives
\begin{equation}
\label{bg_GBH}
    \Bar{P}_n' + (2n+3) \mathcal{H} \Bar{P}_n - (2n-1)\mathcal{H}  \Bar{P}_{n+1} = \frac{a\Gamma}{9}\Big(2n(2n+1)m \Bar{N}_{n-1}  -(8n^2-3)m \Bar{N}_n + 2n (2n-1) m\Bar{N}_{n+1}\Big)\ ,
\end{equation}
where $N_n$'s are the higher moment analogs of the number density, defined via
\begin{equation}
\label{N_n_definition}
    \Bar{N}_n = 4\pi \int_0^{\infty} dq\ q^2\Big(\frac{q}{E}\big)^{2n} f_0\ .
\end{equation}
$\Bar{N}_0=\Bar{n}_{dm}$ is the background number density. These number density moments arise from the diffusion term. Now in order to have a closed system of equations, one might try to find the evolution equations for $\Bar{N}_n$, but those will also contain integrals that do not simplify to either $\Bar{P}_n$ or $\Bar{N}_n$. Instead, similar to \eqref{mn=rho-3/2P}, we expand $\Bar{N}_n$ in terms of higher $\Bar{P}_n$ moments. To do so, first use $E^2=m^2+q^2$ to write
\begin{equation}
\label{m_rel_expansion}
    m = -E \sum_{j=0}^{\infty}  \frac{(2j-3)!!}{j!}  \left(\frac{q^2}{2E^2}\right)^j.
\end{equation}
We plug this into \eqref{N_n_definition} to find
\begin{equation}
\label{N_n_series}
    m\Bar{N}_n=-3 \sum_{j=0}^{\infty}  \frac{(2j-3)!!}{2^jj!}  \Bar{P}_{n+j}\ .
\end{equation}
The equations \eqref{bg_GBH} and \eqref{N_n_series} together give an infinite hierarchy of equations for the background. In principle, by including high enough moments in this hierarchy, one could account for relativistic contributions to an arbitrary precision. In practice, we set all the pressure moments beyond some $n_{max}$ to zero.

In the non-relativistic approximation, though, one could directly use the approximate solution \eqref{background_dist_solution} to find all higher-order pressure moments. The result turns out to be
\begin{equation}
\label{w_n_in_terms_of_w}
    w_n\simeq \frac{(2n+1)!!}{3}w_{1}^n\ ,
\end{equation}
where $w_{1}$ is given by \eqref{w_dm_approx}. We use CDM as the initial condition for the stochastic dark matter; therefore, at any non-zero scale factor, we use \eqref{w_dm_approx} and \eqref{w_n_in_terms_of_w} for setting the initial conditions of the hierarchy.

For perturbations, one could similarly define the higher-order density perturbations as~\citep{de2021generalized}:
\begin{equation}
    \delta P_n = \frac{2\pi}{3} \int_0^\infty dq\int_{-1}^1 d\mu\ q^2E\Big(\frac{q}{E}\Big)^{2n} F\  ,
\end{equation}
where the total distribution function has been expanded as $f=f_0+F$, and $\mu=\hat{k}.\hat{q}$, $k$ being the wavevector.
In addition, at the perturbations level, the angular moments of the distribution function become important as well \citep{de2021generalized}: 
\begin{equation}
    (\Bar{\rho}_{dm}+\Bar{P}_{dm})f_{n,\ell} = \frac{2\pi\ell!i^\ell}{(2\ell-1)!!} \int_0^\infty dq\int_{-1}^1 d\mu\ P_\ell(\mu)q^2E\Big(\frac{q}{E}\Big)^{2n+\ell} F \ .
\end{equation}
Therefore, we have a hierarchy characterized by two indices $(n,\ell)$, the Generalized Boltzmann hierarchy. Define the $n$-th order density contrast and velocity divergence as $\delta_n=\delta P_n/\Bar{\rho}_{dm}$ and $kf_{n,1}=\theta_n$, respectively. The lowest order moments correspond to the fluid quantities as follows
\begin{align}
    \delta_0&=\frac{1}{3}\delta_{dm},\ \  \delta_1=\frac{\delta P}{\Bar{\rho}_{dm}}=c_s^2\delta_{dm}\\
    &kf_{0,1}=\theta,\ \  kf_{0,2}=\sigma\ .
\end{align}
To find the coupled equations, we first need to write down the covariant diffusion equation \eqref{SW_Main_Equation} for the perturbed distribution $F$ in FRW:
\begin{align}
\label{sw_pert_spherical}
    F'-&q\mathcal{H}\frac{\partial F}{\partial q}+i\frac{q}{E}k\mu F+\frac{\partial f_0}{\partial q}\big(q\phi'-iEk\mu\psi\big)\nonumber\\&=\frac{a\Gamma m}{3}\frac{1}{q^2}\Big[\psi\frac{\partial}{\partial q}\Big(Eq^2\frac{\partial}{\partial q}f_0\Big)+\frac{\partial}{\partial q}\Big(Eq^2\frac{\partial}{\partial q}F\Big)+\frac{m^2}{E}\frac{1}{\sin\vartheta}\frac{\partial}{\partial \vartheta}\Big(sin\vartheta\frac{\partial}{\partial \vartheta}F\Big)\Big]\ ,
\end{align}
where $\mu=cos\vartheta$. By integrating this equation with suitable pre-factors, we get the evolution equations for $\delta_n$ and $f_{n,\ell}$. Similar to what happened in the background, here the right-hand side of the above equation gives integrals that are not readily in the form of either $\delta_n$ or $f_{n,\ell}$. To have a closed system, we need to insert expansions similar to \eqref{m_rel_expansion} into such integrals. After some algebra, we find 
\begin{align}
\label{delta_n_evol_eq}
    \delta_n' +& \frac{1}{3}(1+w_1)\theta_n + (2n-3w_1)\mathcal{H}\delta_n - (2n-1)\mathcal{H}\delta_{n+1} -\Big( (2n+3)w_n - (2n-1)w_{n+1} \Big)\phi'\nonumber \\&= \frac{a\Gamma}{3} \left(\sum_{j=-1}^{\infty} c_j^{(n)} w_{n+j} \psi- 3\sum_{j=-1}^{\infty} c_j^{(0)} w_j \delta_n + \sum_{j=-1}^{\infty} c_j^{(n)} \delta_{n+j} \right)\ ,
\end{align}
and
\begin{align}
\label{f_nl_evol_eq}
    f&_{n,\ell}' + \Big(2n+\ell+\frac{-5w_1+w_2}{1+w_1}\Big)\mathcal{H}f_{n,\ell} - (2n+\ell-1)\mathcal{H}f_{n+1,\ell} - \frac{\ell^2}{4\ell^2-1}kf_{n+1,\ell-1} +kf_{n,\ell+1}\nonumber \\&-\delta_{\ell,1}\Big(\frac{ (2n+3)w_n - (2n-1)w_{n+1} }{1+w_1}\Big)k\psi= a\Gamma \left(-\sum_{j=0}^{\infty} (c_j^{(0)}+\frac{1}{3}c_{j-1}^{(1)}) \frac{w_j}{1+w_1} f_{n,\ell} + \sum_{j=-1}^{\infty} e_j^{(n,\ell)} f_{n+j,\ell} \right)\ ,
\end{align}
Note that $\delta_{\ell,1}$ is a Kronecker delta. The coefficients are defined by
\begin{equation}
    e_j^{(n,\ell)}=\frac{1}{3}c_j^{(n+\ell/2)}-\ell(\ell+1)\frac{(2j-3)!!   }{(j+1)!2^{j+1}}\ ,
\end{equation}
together with
\begin{equation}
    c^{(n)}_j = -2n(2n+1)\frac{(2j-1)!!}{2^{j+1} (j+1)!} + (8n^2-3)\frac{(2j-3)!!}{2^jj!} - 2n(2n-1)\frac{(2j-5)!!}{2^{j-1} (j-1)!}\ ,
\end{equation}
with the exception that for $j=-1$, only the first and for $j=0$ only the first two terms on the right should be taken into account, and that $(-5)!!=1/3$ and $(-3)!!=-1$.

Whenever we truncate this hierarchy of equations at $(n_{max},\ell_{max})$, we should use an ansatz for higher-order moments. We follow the truncation scheme of \citep{de2021generalized}. As for the initial conditions, we use the same initial conditions as CDM, setting higher moments to zero at the start of numerical integration. Strictly speaking, one should find the superhorizon solution to the distribution function and use it it set the initial conditions as we did for background quantities; See Appendix \ref{app:cs2_super_sol} for the case of $\delta_1$. However, it turns out that even by setting the higher moments to zero, they quickly catch up with their attractor solution, see Figure~\ref{w_plot}.

Finally, one might be worried that the diffusion source term on the right-hand side of \eqref{delta_n_evol_eq} and \eqref{f_nl_evol_eq} that are now appearing as a series of higher-order moments would instantly couple lower-order moments to all of the moments, causing fast propagation of numerical errors. However, as long as the stochastic dark matter particles remain non-relativistic, these higher moments are suppressed, and even restricting oneself to the first two terms of each of these infinite sums results in the same numerical solutions.

\subsection{Superhorizon Solution}
\label{app:cs2_super_sol}
As a simple demonstration of the perturbation equations, let us solve for the perturbed pressure density for small $\Gamma$ in the superhorizon limit $k\ll \mathcal{H}$. $\delta_1$ is already of first order in a $\Gamma/H_0$ expansion. If we are interested in $c_s^2=\delta_1/\delta$, we only need $\delta$ up to zeroth order in $\Gamma$, which is well-known:
\begin{equation}
    \delta'-3\phi'=0\quad\rightarrow\quad \delta=3\phi+const.
\end{equation}
Using adiabatic initial conditions of CDM for stochastic dark matter and ignoring the early-time contribution of neutrinos, the above equation gives the frozen superhorizon solution
\begin{equation}
    \delta=-\frac{3}{2}\phi.
\end{equation}
The evolution equation for $\delta_1$, up to first order in $\Gamma/H_0$, is
\begin{equation}
    \delta_1'+2\mathcal{H}\delta_1=\frac{2}{9}a\Gamma\delta.
\end{equation}
Changing to the $\mathcal{T}$ time variable \eqref{time_variable}, this can be solved:
\begin{equation}
    \delta_1=\frac{2}{9}\frac{\Gamma\mathcal{T} }{a^2}\delta.
\end{equation}
Therefore, comparing to \eqref{w_dm_approx}, the superhorizon $c_s^2$ is
\begin{equation}
\label{cs2=2/9gT/a2}
    c_s^2=\frac{2}{9}\frac{\Gamma\mathcal{T} }{a^2}=\frac{1}{3}w.
\end{equation}
We can see this behavior in Figure \ref{w_plot} briefly before the mode crosses the horizon. Finally, note that for stochastic dark matter, the effective sound speed $c_s^2$ does not have its usual interpretation of the speed of propagation of acoustic waves through dark matter; it is only a measure of typical velocities of dark matter particles at a given time as a result of diffusion. 

\section{Dark Energy Equations}
\label{DEappendix}

As discussed in section \ref{DEmaintext}, at the background level we have $\Bar{P}_x=-\Bar{\rho}_x$. All other imperfect fluid quantities are 1st order in perturbations. Now the balance equation \eqref{balance eq} gives
\begin{equation}
    \Bar{\rho}_x'=-a\Gamma\Bar{\rho}_{dm},
\end{equation}
where a term proportional to $\Gamma w$ is omitted because $w$ itself is linear in $\Gamma$, and terms proportional to $\Gamma^2$ can be ignored in the background since $\Gamma\ll H_0$. Assuming $\Bar{P}_{dm}\ll \Bar{\rho}_{dm}$, one can check that the following is the solution to the above equation in the matter-dark energy era:
\begin{equation}
    \Bar{\rho}_x=\Bar{\rho}_x^0+\frac{2}{3}\Gamma(H-H_0).
\end{equation}
This explicitly confirms that when $\Gamma\ll H_0$, the deviation from a constant dark energy density $\Bar{\rho}_x^0$ is negligible.

For the perturbations, it is crucial to note that the relation $c_x^2=\delta P_x/\delta\rho_x$ holds only in the rest frame of the dark energy. In general, one should account for the velocity of the dark energy fluid \citep{valiviita2008large}
\begin{equation}
    \delta P_x=c_x^2\delta\rho_x+(c_x^2-c_a^2)V_x\Bar{\rho}_x',
\end{equation}
where $c_a^2=w_x=-1$ defines the adiabatic sound speed of dark energy, and $V_x$ appears in the velocity vector of dark energy by $U^\mu=\frac{1}{a}(1-\psi,\partial_iV_x)$. Using our assumption $c_x^2=1$ and the tight coupling between dark matter and dark energy,
\begin{equation}
    \delta P_x=\delta\rho_x+2a\Gamma\Bar{\rho}_{dm}\frac{\theta}{k^2}.
\end{equation}
Furthermore, we write the heat flux vector as $\mathcal{Q}_x^\mu=\frac{1}{a}(0,\Bar{\rho}_x\beta^i)$, and we define $\beta=\partial_i\beta^i$. Note that this is a perturbation quantity. Now, the zeroth component of the balance equation~\eqref{balance eq} gives
\begin{equation}
    \delta_x'+6\mathcal{H}\delta_x+\beta=a\Gamma\frac{\Bar{\rho}_{dm}}{\Bar{\rho}_x}\Big(\delta_x-\delta_{dm}-\psi-6\mathcal{H}\frac{\theta}{k^2}\Big).
\end{equation}
The spatial components of the balance equation determine the dynamics of the heat flux:
\begin{equation}
    \beta'+4\mathcal{H}\beta-k^2\delta_x=a\Gamma\frac{\Bar{\rho}_{dm}}{\Bar{\rho}_x}\Big(\beta+\theta\Big).
\end{equation}
The heat flux of dark energy also enters into Einstein's equations. For example, the $0i$ components of the Einstein equations give
\begin{equation}
    k^2(\phi'+\mathcal{H}\psi)=\frac{3}{2}\mathcal{H}^2\Big(\frac{\Bar{\rho}_x}{\Bar{\rho}_{tot}}\beta+\sum_n\frac{\Bar{\rho}_n}{\Bar{\rho_{tot}}}(1+w_n)\theta_n\Big).
\end{equation}
The summation is over all species. The relativistic Poisson equation also gets modified accordingly \citep{zimdahl2019matter}.

\section{Derivation of Semi-Analytic Solutions in FRW}\label{Appendix_Analytic_sol}
Starting from the covariant diffusion equation \eqref{SW_Main_Equation}, the modified Boltzmann equation for $F$ is
\begin{align}
    \frac{\partial F}{\partial\eta}-&q\mathcal{H}\frac{\partial F}{\partial q}+ik\frac{q}{E}\mu F+\frac{\partial f_0}{\partial q}\Big(q\phi'-iEk\mu\psi\Big)\nonumber\\&=\frac{a\Gamma m^3}{3}\Big[\psi\frac{1}{m^2q^2}\frac{\partial}{\partial q}\big(Eq^2\partial_q f_0\big)+\frac{\partial}{\partial q^j}\Big(\big(\frac{\delta^{ij}}{E}+\frac{1}{Em^2}q^iq^j\big)\frac{\partial}{\partial q^i}F\Big)\Big].
\end{align}
This is the same equation as \eqref{sw_pert_spherical}, but this time the hyperbolic Laplacian has been expanded in Cartesian momentum coordinates instead of spherical. The most important approximation is to assume that the stochastic dark matter particles remain non-relativistic up to the present time. We apply this by writing $E\simeq m$. By ignoring the second term in the inverse metric on the momentum space (which is suppressed by $(q/E)^2$), the second term on the right-hand side simply becomes proportional to the flat Laplacian in the momentum space. Also, our experience with numerical solutions of the generalized Boltzmann hierarchy shows that the first term on the right is negligible compared to the second. Moreover, we focus on the small-scale solutions where $k\gg \mathcal{H}$. Therefore, the term $q\phi'$ is suppressed by both $q/E$ and $\mathcal{H}/k$ compared to $-iEk\mu\psi$, and we can safely ignore it. Now, using the comoving momentum $\Tilde{q}=aq$ and comoving energy $\epsilon=\sqrt{\Tilde{q}^2+m^2a^2}$, the Hubble friction term gets absorbed into the time derivative, and we find
\begin{equation}
    \frac{1}{a^3}\frac{\partial F}{\partial\eta}+\frac{1}{a^4m}i\Vec{k}.\Vec{\Tilde{q}} F-\frac{1}{a^2} \frac{\partial f_0}{\partial \Tilde{q}}im k\mu\psi=\frac{\Gamma m^2}{3}\delta^{ij}\frac{\partial^2}{\partial\Tilde{q}^i\partial\Tilde{q}^j}F.
\end{equation}

The second term on the left-hand side can be absorbed through a redefinition of $F$
\begin{equation}
    \Tilde{F}\coloneqq e^{-ik\mu y(\Tilde{q},\eta)}F.
\end{equation}
Here $y$ is the free streaming length: the comoving length traveled by a particle with comoving momentum $\Tilde{q}$ from $\eta$ to some reference conformal time $\eta_f$. It is only defined through its derivative:
\begin{equation}
    \frac{dy}{d\eta}(\Tilde{q},\eta)=-\frac{\Tilde{q}}{m}\frac{1}{a(\eta)}.
\end{equation}
More on the choice for $\eta_f$ will be said later.
If we also change the time variable to $d\mathcal{T}=a^3d\eta$, the equation for $\Tilde{F}$ becomes
\begin{equation}
    \frac{\partial \Tilde{F}}{\partial\mathcal{T}}-e^{-ik\mu y}\frac{\partial f_0}{\partial\Tilde{q}}ik\mu\frac{m}{a^2}\phi=\frac{\Gamma m^2}{3}\Big(\delta^{ij}\frac{\partial^2}{\partial\Tilde{q}^i\partial\Tilde{q}^j}\Tilde{F}+2i\frac{y}{\Tilde{q}}k^i\frac{\partial}{\partial\Tilde{q}^i}\Tilde{F}-\big(\frac{y}{\Tilde{q}}\big)^2k^2\Tilde{F}\Big).
\end{equation}
Now, to eliminate the term $\partial_{\Tilde{q}^i}\Tilde{F}$ on the right, we introduce yet another momentum variable
\begin{equation}
    \Vec{Q}\coloneqq\Vec{\Tilde{q}}-\frac{2}{3}\Gamma ms(\mathcal{T})i\Vec{k},
\end{equation}
such that $s$ is a time function that is defined by
\begin{equation}
\label{b-definition}
    \frac{ds}{d\mathcal{T}}=-\mathcal{Y}(\mathcal{T}),\ \ \ \ \mathcal{Y}(\mathcal{T})\coloneqq\frac{m}{\Tilde{q}}y.
\end{equation}
Writing $\Tilde{F}$ as a function of $Q$ rather than $\Tilde{q}$ simplifies its equation to 
\begin{equation}
    \frac{\partial \Tilde{F}}{\partial\mathcal{T}}-e^{-ik\mu y}\frac{\partial f_0}{\partial\Tilde{q}}ik\mu\frac{m}{a^2}\phi=\frac{\Gamma m^2}{3}\Big(\delta^{ij}\frac{\partial^2}{\partial Q^i\partial Q^j}\Tilde{F}-\frac{k^2}{m^2}\mathcal{Y}^2\Tilde{F}\Big).
\end{equation}
The last term on the right can be absorbed into the time derivative with yet another redefinition
\begin{equation}
    \doubletilde{F}(Q,\mathcal{T})\coloneqq \Tilde{F}(Q,\mathcal{T})e^{-\frac{1}{3}\Gamma k^2\mathcal{G}(\mathcal{T})},
\end{equation}
where $\mathcal{G}$ is also a time function only defined by its derivative
\begin{equation}
\label{h-defintion}
    \frac{d\mathcal{G}}{d\mathcal{T}}(\mathcal{T})=-\mathcal{Y}(\mathcal{T})^2.
\end{equation}
We also define $\mathcal{G}(\mathcal{T},\mathcal{T}')\equiv -\mathcal{G}(\mathcal{T})+\mathcal{G}(\mathcal{T}')$. For $\doubletilde{F}$ we have
\begin{equation}
\label{Heateq_with_inhom}
    \frac{\partial \doubletilde{F}}{\partial\mathcal{T}}+\mathscr{H}(\mathcal{T},Q)=\frac{\Gamma m^2}{3}\delta^{ij}\frac{\partial^2}{\partial Q^i\partial Q^j}\doubletilde{F},
\end{equation}
This is a simple heat equation with an additional non-homogeneous term:
\begin{equation}
    \mathscr{H}=-e^{-ik\mu y}\frac{\partial f_0}{\partial\Tilde{q}}ik\mu\frac{m}{a^2}\phi.
\end{equation}It can be solved using Duhamel's principle \citep{kadivar2018techniques} by first solving for the homogeneous equation with an arbitrary initial condition:
\begin{equation}
    \begin{aligned}
        \frac{\partial G}{\partial \mathcal{T}} &= \frac{\Gamma m^2}{3} \delta^{ij} \frac{\partial^2}{\partial Q^i \partial Q^j}G, \\
        G\big|_{\mathcal{T}_1}&=\zeta(Q).
    \end{aligned}
\end{equation}
Let us call this solution $G[\mathcal{T},Q;\zeta,\mathcal{T}_1]$. Using the flat heat kernel, this can be written as
\begin{equation}
    G\big[\mathcal{T},Q;\zeta,\mathcal{T}_1\big]=\int d^3Q'\ \frac{e^{-\frac{|Q-Q'|^2}{\frac{4}{3}\Gamma m^2 (\mathcal{T}-\mathcal{T}_1)}}}{\left(\frac{4}{3}\pi\Gamma m^2 (\mathcal{T}-\mathcal{T}_1)\right)^{3/2}} \zeta(Q').
\end{equation}
Duhamel's principle says that the solution to \eqref{Heateq_with_inhom} is the following:
\begin{equation}
    \doubletilde{F}(\mathcal{T},Q)=G\Big[\mathcal{T},Q;\doubletilde{F}\big|_{\mathcal{T}_i},\mathcal{T}_i\Big]+\int _{\mathcal{T}_i}^{\mathcal{T}}d\mathcal{T}'\ G\Big[\mathcal{T},Q;\mathscr{H}\big|_{\mathcal{T}'},\mathcal{T}'\Big].
\end{equation}
Using the heat kernel solution to $G$, one can recover the solution \eqref{integral_solution} to $F$:
\begin{align}
\label{integral_solution_appendix}
F&(\mathcal{T}, k, Q) = \int d^3Q' \left( e^{-\frac{1}{3}\Gamma k^2\mathcal{G}(\mathcal{T},\mathcal{T}_i)} \frac{e^{-\frac{|Q-Q'|^2}{\frac{4}{3}\Gamma m^2 (\mathcal{T}-\mathcal{T}_i)}}}{\left(\frac{4}{3}\pi\Gamma m^2 (\mathcal{T}-\mathcal{T}_i)\right)^{3/2}} e^{-i\Vec{k}.\Vec{y}(Q',\mathcal{T}_i)+i\Vec{k}.\Vec{y}(Q,\mathcal{T})} F(\mathcal{T}_i, k, Q') \right)\nonumber\\ &
+ \int_{\mathcal{T}_i}^{\mathcal{T}} d\mathcal{T}' \int d^3Q' \left( e^{-\frac{1}{3}\Gamma k^2\mathcal{G}(\mathcal{T},\mathcal{T}')} \frac{e^{-\frac{|Q-Q'|^2}{\frac{4}{3}\Gamma m^2 (\mathcal{T}-\mathcal{T}')}}}{\left(\frac{4}{3}\pi\Gamma m^2 (\mathcal{T}-\mathcal{T}')\right)^{3/2}} e^{-i\Vec{k}.\Vec{y}(Q',\mathcal{T}')+i\Vec{k}.\Vec{y}(Q,\mathcal{T})} \frac{m}{a(\mathcal{T}')^2}i\Vec{k}.\Hat{q}' \frac{\partial f_0}{\partial \Tilde{q}} \phi_k(\mathcal{T}')\right).
\end{align}

This might seem like an intractable integral solution. However, the $Q'$ integrals are Gaussian and can be performed exactly; remember that in the non-relativistic approximation, $f_0$ too is Gaussian~\eqref{background_dist_solution}.

Note that in the limit of negligible diffusion, $\Gamma\rightarrow0$, the heat kernel becomes a delta function, identifying $Q'=Q$. Hence, one would recover the integral solution to the CDM distribution function (see \citep{boyanovsky2011small}). For the rest of this appendix, we are interested in the opposite limit: The strong diffusion regime. In this regime, $\Gamma k^2\mathcal{G}(\mathcal{T})\gg1$. As a result, the first term in \eqref{integral_solution_appendix} would be exponentially suppressed. This is just the mathematical formulation that when the diffusion is strong, the initial conditions become unimportant and the dynamics of the stochastic dark matter particles becomes irreversible. Besides, the integrand of the remaining time integral in \eqref{integral_solution_appendix} becomes highly peaked around $\mathcal{T}'\sim\mathcal{T}$, so that one could use a combination of saddle point approximation and numerical techniques to tackle the integral.

\begin{figure}[h!]
    \centering
    \begin{subfigure}[b]{0.49\textwidth}
        \centering
        \includegraphics[width=\textwidth]{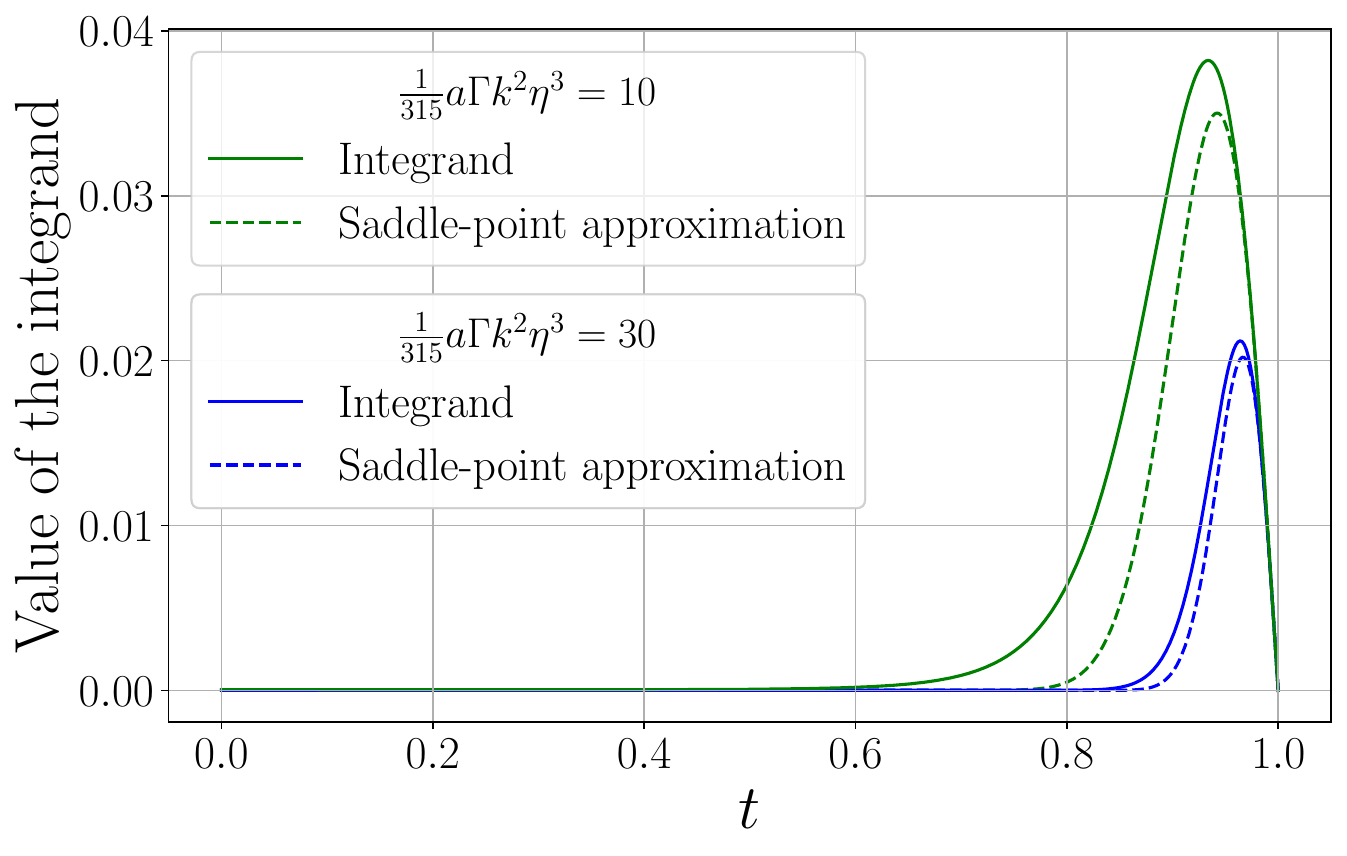}
        \caption{}       \label{fig:delta_integrand}
    \end{subfigure}
    \hfill
    \begin{subfigure}[b]{0.49\textwidth}
        \centering
        \includegraphics[width=\textwidth]{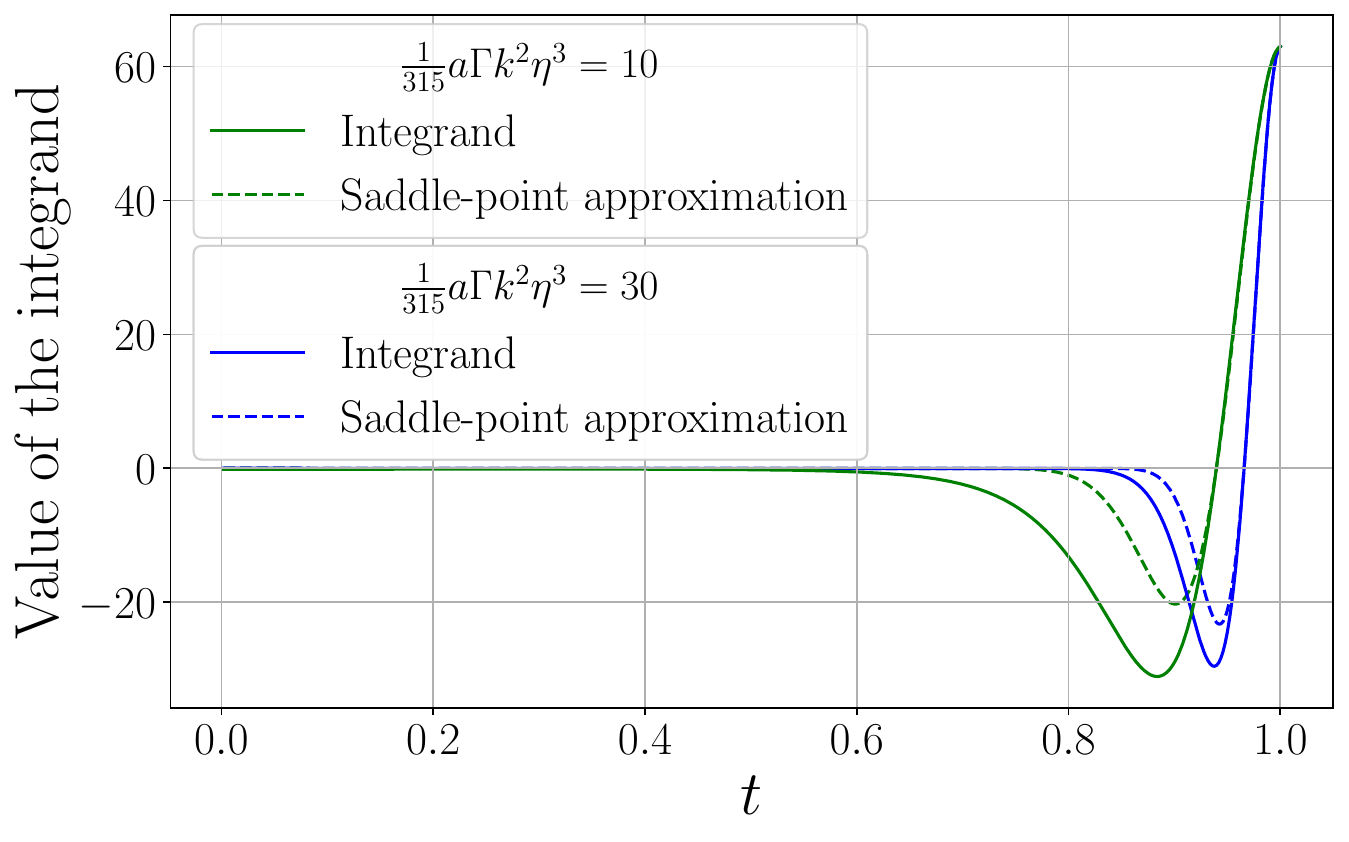}
        \caption{}\label{fig:theta_integrand}
    \end{subfigure}
    \caption{\fontsize{9}{11}\selectfont Comparison of the integrands in \eqref{delta_u_integral} and \eqref{theta_u_integral} with their saddle-point approximations for two different values of $a\Gamma k^2\eta^3$. Note that $t=1$ is not necessarily the saddle point of the whole integrand but just the exponential factor. So what we mean by the saddle-point approximation of the integral is in fact just using the 2nd order Taylor expansion of the exponent $1-6t^5+5t^6\simeq 15(1-t)^2$.}
    \label{fig:integrands}
\end{figure}

Before performing the integrals, we determine the behavior of the time functions $\mathcal{T}$, $\mathcal{Y}$, $s$, and $\mathcal{G}$ during matter domination. Integrating $d\mathcal{T}=a^3d\eta$, one finds
\begin{equation}
       \mathcal{T}=\frac{1}{7}a^3\eta. 
\end{equation}
For simplicity, we set $\eta_f\rightarrow\infty$ to get
\begin{equation}
    y(\Tilde{q},\eta)=\frac{\Tilde{q}}{m}\frac{\eta}{a}.
\end{equation}
Therefore, $\mathcal{Y}(\eta)=\eta/a$. Finally, integrating \eqref{b-definition} and \eqref{h-defintion}, one finds
\begin{equation}
    s(\eta)=-\frac{1}{6}a^2\eta^2,\ \ \ \mathcal{G}(\eta)=-\frac{1}{5}a\eta^3.
\end{equation}
Now everything is ready to perform the integrations. First, we look at the energy density
\begin{equation}
    \delta\rho(\eta,k)=\frac{m}{a^3}\int d^3\Tilde{q}\ F(\eta,k,\Tilde{q}).
\end{equation}
After performing the Gaussian $Q'$ integrals implicit in $F$, the $\Tilde{q}$ integral also turns out to be Gaussian and can be done exactly. Some algebra reveals that
\begin{equation}
\label{delta_ISW_integral}
    \delta(\eta,k)=-k^2\int_{\eta_i}^\eta d\eta'\ \eta'\big(1-\frac{\eta'}{\eta}\big)\phi(\eta',k)e^{-\frac{1}{315}a\Gamma k^2\eta^3\big(1-6(\frac{\eta'}{\eta})^5+5(\frac{\eta'}{\eta})^6\big)}.
\end{equation}
This is where the strong diffusion regime helps us to approximate the integral. It is defined by
\begin{equation}
    \frac{1}{315}a\Gamma k^2\eta^3\gg1.
\end{equation}
In this regime, almost all of the contribution to the integral comes from the region where $\eta'\sim\eta$.
Besides, one can numerically check that $\phi\sim1/\eta$ is a good approximation for the gravitational potential in the strong diffusion regime. Note that this is unlike the usual case where $\phi\sim const$ approximates the behavior of the gravitational potential during matter domination. Using this approximate solution, \eqref{delta_ISW_integral} can be approximated by
\begin{equation}
\label{delta_u_integral}
    \delta(\eta,k)=-(k\eta)^2\phi(\eta,k)\int_{0}^1 dt\ (1-t)e^{-\frac{1}{315}a\Gamma k^2\eta^3\big(1-6t^5+5t^6\big)}.
\end{equation}
Now one can either do a saddle point approximation $1-6t^5+5t^6\simeq 15(1-t)^2$ around $t=1$ and perform a Gaussian integral or instead perform the integral in \eqref{delta_u_integral} numerically for different values of $a\Gamma k^2\eta^3$ and find the power law dependence for large values of $a\Gamma k^2\eta^3$. The result of the latter is 
\begin{equation}
     \delta(\eta,k)\approx\frac{-11.5}{a\Gamma\eta}\phi(\eta,k).
\end{equation}
Notice that this is more accurate than \eqref{delta_dm_asymptotic} which uses the saddle-point approximation. The difference is visible in Figure~\ref{fig:delta_integrand}. By increasing $a\Gamma k^2\eta^3$, one can see that the integrand gets increasingly concentrated at its endpoint, and furthermore, the saddle-point approximation becomes better.

The same numerical approach can be applied to the perturbed pressure 
\begin{equation}
    \delta P =\frac{1}{3m}\frac{1}{a^5}\int d^3\Tilde{q}\ \tilde{q}^2F
\end{equation}
to find
\begin{equation}
    \frac{\delta P}{\Bar{\rho}}\simeq -\Big(1.06+\mathcal{O}\big(\frac{1}{\sqrt{a\Gamma k^2\eta^3}}\big)\Big)\phi.
\end{equation}
Perhaps the most illuminating case is that of the perturbed velocity. Starting from
\begin{equation}
    (\Bar{\rho}+\Bar{P})\theta=\frac{1}{a^4}\int d^3\Tilde{q}\ i(\Vec{k}.\tilde{q})F,
\end{equation}
and performing the Gaussian $Q'$ and $\Tilde{q}$ integrals, one finds
\begin{equation}
\label{theta_u_integral}
    \big(1+w(\eta)\big)\frac{\theta(\eta,k)}{k}=0.016 k\eta\phi(\eta,k)\int_0^1 dt\ t\Big(63-a\Gamma k^2\eta^3\big(\frac{1}{t}-1-t^5+t^6\big)\Big)e^{-\frac{1}{315}a\Gamma k^2\eta^3\big(1-6t^5+5t^6\big)}.
\end{equation}
The interesting thing about this integrand is that due to the polynomial prefactor, it changes sign once near $t=1$. The result of the integral lies in the fine difference between the integral of the positive and negative parts of the integrand. Therefore, if one adopts the saddle point approximation, this fine difference gets lost, and one would even get the sign of the integral wrong. This can be clearly seen in Figure~\ref{fig:theta_integrand}. The true integrand has a larger tail in its negative region, while the saddle-point approximation cannot capture this behavior. As a result, in this case, one has no choice but to use the numerical approach. The result is 
\begin{equation}
    (1+w)\frac{\theta}{k}\approx-48.6\frac{1}{a\Gamma k\eta^2}\phi\ .
\end{equation}

\section{Full Posteriors}
\label{full_posterior_appendix}
\begin{figure}[h!]
    \centering
   \makebox[\textwidth][c]{\includegraphics[width=1.12\textwidth]{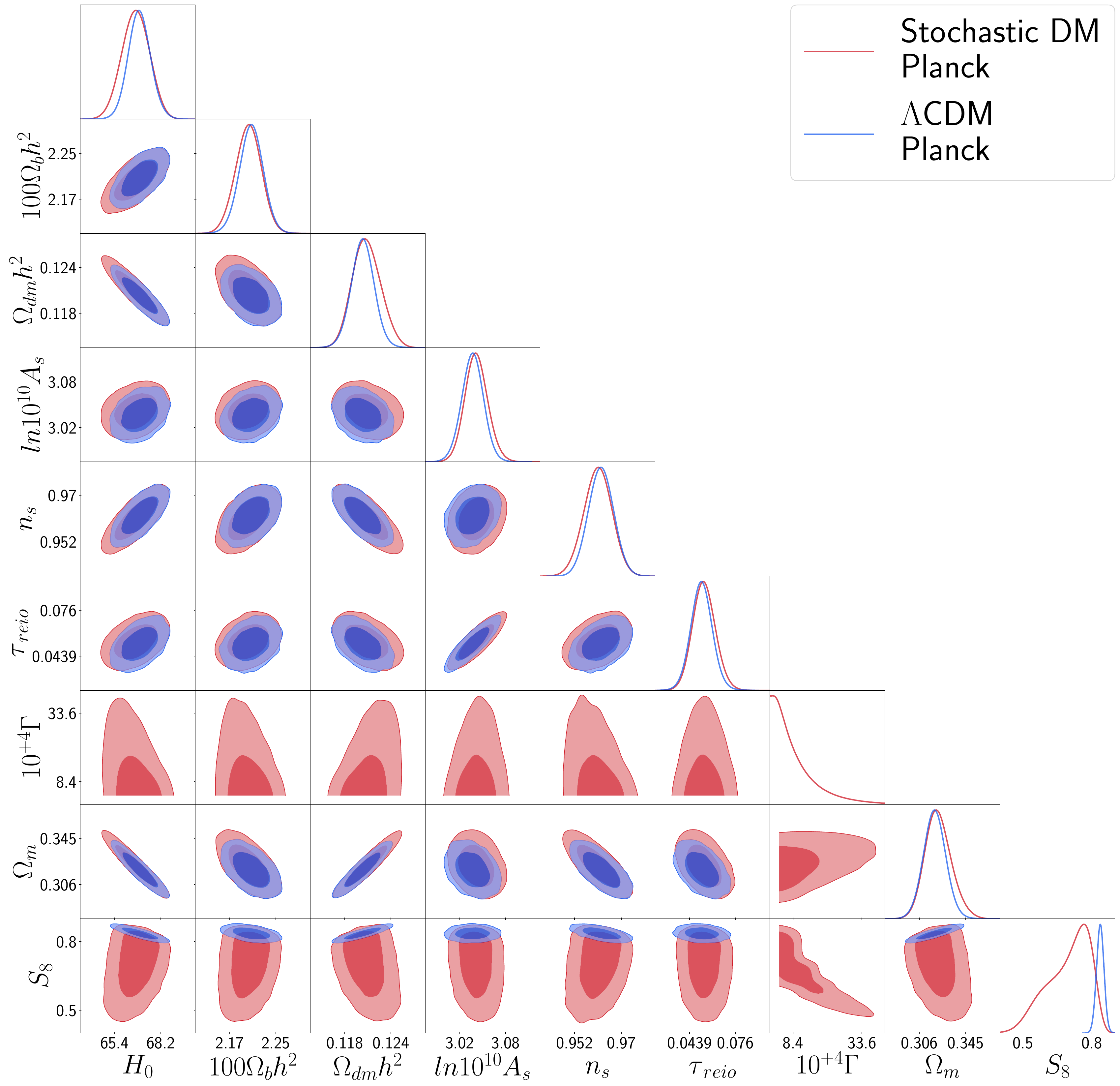}}
    \caption{\fontsize{9}{11}\selectfont 68\% and 95\% confidence-level posterior distributions of all cosmological parameters for stochastic dark matter and $\Lambda$CDM from Planck. $\Gamma$ specifically target structures at small scales, giving a very wide $S_8$ posterior while leaving other cosmological parameters with roughly the same constraints.}
    \label{fig:Swerves_LCDM_Planck}
\end{figure}

\begin{figure}[h!]            \centering
\makebox[\textwidth][c]{\includegraphics[width=1.12\textwidth]{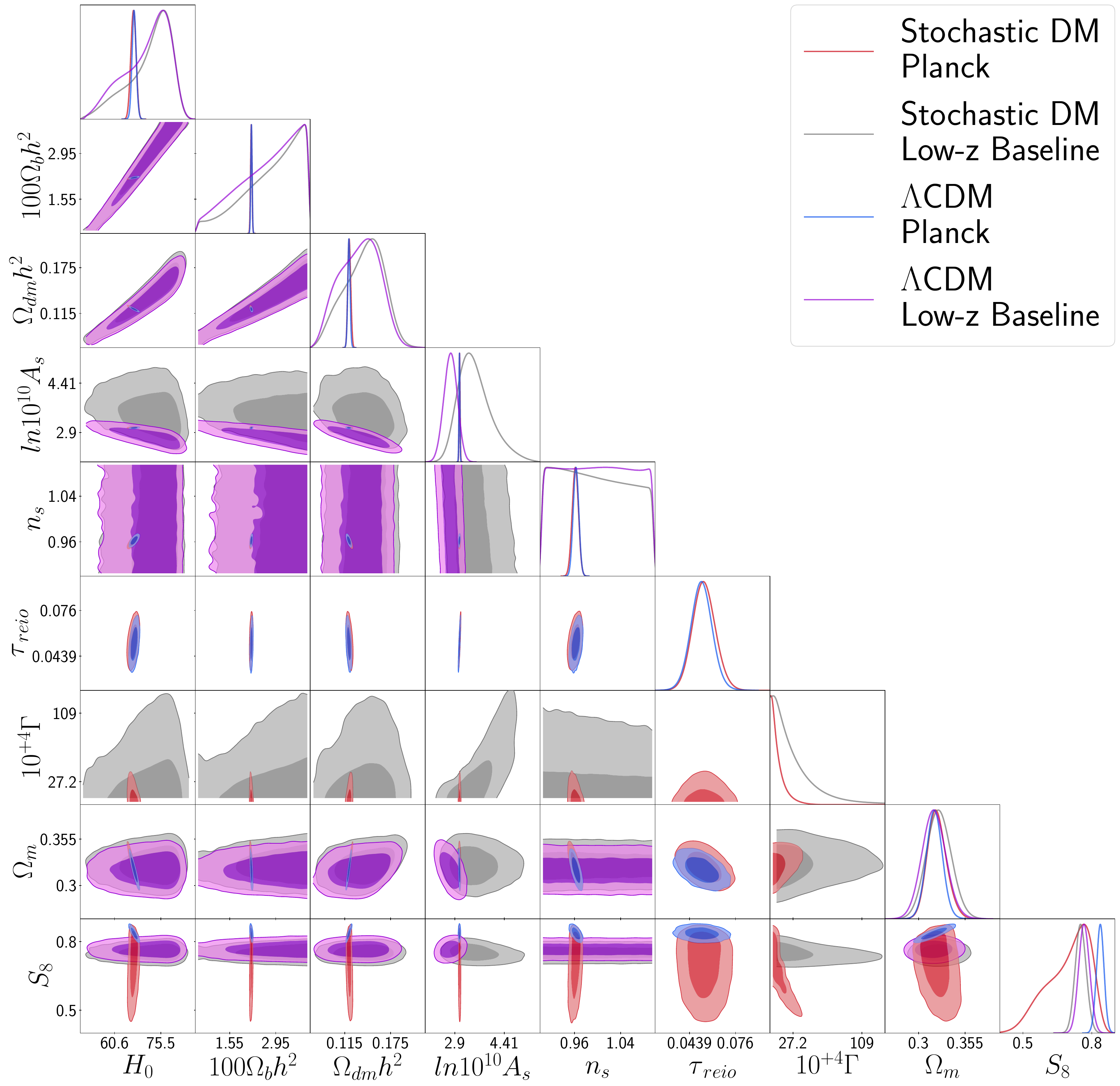}
}
    \caption{\fontsize{9}{11}\selectfont 68\% and 95\% confidence-level posterior distributions of all cosmological parameters for stochastic dark matter and $\Lambda$CDM, from Planck and low-z Baseline. The constraints on the amplitude of primordial fluctuations, $A_s$, are very loose from the low-z Baseline. As a result, for stochastic dark matter, $\Gamma$ and $A_s$ are positively correlated: larger $\Gamma$ strongly dampens the fluctuations in dark matter, which gets compensated by a larger primordial amplitude $A_s$ to yield the same $S_8$. So $S_8$ has no correlation with $A_s$ or $\Gamma$. In contrast, Planck data puts a strong constraint on $A_s$ and therefore yields a stronger constraint on $\Gamma$ as well. Since these values of $\Gamma$ cannot get compensated by $A_s$, they still significantly decrease $S_8$, leading to a very broad posterior for $S_8$.}
    \label{fig:StDM_LCDM_Lowz_Planck}
\end{figure}

\begin{figure}[h!]
    \centering
    \makebox[\textwidth][c]{\includegraphics[width=1.12\textwidth]{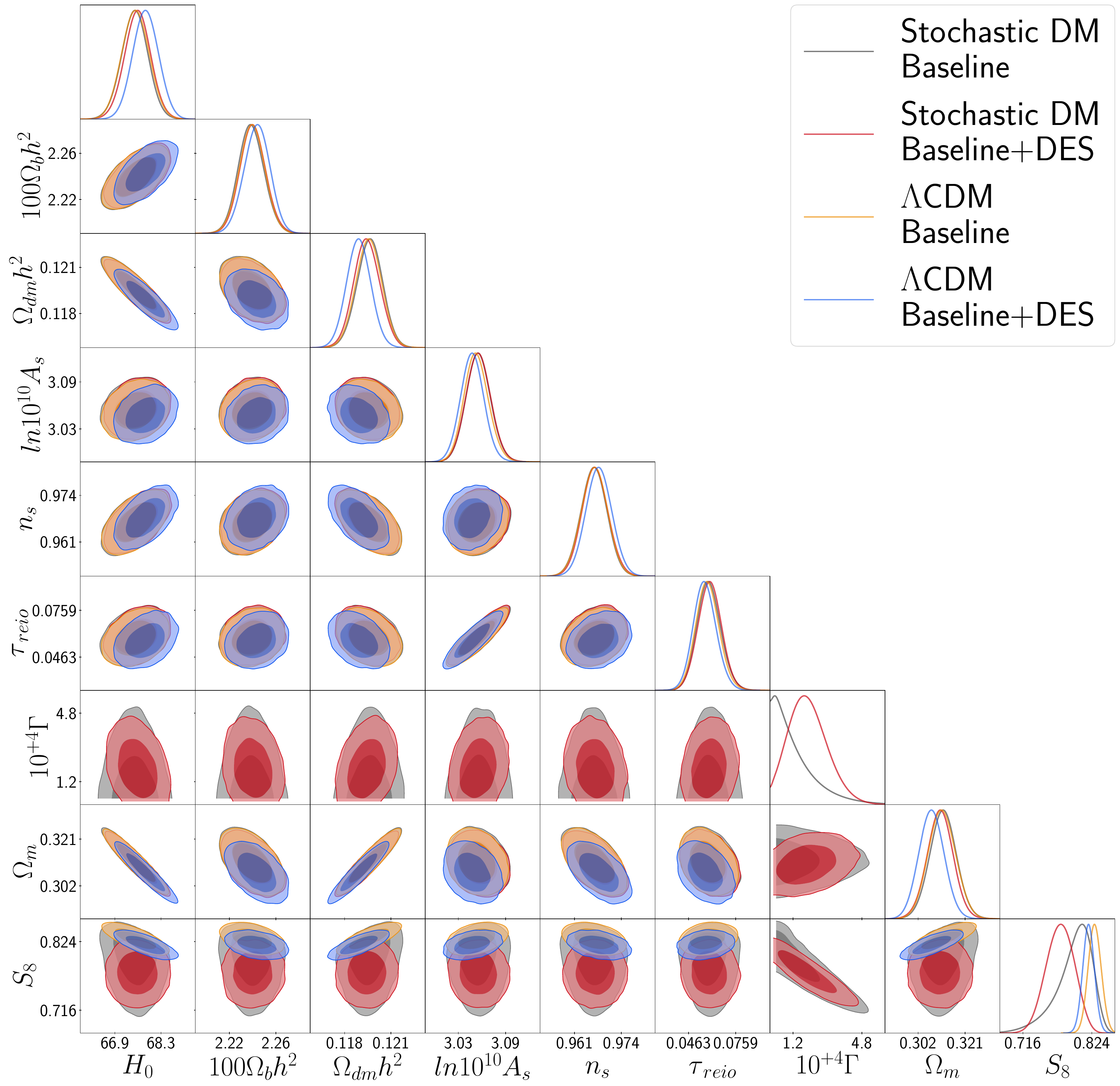}}
    \caption{\fontsize{9}{11}\selectfont 68\% and 95\% confidence-level posterior distributions of all cosmological parameters for stochastic dark matter and $\Lambda$CDM, from Baseline and Baseline+DES. In the case of CDM, decreasing $S_8$ to satisfy the DES prior is obtained by very slight shifts in $\Omega_m$ and $A_s$ to smaller values; CDM provides no other mechanism for less structure formation. The stochastic dark matter naturally does this by a non-zero $\Gamma$.}
    \label{fig:StDM_LCDM_Baseline_All}
\end{figure}

We present the 68\% and 95\% confidence levels of marginalized posteriors for the 7 main parameters used in our MCMC analysis, alongside $\Omega_m$ and $S_8$ for illustrative purposes. The plots include both stochastic dark matter and $\Lambda$CDM. Figure~\ref{fig:Swerves_LCDM_Planck} is based on Planck2018 data only. Figure~\ref{fig:StDM_LCDM_Lowz_Planck} compares the Planck2018 posteriors to the Low-z Baseline. Figure~\ref{fig:StDM_LCDM_Baseline_All} contrasts the Baseline with Baseline+DES, thereby highlighting the effect of the $S_8$ prior from DES.

\newpage
\newpage
\newpage

\end{appendices}

\bibliographystyle{jhep}
\bibliography{main.bib}
\end{document}